\documentclass[11pt,a4paper]{article}
\usepackage{jheppub}
\usepackage{bbm}
\usepackage{pdfpages}

\input pix.sty

\newcommand{\sumint}[1]{\mbox{$\sum$}\!\!\!\!\!\!\!\int_{#1}}

\newcommand{\dd}{\mathrm{d}}
\newcommand{\tinymsbar}{{\overline{\mbox{\tiny\rm{MS}}}}}

\newcommand{\Nc}{N_{\rm c}}

\newcommand{\gB}{g_\rmii{B}}

\newcommand{\rmO}{{\mathcal{O}}}
\newcommand{\bmu}{\bar\Lambda} 

\def\lsi{\raise0.3ex\hbox{$<$\kern-0.75em\raise-1.1ex\hbox{$\sim$}}}
\def\gsi{\raise0.3ex\hbox{$>$\kern-0.75em\raise-1.1ex\hbox{$\sim$}}}

\newcommand{\rmii}[1]{{\mbox{\tiny\rm{#1}}}}

\newcommand{\im}{\mathop{\mbox{Im}}}

\newcommand{\Tint}[1]{{\hbox{$\sum$}\!\!\!\!\!\!\!\int\,}_{\!\!\!\!\raise-0.9ex\hbox{$\scriptstyle{#1}$}}}
\newcommand{\Tinti}[1]{{{\Sigma}\!\!\!\!\raise0.3ex\hbox{$\int$}_\rmii{${#1}$}}}
\renewcommand{\Tint}[1]{\sumint{#1}}

\newcommand{\bi}{\begin{itemize}}
\newcommand{\ei}{\end{itemize}}

\newcommand{\hide}[1]{ }

\renewcommand{\i}{i}
\renewcommand{\v}{ii} 
\newcommand{\iv}{iii} 

\newcommand{\Jt}[2]{\mathcal{J}_\rmii{#1}^\rmii{#2}}
\newcommand{\It}[2]{\mathcal{I}_\rmii{#1}^\rmii{#2}}

\allowdisplaybreaks

\title{The shear channel spectral function in hot Yang-Mills theory}

\author[1]{Yan Zhu}
\author[1]{and Aleksi Vuorinen}

\affiliation[1]{Faculty of Physics, University of Bielefeld, D-33501 Bielefeld, Germany}

\emailAdd{yzhu@physik.uni-bielefeld.de}
\emailAdd{vuorinen@physik.uni-bielefeld.de}

\abstract{We determine a next-to-leading order result for the shear channel thermal spectral function in SU($N$) Yang-Mills theory, working in the limit of vanishing external three-momentum. The result is subsequently applied to the evaluation of the corresponding imaginary time correlator, and its use in the context of sum rules is discussed. Our hope is that the calculation will eventually find use in the nonperturbative determination of the shear viscosity of the theory.}


\begin{document}

\rightline{BI-TP 2012/52}

\rightline{INT-PUB-12-063}

\maketitle

\section{Introduction}\label{intro}

In recent years, a lot of effort has been put in the determination of various transport coefficients of the quark gluon plasma (QGP), characterizing its relaxation towards thermal equilibrium. This work has been largely motivated by the experimental heavy ion programs taking place at RHIC and the LHC \cite{Tannenbaum,Muller}, which have highlighted the importance of not only understanding the bulk equilibrium properties of the produced matter but also its hydrodynamic behavior. One particularly prominent observation in this direction has been the discovery of the sizable effect that the small, yet finite value of the shear viscosity $\eta$ has on the flow properties of the plasma \cite{Romatschke:2009im,Shen:2011zc}.

Unfortunately, the first principles determination of transport coefficients in an interacting quantum field theory is a notoriously hard problem. In the case of heavy ion physics, a further complication arises from the fact that the matter produced at RHIC has been observed to behave more like a strongly coupled liquid than a gas of weakly interacting quasiparticles, clearly invalidating its description in terms of perturbative methods (which furthermore are extremely difficult in this context, cf.~e.g.~\cite{Aarts:2002cc,Arnold:2003zc}). At the same time, a reliable nonperturbative determination of transport coefficients, in principle available from the zero frequency limit of the corresponding (Minkowskian) spectral functions, has proven extremely challenging, as the lattice approach is restricted to the Euclidean formulation of the theory \cite{Meyer:2007ic,Meyer:2007dy,Meyer:2011gj}. As a result, a considerable amount of attention has shifted towards the gauge/gravity duality, which has in particular led 
to the famous claim of the QGP being a nearly `ideal' fluid, with a shear viscosity to entropy ratio close to the conjectured limit $\eta/s\geq 1/(4\pi)$ \cite{Kovtun:2004de} (see also \cite{Kovtun:2011np,Rebhan:2011vd} for some interesting recent developments).

Clearly, it would be desirable to be able to determine parameters such as the shear viscosity of the QGP using nonperturbative tools but staying within the physical theory. To this end, an ambitious program was initiated in \cite{Burnier:2011jq}, where the authors tested a numerical recipe for the analytic continuation of Euclidean lattice data to Minkowskian signature. This method in practice amounts to the inversion of an integral relation between the spectral function $\rho(\omega)$ and the corresponding imaginary time correlator $G(\tau)$,
\begin{equation}
 G(\tau) =
 \int_0^\infty
 \frac{{\rm d}\omega}{\pi} \rho(\omega)
 \frac{\cosh\Big[\! \left(\frac{\beta}{2} - \tau\right)\omega\Big]}
 {\sinh\frac{\beta \omega}{2}}\, ,\quad \quad 0<\tau <\beta \, ,  \la{int_rel}
\end{equation}
constituting an in principle well defined but numerically extremely challenging task. While early results have been encouraging (see also \cite{Burnier:2012ts,Burnier:2012ze}), they have also highlighted the necessity of subtracting short-distance divergences from the Euclidean correlators and in general obtaining as much analytic information on their behavior as possible. This includes not only determining the temperature independent ultraviolet (UV) singularities, available from $T=0$ perturbation theory (see e.g.~\cite{Zoller:2012qv}), but also the leading temperature dependent terms. The latter quantities are most conveniently obtained from the behavior of the corresponding spectral functions at $\omega\sim \pi T$, computable using finite-temperature perturbation theory.

For the bulk channel of SU($N$) Yang-Mills theory, including operators such as the trace of the energy momentum tensor, the task of perturbatively determining the spectral function was undertaken in \cite{Laine:2011xm}. There, the quantity was evaluated to next-to-leading order (NLO) in the limit of vanishing external three-momentum, building on the earlier work of \cite{Laine:2010fe,Laine:2010tc}. In the paper at hand, our goal is to extend this result to the perhaps even more interesting but also technically more complicated shear channel, which necessitates the generalization of the methods of \cite{Laine:2011xm} to a number of new master integrals, involving components of the loop three-momenta in the numerator. As a preparation for this, the UV limit of the shear correlator was analyzed in terms of an Operator Product Expansion (OPE) in \cite{Schroder:2011ht}, leading to results compatible with known sum rules \cite{Romatschke:2009ng,Meyer:2010gu} as well as the arguments of \cite{CaronHuot:2009ns}. Our 
hope is that the full NLO thermal spectral function will eventually find use in the extraction of the shear viscosity, 
\ba
\eta &=& \lim_{\omega\to 0} \fr{\rho_\eta(\omega)}{\omega}\, ,
\ea
from lattice data. In addition, our results can immediately be used to evaluate the Euclidean imaginary time correlator, a quantity that has been addressed in several recent lattice and AdS/CFT calculations in both the shear and bulk channels (see e.g.~\cite{Huebner:2008as,Iqbal:2009xz,Springer:2010mf,Springer:2010mw,Kajantie:2010nx,Kajantie:2011nx}, some of which also address spatial correlators).

The outline of our paper is as follows. In Section 2, we present our setup and introduce the necessary notations, while Section 3 contains an overview of how our computation was performed. Section 4 then lays out our result for the spectral function and discusses its implications for the Euclidean imaginary time correlator and sum rules, after which we draw our conclusions in Section 5. Most details of our calculation, including the definitions of all master integrals, can be found from appendices \ref{masters} and \ref{details}.

\section{Setup}\label{setup}

The setting of our calculation is very similar to that introduced in detail in \cite{Laine:2011xm,Schroder:2011ht}, and for simplicity we adopt our notation from these references. We work within pure SU($N$) Yang-Mills theory at a nonzero temperature $T$, defined by the dimensionally regularized Euclidean action
\ba
 S_\mathrm{E} &=& \int_{0}^{\beta} \! \dd \tau \int \! {\rm d}^{3-2\epsilon}\vec{x}
 \, \frac{1}{4} F^a_{\mu\nu} F^a_{\mu\nu} \, \equiv \, \int_x \, \frac{1}{4} F^a_{\mu\nu} F^a_{\mu\nu} \,, \label{action}
\ea
with $\beta\equiv 1/T$ and
\ba
F^a_{\mu\nu} &\equiv&  \partial_\mu A^a_\nu - \partial_\nu A^a_\mu + \gB f^{abc} A^b_\mu A^c_\nu\, .
\ea
It is straightforward to see from here that the energy-momentum tensor of the theory takes the form
\ba\la{eq:T}
 T_{\mu\nu} &=& \frac{1}{4} \delta_{\mu\nu} F^a_{\alpha\beta} F^a_{\alpha\beta} -F^a_{\mu\alpha} F^a_{\nu\alpha} \, ,
\ea
the (connected) correlation functions of which we set out to compute
\ba
 G_{\mu\nu,\alpha\beta}(x)  &\equiv& \langle T_{\mu\nu}(x)\: T_{\alpha\beta}(0)\rangle_c\, .
\label{eq:corr_def_coord}
\ea
In momentum space, the corresponding Euclidean correlator takes the form
\ba
 \tilde G_{\mu\nu,\alpha\beta}(P) &\equiv& \int_x e^{-iP\cdot x} G_{\mu\nu,\alpha\beta}(x) \, ,
\label{eq:corr_def_mom}
\ea
where the Fourier transform is taken in $D=4-2\epsilon$ dimensions and from which the desired spectral functions can be obtained,
\ba
 \rho_{\mu\nu,\alpha\beta} (\omega) 
 &\equiv& \im \Bigl[ \widetilde{G}_{\mu\nu,\alpha\beta}(P) 
 \Bigr]_{P \to (-i[\omega + i 0^+],\vec{0})}
 \;. \label{rho_general}
\ea

As discussed in some length in \cite{Schroder:2011ht}, to evaluate the momentum space shear correlator $\tilde G_{12,12}(P)$ within dimensional regularization, it is convenient to introduce the projector
\ba
X_{\mu\nu,\alpha\beta} &\equiv& P_{\mu\nu}^{T}P_{\alpha\beta}^{T}
	-\frac{D-2}{2}(P_{\mu\alpha}^{T}P_{\nu\beta}^{T}+P_{\mu\beta}^{T}P_{\nu\alpha}^{T}) \,,\label{eq:xproj}
\ea
where $P_{\mu\nu}^{T}(P)$ is a symmetric transverse projector orthogonal to the four-vectors $P$ and $U\equiv(1,\mathbf{0})$, satisfying $P_{00}^T=0=P_{0i}^T$, $P_{ij}^T = \delta_{ij}-\frac{p_i p_j}{\mathbf{p}^2}$ (for convenience, we keep $\mathbf{p}$ nonzero here). Using this definition and choosing the spatial momentum $\mathbf{p}$ to point in the $x_{D-1}$ direction, we obtain
\begin{equation}
    X_{\mu\nu,\alpha\beta}\: \tilde G_{\mu\nu,\alpha\beta}(P) = -D(D-2)(D-3)\: \tilde G_{12,12}(P) \,.
\end{equation}
This suggests that if we define the momentum space shear channel correlator through
\ba
 \tilde G_\eta(P) \equiv 2 X_{\mu\nu,\alpha\beta} \, \tilde G_{\mu\nu,\alpha\beta}(P)\, , \label{eq:g_eta_p}
\ea
then in the $\epsilon \to 0$  limit the corresponding ($\mathbf{p}= 0$) spectral function $\rho_\eta(\omega)$ satisfies
\ba
\rho_\eta(\omega)&=&-16\,\rho_{12,12}(\omega)\, . \label{sheardef}
\ea

Finally, we note that our calculation will be carried out within the $\msbar$ renormalization scheme. The sum-integral measure we use reads
\ba
\Tint{Q} &\equiv& T \sum_{q_0} \int_\vec{q}, \quad \int_\vec{q}\, \equiv \,\int\! \frac{{\rm  d}^{D-1}\vec{q}}{(2\pi)^{D-1}} \,=\, \Lambda^{-2\epsilon} \left( \frac{{\rm e}^{\gamma_E}\bar\Lambda^2}{4\pi}\right)^{\epsilon}
\int\! \frac{{\rm  d}^{D-1}\vec{q}}{(2\pi)^{D-1}} \; ,
\ea
where $\Lambda$ and $\bar\Lambda$ stand for the renormalization scales in the MS and $\msbar$ schemes, respectively.

\section{Calculations}\label{calculations}

\begin{figure}
\centering
\includegraphics[width=10cm]{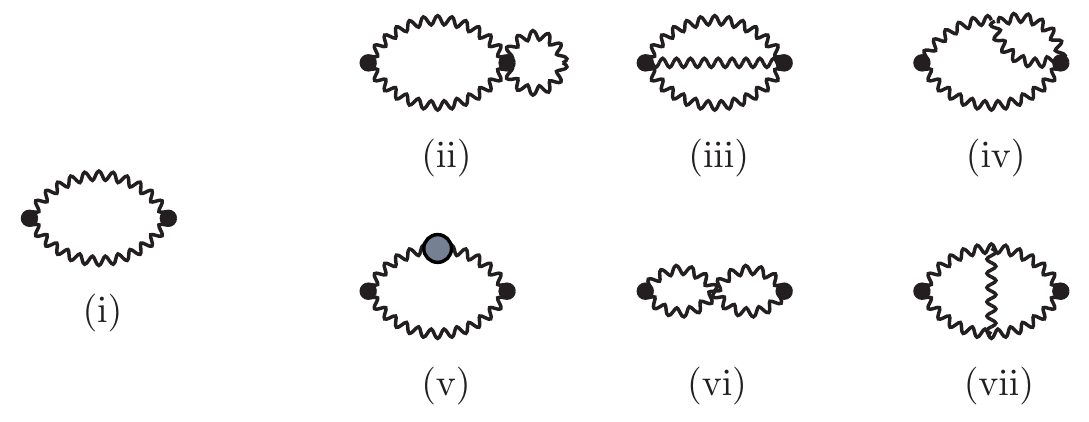}
\caption{The one- and two-loop Feynman diagrams contributing to the NLO shear spectral function in SU($N$) Yang-Mills theory, with the curly line corresponding to the gluon field. The grey blob in (v) denotes the one-loop gluon self energy.}
\label{figs}
\end{figure}

We begin the determination of the NLO spectral function $\rho_\eta(\omega)$ from the Euclidean momentum space correlator $\tilde{G}_\eta(P)$, which can be seen to obtain contributions from the one- and two-loop graphs displayed in fig.~\ref{figs}. Taking advantage of eq.~(3.2) of \cite{Schroder:2011ht} and dropping contributions that depend on the external momentum in a trivial way and thus don't contribute to the spectral function, we see that this quantity can be written in the form
\be
\frac{\rho_\eta (\omega)}{4 d_A \Lambda^{2\epsilon}} =
\rho^{ }_{\Jt{b}{}} (\omega) + \gB^2 \Nc\big\{\rho^{ }_{\It{b}{ }} (\omega)+ \rho^{ }_{\It{d}{ }} (\omega)+ \rho^{ }_{\It{f}{ }} (\omega)+ \rho^{ }_{\It{h}{ }} (\omega)+ \rho^{ }_{\It{i}{ }} (\omega)+ \rho^{ }_{\It{j}{ }} (\omega)\big\}\, , \label{rhoeta1}
\ee
where $d_A = N_c^2-1$ and we have denoted
\ba
\rho^{ }_{\Jt{b}{ }} (\omega) &\equiv& -\fr{D(D-2)(D-3)}{8}\rho^{ }_{\Jt{b}{0}} (\omega) -D(D-3)\rho^{ }_{\Jt{b}{1}} (\omega) \nn
&-& (D-2)(D-3)\rho^{ }_{\Jt{b}{2}} (\omega)\;,
\\
\rho^{ }_{\It{b}{ }} (\omega)&\equiv&\fr{D(D-2)(D-3)}{2}\bigg(2 \rho^{ }_{\It{b}{1}} (\omega) +  \rho^{ }_{\It{b}{2}} (\omega) \bigg)
\;,
\\
\rho^{ }_{\It{d}{ }} (\omega)&\equiv&D(D-2)(D-3)\bigg( \rho^{ }_{\It{d}{1}} (\omega) + 3 \rho^{ }_{\It{d}{2}} (\omega) \bigg)
+ \fr{(D-2)^2(D-3)}{4}
\nn&\times&\bigg( D \rho^{ }_{\It{d}{0}} (\omega) + 8 \rho^{ }_{\It{d}{3}} (\omega)\bigg)
\;, \\
\rho^{ }_{\It{f}{ }} (\omega)&\equiv&-\fr{3D(D-2)(D-3)}{4} \rho^{ }_{\It{f}{1}} (\omega)
\;, \\
\rho^{ }_{\It{h}{ }} (\omega)&\equiv&-\fr{D(D-2)(D-3)}{4}\bigg(4 \rho^{ }_{\It{h}{0}} (\omega) + 4 \rho^{ }_{\It{h}{1}} (\omega) -  \rho^{ }_{\It{h}{3}} (\omega) \bigg) 
+ \fr{D(D-2)}{2}
\nn &\times& \bigg( 2 \rho^{ }_{\It{h}{4}} (\omega) + \rho^{ }_{\It{h}{5}} (\omega) - 4 \rho^{ }_{\It{h}{6}} (\omega) - 2 \rho^{ }_{\It{h}{7}} (\omega)\bigg) -4D(D-3)\rho^{ }_{\It{h}{2}} (\omega)
\;,
\\
\rho^{ }_{\It{i}{ }} (\omega)&\equiv&-\fr{D(D-2)(D-3)}{2}\bigg( 2 \rho^{ }_{\It{i}{1}} (\omega) + 2 \rho^{ }_{\It{i}{2}} (\omega) +  \rho^{ }_{\It{i}{3}} (\omega) \bigg) \;,
\\
\rho^{ }_{\It{j}{ }} (\omega)&\equiv&\fr{D(D-2)(D-3)}{4}\rho^{ }_{\It{j}{0}} (\omega)
+D(D-3)\bigg(2\rho^{ }_{\It{j}{1}} (\omega)+\rho^{ }_{\It{j}{2}} (\omega)\bigg) 
 \nn   
&+&\fr{12-16D+3D^2}{2}\bigg(2\rho^{ }_{\It{j}{3}} (\omega)+\rho^{ }_{\It{j}{4}} (\omega)\bigg)-D(D-6)\bigg(\rho^{ }_{\It{j}{5}} (\omega)+2\rho^{ }_{\It{j}{6}} (\omega)\bigg)
  . \;\; \label{Rhomasters}
\ea
The definitions of the various `master spectral functions' appearing here can be read off from appendix \ref{definitions}; note that our notation for them differs somewhat from that of \cite{Schroder:2011ht}. The primary computational task of the present paper will be to derive explicit results for these quantities, evaluating the divergent terms analytically and the most complicated finite parts numerically.

The general techniques required in the computation of the above masters were developed and rather thoroughly explained in \cite{Laine:2011xm}; in fact, many of the results needed in the present calculation can be directly read off from this reference. The main complication involved in the evaluation of the masters only appearing in the shear channel is related to how one should deal with the projection operators $P_T(Q)\equiv Q_\mu Q_\nu P^T_{\mu\nu}(P)$, which appear frequently in the definitions of appendix \ref{definitions}. Here, a crucial simplification appears due to setting the external three-momentum $\mathbf p$ to zero, which allows us to average the results over its direction, applying the simple relations 
\ba
\left<\mathbf{\hat p}_i\mathbf{\hat p}_j\right>_\mathbf{\hat p}&=&\frac{1}{D-1}\delta_{ij}\, , \\
\left<\mathbf{\hat p}_i\mathbf{\hat p}_j\mathbf{\hat p}_k\mathbf{\hat p}_l\right>_\mathbf{\hat p}&=&\frac{1}{D^2-1}\left(\delta_{ij}\delta_{kl} +\delta_{ik}\delta_{jl} +\delta_{il}\delta_{jk}\right)\, .
\ea
This leads to the replacement rules (with $Q$ and $R$ standing for $D$-dimensional loop momenta here)
\ba
P_T(Q)\rightarrow\frac{D-2}{D-1}q^2\,,\hspace{2em} P^2_T(Q)\rightarrow\frac{D(D-2)}{D^2-1}q^4\,,\nn
P_T(Q)P_T(R)\rightarrow\frac{D^2-2 D-2}{D^2-1}q^2r^2+\frac{2}{D^2-1}(\mathbf{q}\cdot\mathbf{r})^2\, , \label{PTrels}
\ea
which will prove extremely useful in the following.

Next, we introduce the notation 
\ba
 \rho^{ }_{\mathcal{J}^{n}_\rmii{x}}(\omega) \equiv 
 \int_{\vec{q}}f_{\mathcal{J}^{n}_\rmii{x}}\;, 
 \hspace{3em}
 \rho^{ }_{\mathcal{I}^{n}_\rmii{x}}(\omega) \equiv 
 \int_{\vec{q,r}}f_{\mathcal{I}^{n}_\rmii{x}}\; , \label{fdef}
\ea
in which the one- and two-loop functions $f_{\mathcal{J}^{0}_\rmii{x}}$ and $f_{\mathcal{I}^{0}_\rmii{x}}$, corresponding to the integrals encountered in the bulk case, can be read off from \cite{Laine:2011xm} and are for completeness reproduced in appendix \ref{fs}. Using eq.~(\ref{PTrels}) then allows us to immediately write down similar expressions for several of our new masters,
\ba
 f_{\mathcal{J}^{1}_\rmii{b}} =
 -\frac{D-2}{D-1}\frac{q^2}{\omega^2}f_{\mathcal{J}^{0}_\rmii{b}}\;,
 &\hspace{2em}&
 f_{\mathcal{J}^{2}_\rmii{b}} =
 \frac{D(D-2)}{D^2-1}\frac{q^4}{\omega^4}f_{\mathcal{J}^{0}_\rmii{b}}\;, \\
 f_{\mathcal{I}^{1}_\rmii{b}} =
 -\frac{D-2}{D-1}\frac{q^2}{\omega^2}f_{\mathcal{I}^{0}_\rmii{b}}\;,
 &\hspace{2em}&
 f_{\mathcal{I}^{2}_\rmii{b}} =
 -\frac{D-2}{D-1}\frac{r^2}{\omega^2}f_{\mathcal{I}^{0}_\rmii{b}}\;,\\
 f_{\mathcal{I}^{1}_\rmii{d}} =
 -\frac{D-2}{D-1}\frac{q^2}{\omega^2}f_{\mathcal{I}^{0}_\rmii{d}}\;,
 &\hspace{2em}&
 f_{\mathcal{I}^{2}_\rmii{d}} =
 -\frac{D-2}{D-1}\frac{r^2}{\omega^2}f_{\mathcal{I}^{0}_\rmii{d}}\;,\\
 f_{\mathcal{I}^{3}_\rmii{d}} =
 \frac{D(D-2)}{D^2-1}\frac{r^4}{\omega^4}f_{\mathcal{I}^{0}_\rmii{d}}\;,
 &\hspace{2em}&
 f_{\mathcal{I}^{1}_\rmii{f}} =
 -\frac{D-2}{D-1}\frac{q^2}{\omega^2}f_{\mathcal{I}^{0}_\rmii{f}}\;,\\
 f_{\mathcal{I}^{1}_\rmii{h}} =
 -\frac{D-2}{D-1}\frac{q^2}{\omega^2}f_{\mathcal{I}^{0}_\rmii{h}}\;,
 &\hspace{2em}&
 f_{\mathcal{I}^{2}_\rmii{h}} =
 -\frac{D-2}{D-1}\frac{r^2}{\omega^2}f_{\mathcal{I}^{0}_\rmii{h}}\;,\\
 f_{\mathcal{I}^{1}_\rmii{j}} =
 -\frac{D-2}{D-1}\frac{q^2}{\omega^2}f_{\mathcal{I}^{0}_\rmii{j}}\;,
 &\hspace{2em}&
 f_{\mathcal{I}^{2}_\rmii{j}} =
 -\frac{D-2}{D-1}\frac{E_{qr}^2-\lambda^2}{\omega^2}f_{\mathcal{I}^{0}_\rmii{j}}\;,\\
 f_{\mathcal{I}^{3}_\rmii{j}} =
 \frac{D(D-2)}{D^2-1}\frac{q^4}{\omega^4}f_{\mathcal{I}^{0}_\rmii{j}}\;,
 &\hspace{2em}&
 f_{\mathcal{I}^{4}_\rmii{j}} =
 \frac{D(D-2)}{D^2-1}\frac{(E_{qr}^2-\lambda^2)^2}{\omega^4}f_{\mathcal{I}^{0}_\rmii{j}}\;,\\
 &&\hspace{-17em}f_{\mathcal{I}^{5}_\rmii{j}} =
 \frac{1}{\omega^4}\left(\frac{D^2-2 D-2}{D^2-1}q^2r^2+\frac{2}{D^2-1}(\mathbf{q}\cdot\mathbf{r})^2\right)f_{\mathcal{I}^{0}_\rmii{j}}\;,\\
 &&\hspace{-17em}f_{\mathcal{I}^{6}_\rmii{j}} =
 \frac{1}{\omega^4}\left(\frac{D^2-2 D-2}{D^2-1}q^2(E_{qr}^2-\lambda^2)+\frac{2}{D^2-1}(\mathbf{q}\cdot(\mathbf{q-r}))^2\right)f_{\mathcal{I}^{0}_\rmii{j}}\, ,
\ea
where $\lambda$ is a fictitious mass parameter that was introduced in \cite{Laine:2011xm} to regulate IR divergences present at intermediate stages of the computation (but which will all cancel in the end, allowing for the $\lambda\to 0$ limit to be taken). 

The remaining master integrals --- $\mathcal{I}^{3}_\rmii{h},\, \mathcal{I}^{4}_\rmii{h},\, \mathcal{I}^{5}_\rmii{h},\, \mathcal{I}^{6}_\rmii{h},\, \mathcal{I}^{7}_\rmii{h},\, \mathcal{I}^{1}_\rmii{i},\, \mathcal{I}^{2}_\rmii{i},\, \mathcal{I}^{3}_\rmii{i}$ --- differ from the ones evaluated in \cite{Laine:2011xm} not only by the presence of projection operators but also because they each contain a squared propagator. To deal with the complications induced by this, we have found it most convenient to introduce a mass parameter $m$ in this special propagator (chosen to carry the loop momentum $R$), evaluate most parts of the integrals with $m\neq 0$, and in the end make use of the identity
\ba  
  \frac{1}{R^4} &=& - \lim_{m\to 0} \Bigg\{ \frac{{\rm d}}{{\rm d}m^2} 
  \frac{1}{R^2+m^2}\Biggr\} \, . \la{rmass}
\ea
Including this mass into the definition of $\mathcal{I}^{0}_\rmii{h}$, we then quickly arrive at the following results for the h type masters:
\ba
 f_{\mathcal{I}^{3}_\rmii{h}} &=&
 -\frac{D-2}{D-1}
 \lim_{m\to 0} 
 \Bigg\{
 \frac{{\rm d}}{{\rm d}m^2}r^2
 f_{\mathcal{I}^{0}_\rmii{h}}
 \Biggr\} \;,
 \\
 f_{\mathcal{I}^{4}_\rmii{h}} &=&
 \frac{D(D-2)}{D^2-1}
 \lim_{m\to 0} 
 \Bigg\{
  \frac{{\rm d}}{{\rm d}m^2}\frac{q^4}{\omega^2}
 f_{\mathcal{I}^{0}_\rmii{h}}
 \Biggr\}\;,
 \\
 f_{\mathcal{I}^{5}_\rmii{h}} &=&
 \frac{D(D-2)}{D^2-1}
 \lim_{m\to 0} 
 \Bigg\{
  \frac{{\rm d}}{{\rm d}m^2}\frac{r^4}{\omega^2}
 f_{\mathcal{I}^{0}_\rmii{h}}
 \Biggr\}\;,
 \\
 f_{\mathcal{I}^{6}_\rmii{h}} &=&
 \lim_{m\to 0} 
 \Bigg\{
  \frac{{\rm d}}{{\rm d}m^2}
 \left(\frac{D^2-2 D-2}{D^2-1}q^2r^2+\frac{2}{D^2-1}(\mathbf{q}\cdot\mathbf{r})^2\right) 
\frac{f_{\mathcal{I}^{0}_\rmii{h}}}{\omega^2} 
 \Biggr\}\; , \\
 f_{\mathcal{I}^{7}_\rmii{h}} &=&
 \lim_{m\to 0} 
 \Bigg\{
  \frac{{\rm d}}{{\rm d}m^2}
 \left(\frac{D^2-2 D-2}{D^2-1}q^2(\mathbf{q-r})^2+\frac{2}{D^2-1}(\mathbf{q}\cdot(\mathbf{q-r}))^2\right) \frac{f_{\mathcal{I}^{0}_\rmii{h}}}{\omega^2}
 \Biggr\}\, .
\ea
An important simplification in the evaluation of these integrals comes from the fact that after the introduction of the parameter $m$, the IR regulator $\lambda$ is no longer needed. A nonzero value of $m$ is seen to regulate all IR divergences in the above expressions and lead to a finite result when the h type masters are added together and the $m\to 0$ limit taken.

Moving finally on to the i type integrals, the introduction of the parameter $m$ in the $R$-propagator of $\mathcal{I}^{0}_\rmii{i}$ leads to the decomposition
\ba
 \mathcal{I}^{0}_\rmii{i} &=& \Tint{Q,R} 
 \frac{4(Q\cdot P)^2 + m^2P^2}{Q^2[R^2+m^2](Q-R)^2(R-P)^2} +
 \Tint{Q,R} 
 \frac{m^2}{Q^2[R^2+m^2](R-P)^2}\nn
 &-&
 \Tint{Q,R} 
 \frac{P^2}{Q^2(Q-R)^2(R-P)^2}
 +
 \Tint{Q,R} 
 \frac{P^2}{Q^2[R^2+m^2](Q-R)^2} \nn
 &+&\Tint{Q,R} 
 \frac{2}{Q^2[R^2+m^2]} -
 \Tint{Q,R} 
 \frac{1}{Q^2R^2}\;, \la{Ii0de}
\ea
where only the first three terms give nonzero contributions to the spectral function. A straightforward calculation now leads to the result given in eq.~(\ref{fIi1}) for $f_{\mathcal{I}^{1}_\rmii{i}}$, while the other i type integrals become
\ba
 f_{\mathcal{I}^{2}_\rmii{i}} &=&
 \lim_{m\to 0} 
 \Bigg\{
  \frac{{\rm d}}{{\rm d}m^2}\omega^2
 f_{\mathcal{I}^{1}_\rmii{i}}
 \Biggr\} \; , \\
f_{\mathcal{I}^{3}_\rmii{i}}
 &=&-\frac{D-2}{D-1}
 \lim_{m\to 0}
 \Bigg\{
  \frac{{\rm d}}{{\rm d}m^2}
 r^2
 f_{\mathcal{I}^{0}_\rmii{i}}
 \Biggr\} \nn
 &=&-\frac{D-2}{D-1}
 \lim_{m\to 0} 
 \Bigg\{
  \frac{{\rm d}}{{\rm d}m^2}
 r^2
 f_{\mathcal{I}^{}_\rmii{i'}}
 \Biggr\}
 - f_{\mathcal{I}^{2}_\rmii{h}} 
 - f_{\mathcal{I}^{2}_\rmii{b}}\; .
\ea

With the above relations, the evaluation of the master integrals  --- and thus the entire NLO spectral function --- becomes an in principle straightforward, though technically quite demanding exercise in integration. In order not to burden the reader extensively, we leave the details of the practical implementation of these integrals to appendix \ref{details}, and instead move directly to the results.

\section{Results}\label{results}

\subsection{Spectral function: Limits and numerical evaluation}

\begin{figure}
\centering
\includegraphics[width=7.3cm]{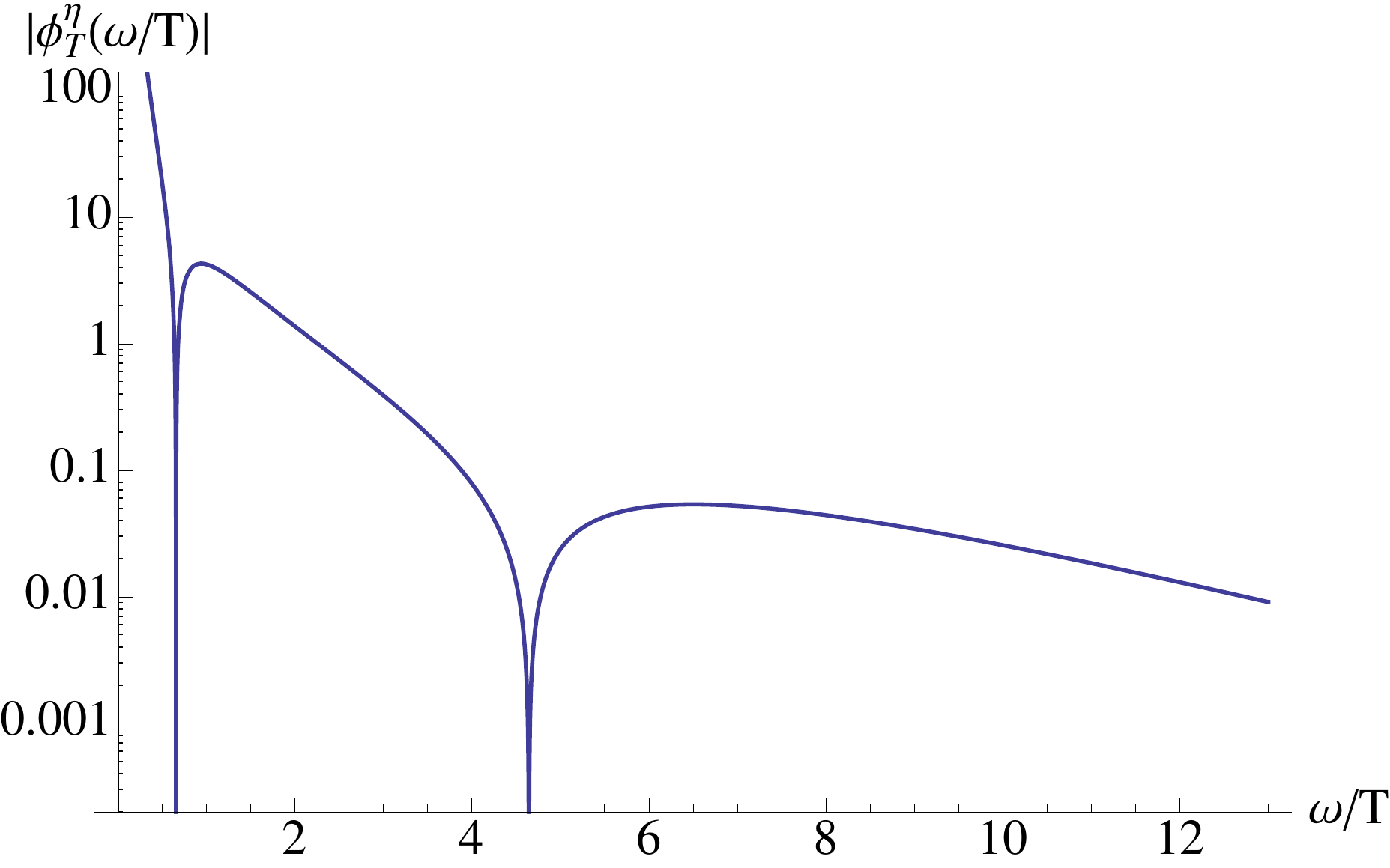}$\;\;\;$\includegraphics[width=7.3cm]{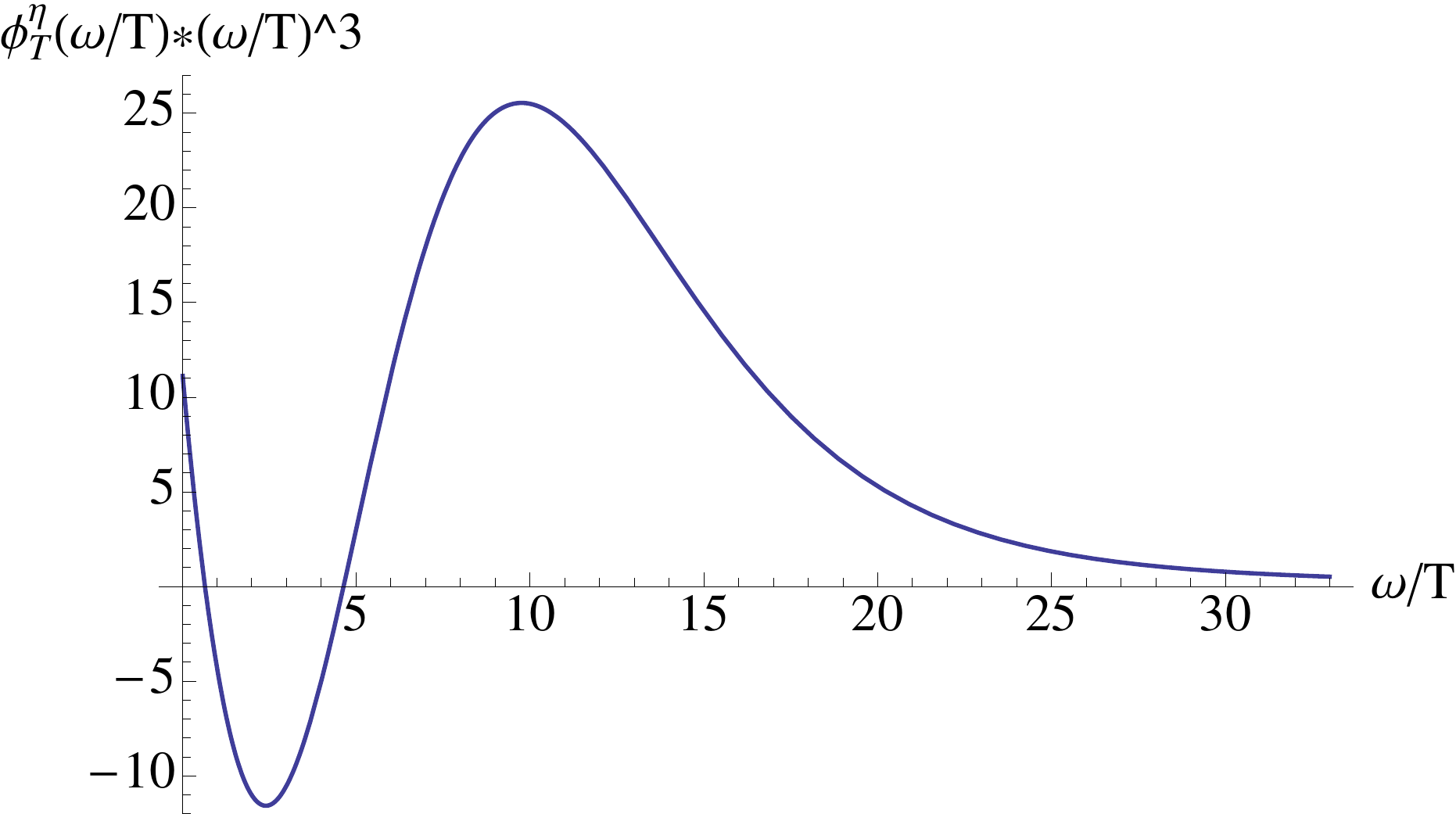}
\caption{The behavior of the function $\phi_T^\eta(\omega/T)$ on a linear and a logarithmic scale, multiplied by $(\omega/T)^3$ in the former case.}
\label{res1}
\end{figure}

Results for each of the functions $\rho^{ }_{\It{x}{ }} (\omega)$, appearing in eq.~(\ref{rhoeta1}), are given in appendix \ref{details}. With the exception of a few very simple cases, they are composed of two parts: An analytic piece containing the possible UV divergences and logarithms of the renormalization scale, and a numerically evaluated finite part denoted by $\tilde{\rho}_{\It{x}{}}$ and consisting of (often rather complicated) one- and two-dimensional integrals. Collecting these results together and inserting them to eq.~(\ref{rhoeta1}), we see that the entire NLO shear spectral function becomes
\ba
\frac{\rho_\eta(\omega)}{4d_A}&=&\frac{\omega^4}{4\pi}\bigl( 1 + 2 n_{\frac{\omega}{2}} \bigr)\Bigg\{-\frac{1}{10}+\frac{g^2N_c}{(4\pi)^2}\bigg(\frac{2}{9}+\phi_T^\eta(\omega/T)\bigg)\Bigg\}\; , \label{result1}
\ea
where $n_{x}\equiv 1/({\rm e}^{\beta x}-1)$ and we have defined the dimensionless function (cf.~eqs.~(4.1)--(4.2) of \cite{Laine:2011xm})
\ba
\phi_T^\eta(\omega/T)&=&-\frac{16\pi^2T^2}{9\omega^2}-\tilde{\rho}_{\It{f}{}} (\omega/T)+\tilde{\rho}_{\It{h}{}} (\omega/T)-\tilde{\rho}_{\It{i}{}} (\omega/T)-\tilde{\rho}_{\It{j}{}} (\omega/T)\; . \label{phiT}
\ea
This function is plotted in fig.~\ref{res1}, and the corresponding lengthy integral expression reproduced in the Mathematica file shearresults.nb, available at \cite{mathfile}.

Before proceeding to plot the above result, let us first perform a number of consistency checks on it. The first thing to note is the automatic cancelation of the $1/\epsilon$ divergences in eq.~(\ref{result1}), as well as the agreement of the constant terms with the $T=0$ results of \cite{Schroder:2011ht,Zoller:2012qv} (note in particular that no renormalization is required at this order). Inspecting next the subleading terms at large $\omega/T$, we note that the functions $\tilde{\rho}_{\It{x}{}} (\omega/T)$ can be analytically expanded in inverse powers of $T^2/\omega^2$. This results in the expressions
\ba
\tilde{\rho}_{\It{f}{}} (\omega/T)&=&\frac{2\pi^2T^2}{3\omega^2}+\frac{8\pi^4T^4}{9\omega^4}+{\mathcal O}(T^6/\omega^6)\; , \\
\tilde{\rho}_{\It{h}{}} (\omega/T)&=&\frac{38\pi^2T^2}{9\omega^2}+\frac{392\pi^4T^4}{135\omega^4}+{\mathcal O}(T^6/\omega^6)\; , 
\\
\tilde{\rho}_{\It{i}{}} (\omega/T)&=&\frac{8\pi^2T^2}{9\omega^2}+\frac{256\pi^4T^4}{135\omega^4}+{\mathcal O}(T^6/\omega^6)\; ,\\
\tilde{\rho}_{\It{j}{}} (\omega/T)&=&\frac{8\pi^2T^2}{9\omega^2}+\frac{16\pi^4T^4}{135\omega^4}+{\mathcal O}(T^6/\omega^6)\; ,
\ea
which --- when inserted into eq.~(\ref{phiT}) --- exactly cancel each other to the order indicated. In fact, it can be shown that in this limit of $\omega\to\infty$ the function $\phi_T^\eta(\omega)$ behaves as
\ba
\phi_T^\eta(\omega/T)&=&\frac{41\pi^6T^6}{3\omega^6} +{\mathcal O}(T^8/\omega^8)\; ,
\ea
which confirms the expectations of \cite{CaronHuot:2009ns}, and in addition constitutes a highly nontrivial cross check of our result.

In the opposite limit of small $\omega/T$, our result is dominated by two integrals, $\tilde{\rho}_{\It{h}{}} (\omega/T)$ and  $\tilde{\rho}_{\It{i}{}} (\omega/T)$, which have the IR expansions
\ba
\tilde{\rho}_{\It{h}{}} (\omega/T)&=&-\frac{7\pi^2 T^3}{9\omega^3}+{\mathcal O}(T^2/\omega^2)\; , \\
\tilde{\rho}_{\It{i}{}} (\omega/T)&=&-\frac{35\pi^2 T^3}{18\omega^3}+{\mathcal O}(T^2/\omega^2)\;.
\ea
This in particular implies that the full function $\phi_T^\eta(\omega)$ behaves at small $\omega$ as 
\ba
\phi_T^\eta(\omega)&=&\frac{7\pi^2T^3}{6\omega^3} +{\mathcal O}(T^2/\omega^2)\; , \label{asym}
\ea
and that the NLO shear spectral function approaches a (positive) constant of ${\mathcal O}(g^2)$ in the $\omega\to 0$ limit. This is in clear contrast with the corresponding limit of the bulk spectral function considered in \cite{Laine:2011xm}, and obviously does not represent the physical IR behavior of the quantity.\footnote{It is nevertheless interesting to compare this result to the findings of ref.~\cite{Kovtun:2011np}, where the authors report that the physical $\omega\to 0$ limit of the spectral function is a constant of ${\mathcal O}(g^8)$.} For $\omega={\mathcal O}(gT)$, a naive perturbative expansion of the type of eq.~(\ref{result1}) namely breaks down, and a resummation of the soft contributions to the spectral function, available through the Hard Thermal Loop (HTL) effective theory, must be performed. For the bulk case, this lengthy exercise was carried out in \cite{Laine:2011xm}, but in our current work, we have decided to refrain from it. The reason for this is that our result is in any case 
expected to 
be trustworthy (and find its applications) in the regime $\omega \gtrsim T$, where the contribution of the extra HTL terms is negligible (cf.~fig.~4 of \cite{Laine:2011xm}). In addition, as will be discussed in the following subsection, the contribution of the perturbative IR sector to quantities such as the imaginary time shear correlator is expected to be numerically rather strongly suppressed.

\begin{figure}
\centering
\includegraphics[width=8.5cm]{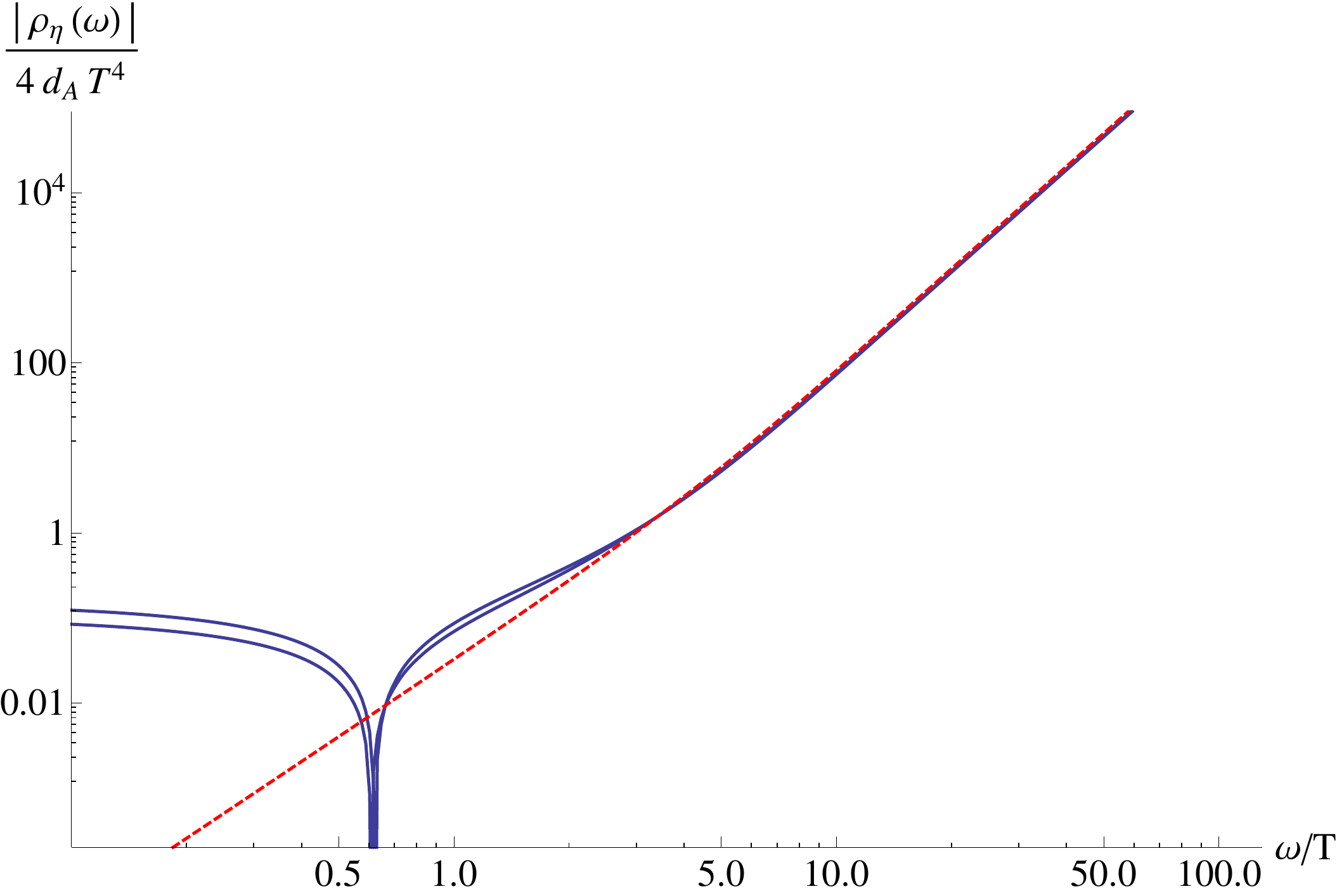}
\caption{The behavior of the absolute value of the shear spectral function, $|\rho_\eta(\omega)|/(4d_A T^4)$, for $T=3T_c$ (corresponding to $3.75\Lambda_\tinymsbar$). The two blue curves stand for the NLO result evaluated with $\bar{\Lambda}=0.5\bar{\Lambda}_\text{opt}$ and $\bar{\Lambda}=2\bar{\Lambda}_\text{opt}$, while the dashed red curve shows the leading order (LO) result. The spikes in the NLO results correspond to a change of sign in the quantity from positive to negative (with increasing $\omega$).}
\label{res2}
\end{figure}

With the above reservation, we now proceed to numerically evaluate the spectral function by inserting the function $\phi_T^\eta(\omega/T)$ into eq.~(\ref{result1}). This leads to the behavior displayed in fig.~\ref{res2}, where we have used a one-loop running coupling, set the number of colors $N_c$ to 3, and varied the renormalization scale $\bar{\Lambda}$ by a factor of two around the `EQCD' value \cite{Kajantie:1997tt}
\ba
\ln\,\frac{\bar{\Lambda}_\text{opt}}{4\pi T}&=&-\gamma_\text{E}-\frac{1}{22} \; .
\ea
As there are no explicit logarithms of the renormalization scale in eq.~(\ref{result1}), an optimization scheme such as the Fastest Apparent Convergence is clearly not available for us. At large values of $\omega$, a choice of the form $\bar{\Lambda}\sim \omega$ might seem more natural than the above, as it would lead to a logarithmically vanishing coupling; as seen from fig.~\ref{res2}, even a scale proportional to $T$, however, leads to a fast convergence of the result towards the free theory limit.

\subsection{Sum rule and imaginary time correlator}

As in the case of the bulk channel \cite{Laine:2011xm}, there are two natural immediate applications of the shear spectral function: The verification of the sum rule of eq.~(4.19) of \cite{Schroder:2011ht} (see also \cite{Romatschke:2009ng} and \cite{Meyer:2010gu}), and the determination of the imaginary time shear correlator, cf.~eq.~(\ref{int_rel}). This time there are, however, two problems preventing a straightforward evaluation of the integrals appearing in these relations. First, in the $\omega\to 0$ limit, the shear spectral function approaches a constant, rendering both of these integrals logarithmically IR divergent. And second, somewhat connected to the first issue, already at LO the spectral function is known to obtain contributions of the form $\omega\delta(\omega)$, which do not show up in our calculation but nevertheless give nonzero contributions to integrals of $\rho_\eta(\omega)/\omega$. At two loop order, some or these terms are in addition proportional to IR divergent momentum integrals, 
further complicating their treatment.

Our approach to resolving the above issues is somewhat different in the two cases. For the sum rule, we expect that a successful verification of the identity will be possible only after extensive further work on the $\omega\to 0$ limit of the spectral function, involving an HTL type resummation. As we do not wish to embark on such a calculation here, we can verify the sum rule only for those master integrals, for which the IR issues do not show up (note that in the form of eq.~(4.19) of \cite{Schroder:2011ht}, the sum rule should indeed hold for each of the master spectral functions separately). Fortunately, this class of masters includes most of the integrals considered in this work, in particular all of the j types and most h's as well. In all of these cases, we have successfully verified the sum rule to the accuracy of our numerical integration, which serves as a powerful check of our results.

\begin{figure}
\centering
\includegraphics[width=8.5cm]{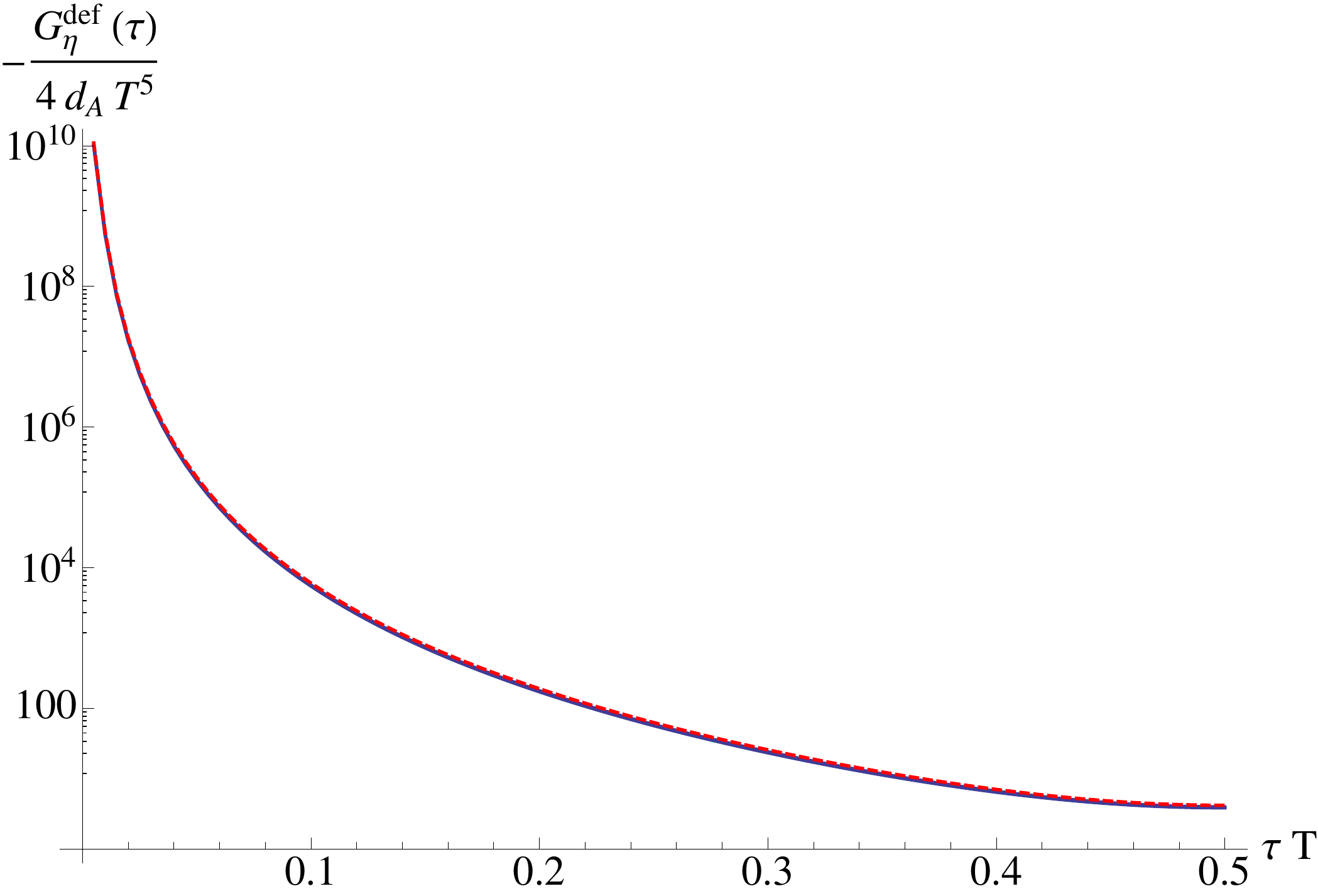}
\caption{The imaginary time correlator $G_\eta^\text{def}(\tau)$, defined by eq.~(\ref{tau1}) and displayed on a logarithmic scale. Just as in fig.~\ref{res2}, altogether three curves are displayed here, but as is apparent form the figure, the two blue NLO curves practically overlap with the dashed red (LO) one.}
\label{res3}
\end{figure}

Moving next on to the imaginary time correlator, we again expect to be able to do the full integral in eq.~(\ref{int_rel}) only after a proper handling of the IR limit of the spectral function. However, our expectation based on fig.~\ref{res2} as well as fig.~4 of \cite{Laine:2011xm} is that the physical contribution of the very softest frequencies to this quantity will most likely be numerically fairly small. Neglecting the constant shift of the correlator produced by the unknown $\omega\delta(\omega)$ contributions, we may thus define a deformed correlation function
\begin{equation}
 G_\eta^\text{def}(\tau) =
 \int_{\omega_0}^\infty
 \frac{{\rm d}\omega}{\pi} \rho_\eta(\omega)
 \frac{\cosh\Big[\! \left(\frac{\beta}{2} - \tau\right)\omega\Big]}
 {\sinh\frac{\beta \omega}{2}}\, ,\quad \quad 0<\tau <\beta \, , \label{tau1}
\end{equation}
where $\omega_0\approx 0.6T$ stands for the frequency, at which the NLO spectral function changes its sign. This is in fact a very natural choice, as below this value the LO spectral function is seen to obtain a relative correction of more than 100$\%$. 

The `deformed' imaginary time correlator is plotted in fig.~\ref{res3}, which displays an exceptionally good agreement between the LO and NLO results. This behavior can be attributed on one hand to the small relative size of the NLO correction in the $T=0$ spectral function, cf.~eq.~(\ref{result1}), and on the other hand to the vanishing of the leading temperature dependent corrections to the quantity at large $\omega$. It would be very interesting to quantitatively compare this behavior to existing lattice and AdS predictions (see e.g.~\cite{Meyer:2007ic,Kajantie:2011nx}), which in fact is a question that will be addressed in the near future \cite{Krssak}.

\section{Conclusions}\label{conclusions}

In the paper at hand, we have performed a NLO calculation of the thermal spectral function in the shear channel of SU($N$) Yang-Mills theory. The main results of the paper were presented in sec.~\ref{results}, with the most important finding being the behavior of the function $\phi_T^\eta(\omega/T)$ that encodes the temperature dependence of the spectral function. In accordance with the general arguments of \cite{CaronHuot:2009ns} as well as the OPE results of \cite{Schroder:2011ht}, we found that this function vanishes in the UV limit like $1/\omega^6$, with a proportionality constant given in eq.~(\ref{asym}). This result was subsequently used to determine the imaginary time shear correlator, in which the NLO term was seen to be highly suppressed.

On the computational level, the derivation of the NLO spectral function necessitated an extensive generalization of the methods developed for the bulk channel calculation of \cite{Laine:2011xm}. Apart from the the results displayed in the previous section, a very important part of our paper can indeed be found from sec.~\ref{calculations} as well as appendix \ref{details}, which detail the techniques we used in our work. Most importantly, these sections contain a detailed explanation of how one can efficiently deal with components of three-momenta as well as squared bosonic propagators in the integrands of two-loop master spectral functions.

In the future, we hope that our results will find important use in the eventual nonperturbative determination of the shear viscosity of Yang-Mills theory, following the rationale of \cite{Burnier:2012ts}. To this end, our finding concerning the small magnitude of the finite temperature corrections to the spectral function --- and in particular to the imaginary time correlator --- is clearly intriguing. It in particular highlights the relative importance of the subleading corrections to the $T=0$ spectral function, which incidentally have recently been determined to an impressively high perturbative order \cite{Zoller:2012qv}. 

Finally, we note that amongst the directions, to which our present calculation should be continued, two stand out quite clearly. First, for the sake of completeness, it would be important to be able to properly determine the IR limit of the NLO shear spectral function, so that all contributions of this order to the sum rule and imaginary time correlator could be accounted for. This would imply both an HTL type resummation at frequencies $\omega\sim gT$ and a successful treatment of the $\omega\delta(\omega)$ contributions to the spectral function. At the same time, it would also be interesting to generalize the present calculation to the presence of (massless) fermions, extending the theory to full QCD. In this case, the result is expected to contain $T$-dependent terms of orders $1/\omega^2$ and $1/\omega^4$ \cite{CaronHuot:2009ns}, so a natural first step might be to generalize the OPE results of \cite{Laine:2010tc,Schroder:2011ht} to the presence of quarks.\footnote{Note that while including the effects 
of fermions, the OPE computation of \cite{Zoller:2012qv} was not able to determine the contribution of the Lorentz non-invariant operator $\epsilon + p$, numerically important at nonzero $T$.}

\section*{Acknowledgments}
We are indebted to Mikko Laine for valuable discussions and advice throughout the project. In addition, we wish to thank Keijo Kajantie, Harvey Meyer, Paul Romatschke, York Schr\"oder, and Mikko Veps\"al\"ainen for useful discussions. The work was supported by the Sofja Kovalevskaja program of the Alexander von Humboldt foundation, as well as by the DFG graduate school \textit{Quantum Fields and Strongly Interacting Matter}. We also acknowledge the hospitality of the Institute for Nuclear Theory, Seattle, and its program \textit{Gauge Field Dynamics In and Out of Equilibrium}, during which part of this work was performed.


\begin{appendix}
\section{Master integrals}\label{masters}

In this first appendix, we first list all master integrals appearing in our calculation, and then recall from \cite{Laine:2011xm} the forms of the integrand functions $f^{ }_{\mathcal{J}^{0}_\rmii{x}}$, $f^{}_{\mathcal{I}^{0}_\rmii{x}}$, defined in eq.~(\ref{fdef}).

\subsection{Definitions} \label{definitions}

The master integrals appearing in our result (\ref{rhoeta1}) for the shear spectral function read
\begin{align}
 \mathcal{J}_\rmi{b}^0 & \equiv
 \Tint{Q} \frac{P^4}{Q^2(Q-P)^2}
 \;, \la{Jb0} \displaybreak[0] \\
 \mathcal{J}_\rmi{b}^1 & \equiv
 \Tint{Q} \frac{P^2}{Q^2(Q-P)^2}P_T(Q)
 \;, \la{Jb1} \displaybreak[0] \\
 \mathcal{J}_\rmi{b}^2 & \equiv
 \Tint{Q} \frac{1}{Q^2(Q-P)^2}P_T(Q)^2
 \;, \la{Jb2}\displaybreak[0] \\
 \mathcal{I}_\rmi{b}^0 & \equiv
 \Tint{Q,R} \frac{P^2}{Q^2R^2(R-P)^2}
 \;,\displaybreak[0] \\
 \mathcal{I}_\rmi{b}^1 & \equiv
 \Tint{Q,R} \frac{1}{Q^2R^2(R-P)^2}P_T(Q)
 \;,\displaybreak[0] \\
 \mathcal{I}_\rmi{b}^2 & \equiv
 \Tint{Q,R} \frac{1}{Q^2R^2(R-P)^2}P_T(R)
 \;,\displaybreak[0] \\
 \mathcal{I}_\rmi{d}^0 & \equiv
 \Tint{Q,R} \frac{P^4}{Q^2R^4(R-P)^2}
 \;, \displaybreak[0] \\
 \mathcal{I}_\rmi{d}^1 & \equiv
 \Tint{Q,R} \frac{P^2}{Q^2R^4(R-P)^2}P_T(Q)
 \;,\displaybreak[0] \\
 \mathcal{I}_\rmi{d}^2 & \equiv
 \Tint{Q,R} \frac{P^2}{Q^2R^4(R-P)^2}P_T(R)
 \;,\displaybreak[0] \\
 \mathcal{I}_\rmi{d}^3 & \equiv
 \Tint{Q,R} \frac{1}{Q^2R^4(R-P)^2}P_T(R)^2
 \;,\displaybreak[0] \\
\mathcal{I}_\rmi{f}^0 & \equiv
 \Tint{Q,R} \frac{P^2}{Q^2(Q-R)^2(R-P)^2}
 \;, \displaybreak[0] \\
 \mathcal{I}_\rmi{f}^1 & \equiv
 \Tint{Q,R} \frac{1}{Q^2(Q-R)^2(R-P)^2}P_T(Q)
 \;,\displaybreak[0] \\
\mathcal{I}_\rmi{h}^0 & \equiv
 \Tint{Q,R} \frac{P^4}{Q^2R^2(Q-R)^2(R-P)^2}
 \;, \la{def_Ih0} \displaybreak[0] \\
\mathcal{I}_\rmi{h}^1 & \equiv
 \Tint{Q,R} \frac{P^2}{Q^2R^2(Q-R)^2(R-P)^2}P_T(Q)
 \;, \la{def_Ih1} \displaybreak[0] \\
\mathcal{I}_\rmi{h}^2 & \equiv
 \Tint{Q,R} \frac{P^2}{Q^2R^2(Q-R)^2(R-P)^2}P_T(R)
 \;, \la{def_Ih2} \displaybreak[0] \\
\mathcal{I}_\rmi{h}^3 & \equiv
 \Tint{Q,R} \frac{P^4}{Q^2R^4(Q-R)^2(R-P)^2}P_T(R)
 \;, \la{def_Ih3} \displaybreak[0] \\
\mathcal{I}_\rmi{h}^4 & \equiv
 \Tint{Q,R} \frac{P^2}{Q^2R^4(Q-R)^2(R-P)^2}P_T(Q)^2
 \;, \la{def_Ih4} \displaybreak[0] \\
\mathcal{I}_\rmi{h}^5 & \equiv
 \Tint{Q,R} \frac{P^2}{Q^2R^4(Q-R)^2(R-P)^2}P_T(R)^2
 \;, \la{def_Ih5} \displaybreak[0] \\
\mathcal{I}_\rmi{h}^6 & \equiv
 \Tint{Q,R} \frac{P^2}{Q^2R^4(Q-R)^2(R-P)^2}P_T(Q)P_T(R)
 \;, \la{def_Ih6} \displaybreak[0] \\
\mathcal{I}_\rmi{h}^7 & \equiv
 \Tint{Q,R} \frac{P^2}{Q^2R^4(Q-R)^2(R-P)^2}P_T(Q)P_T(Q-R)
 \;, \la{def_Ih7} \displaybreak[0] \\
 \mathcal{I}_\rmi{i}^0 & \equiv
 \Tint{Q,R} \frac{(Q-P)^4}{Q^2R^2(Q-R)^2(R-P)^2}
 \;,\displaybreak[0] \\
 \mathcal{I}_\rmi{i}^1 & \equiv
 \Tint{Q,R} \frac{(Q-P)^2}{Q^2R^2(Q-R)^2(R-P)^2}P_T(Q)
 \;,\displaybreak[0] \\
 \mathcal{I}_\rmi{i}^2 & \equiv
 \Tint{Q,R} \frac{P^2(Q-P)^2}{Q^2R^4(Q-R)^2(R-P)^2}P_T(Q)
 \;,\displaybreak[0] \\
 \mathcal{I}_\rmi{i}^3 & \equiv
 \Tint{Q,R} \frac{(Q-P)^4}{Q^2R^4(Q-R)^2(R-P)^2}P_T(R)
 \;,\displaybreak[0] \\
 \mathcal{I}_\rmi{i'} & \equiv
 \Tint{Q,R} \frac{4(Q\cdot P)^2}{Q^2R^2(Q-R)^2(R-P)^2}
 \;, \displaybreak[0]  \label{ii'}\\
 \mathcal{I}_\rmi{j}^0 & \equiv
 \Tint{Q,R} \frac{P^6}{Q^2R^2(Q-R)^2(Q-P)^2(R-P)^2}
 \;,\displaybreak[0] \\
 \mathcal{I}_\rmi{j}^1 & \equiv
 \Tint{Q,R} \frac{P^4}{Q^2R^2(Q-R)^2(Q-P)^2(R-P)^2}P_T(Q)
 \;,\displaybreak[0] \\
 \mathcal{I}_\rmi{j}^2 & \equiv
 \Tint{Q,R} \frac{P^4}{Q^2R^2(Q-R)^2(Q-P)^2(R-P)^2}P_T(Q-R)
 \;,\displaybreak[0] \\
 \mathcal{I}_\rmi{j}^3 & \equiv
 \Tint{Q,R} \frac{P^2}{Q^2R^2(Q-R)^2(Q-P)^2(R-P)^2}P_T(Q)^2
 \;,\displaybreak[0] \\
 \mathcal{I}_\rmi{j}^4 & \equiv
 \Tint{Q,R} \frac{P^2}{Q^2R^2(Q-R)^2(Q-P)^2(R-P)^2}P_T(Q-R)^2
 \;,\displaybreak[0] \\
 \mathcal{I}_\rmi{j}^5 & \equiv
 \Tint{Q,R} \frac{P^2}{Q^2R^2(Q-R)^2(Q-P)^2(R-P)^2}P_T(Q)P_T(R)
 \;,\displaybreak[0] \\
\mathcal{I}_\rmi{j}^6 & \equiv
 \Tint{Q,R} \frac{P^2}{Q^2R^2(Q-R)^2(Q-P)^2(R-P)^2}P_T(Q)P_T(Q-R)
 \;.
\end{align}
In these expressions, we have defined $P_T(Q) \equiv Q_\mu Q_\nu P^T_{\mu\nu}(P) = \mathbf{q}^2-(\mathbf{q\cdot \hat p})^2$, where $P^T_{\mu\nu}(P)$ is the transverse projection operator discussed in section \ref{setup}.

\subsection{Integrands \label{fs}}

In section \ref{calculations}, we reported a number of relations between the integrands $f_{\mathcal{I}^{n}_\rmii{x}}$ of different master spectral functions, relating them to cases encountered in \cite{Laine:2011xm} (albeit with a different IR regulator). Using a notation, where $E_r\equiv \sqrt{q^2+m^2}$, $E_{qr}\equiv |\mathbf{q}-\mathbf{r}|$, and $n_{qr}\equiv n_{E_{qr}}$, the latter read:
\ba
 f^{ }_{\mathcal{J}^{0}_\rmii{b}} &=&  
  \frac{\omega^4\pi}{4 q^2}
 \Bigl[ \delta(\omega - 2 q) - \delta(\omega + 2 q) \Bigr]
 (1+2 n_q)
 \;, \la{fJb0} \\
 f^{ }_{\mathcal{I}^{0}_\rmii{b}} &=&  
  -\frac{\omega^2\pi}{8 q r^2}
 \Bigl[ \delta(\omega - 2 r) - \delta(\omega + 2 r) \Bigr]
 (1+2 n_q) (1+2 n_r)
 \;, \la{fIb0} \\
 f^{ }_{\mathcal{I}^{0}_\rmii{d}} &=&
 -\fr12\lim_{m\to 0} 
 \Bigg\{\frac{{\rm d}}{{\rm d}m^2} \frac{\omega^4\pi}{8 q E_r^2}
 \Bigl[ \delta(\omega - 2 E_r) - \delta(\omega + 2 E_r) \Bigr]
 (1+2 n_q) \bigl( 1+2 n_{E_r}\, \bigr)\Biggr\}
 \;, \la{fId0} \\
f^{ }_{\mathcal{I}^{0}_\rmii{f}} &=& 
 \frac{\omega^2 \pi }{8 q r E_{qr}} \Bigg\{ 
 \la{If} \\
 \nn & - & \!\!
 \Bigl[\delta(\omega - q - r -E_{qr}) - \delta(\omega+q+r+E_{qr}) \Bigr]
 \Bigl[ {(1+n_{qr})(1+n_q+n_r)+n_q n_r} \Bigr]
 \nn & - &  \!\!
 \Bigl[\delta(\omega-q-r+E_{qr}) - \delta(\omega + q + r -E_{qr}) \Bigr]
 \Bigl[ {n_{qr}(1+n_q + n_r )-n_qn_r} \Bigr]
 \nn & - &  \!\!
 \Bigl[\delta(\omega - q + r -E_{qr}) - \delta(\omega+q-r+E_{qr}) \Bigr]
 \Bigl[ {n_r(1+n_q+n_{qr})-n_q n_{qr}} \Bigr]
 \nn & - &  \!\!
 \Bigl[\delta(\omega + q - r -E_{qr}) - \delta(\omega-q+r+E_{qr}) \Bigr]
 \Bigl[ {n_q(1+n_r+n_{qr})-n_r n_{qr}} \Bigr]
 \Biggr\} \;, \la{fIf0} \nn
 f_{\mathcal{I}^{0}_\rmii{h}} &=& 
 \frac{\omega^4 \pi }{8 q r E_{qr}} \Bigg\{  \la{fIh0} \\ 
 & & \!\!
 \frac{1}{2E_r} 
 \Bigl[\delta(\omega - E_{r} - r) - \delta(\omega + E_{r}+r) \Bigr]
 \times 
 \nn & & \times \biggl[
 \biggl( 
 \frac{1}{q+E_{r}-E_{qr}} +  
 \frac{1}{q-E_{r}-E_{qr}} 
 \biggr)
 (1+n_{E_r}+n_r)(n_{qr}-n_q)
 \nn & & \;\; 
 +
 \biggl(
 \frac{1}{q+E_{r}+E_{qr}} +  
 \frac{1}{q-E_{r}+E_{qr}} 
 \biggr)
 (1+n_{E_r}+n_r)(1 + n_{qr}+n_q)
 \biggr]
 \nn &+& \!\!
 \frac{1}{2E_r} 
 \Bigl[\delta(\omega + E_{r} - r) - \delta(\omega - E_{r}+r) \Bigr]
 \times 
 \nn & & \times \biggl[
 \biggl( 
 \frac{1}{q+E_{r}-E_{qr}} +  
 \frac{1}{q-E_{r}-E_{qr}} 
 \biggr)
 (n_{E_r}-n_r)(n_{qr}-n_q)
 \nn & & \;\; 
 +
 \biggl(
 \frac{1}{q+E_{r}+E_{qr}} +  
 \frac{1}{q-E_{r}+E_{qr}} 
 \biggr)
 (n_{E_r}-n_r)(1 + n_{qr}+n_q)
 \biggr]
 \nn & - & \!\!
 \Bigl[\delta(\omega - q - r -E_{qr}) - \delta(\omega+q+r+E_{qr}) \Bigr]
 \frac{(1+n_{qr})(1+n_q+n_r)+n_q n_r}
      {(q+E_{r}+E_{qr})(q-E_{r}+E_{qr})}
 \nn & - &  \!\!
 \Bigl[ \delta(\omega-q-r+E_{qr}) - \delta(\omega + q + r -E_{qr}) \Bigr]
 \frac{n_{qr}(1+n_q + n_r )-n_qn_r}
      {(q+E_{r}-E_{qr})(q-E_{r}-E_{qr})}
 \nn & - &  \!\!
 \Bigl[\delta(\omega - q + r -E_{qr}) - \delta(\omega+q-r+E_{qr}) \Bigr]
 \frac{n_r(1+n_q+n_{qr})-n_q n_{qr}}
      {(q-E_{r}+E_{qr})(q+E_{r}+E_{qr})}
 \nn & - &  \!\!
 \Bigl[\delta(\omega + q - r -E_{qr}) - \delta(\omega-q+r+E_{qr}) \Bigr]
 \frac{n_q(1+n_r+n_{qr})-n_r n_{qr}}
      {(q-E_{r}-E_{qr})(q+E_{r}-E_{qr})}
 \Biggr\}
 \;,  \nonumber
 \\
 f_{\mathcal{I}^{}_\rmii{i'}}
 &=& 
 \frac{\omega^2 \pi q }{2r E_{qr}} \Bigg\{  \la{fIip}
 \\ 
 & & \!\!
 \frac{1}{2E_r} 
 \Bigl[\delta(\omega - E_{r} - r) - \delta(\omega + E_{r}+r) \Bigr]
 \times 
 \nn & & \times 
 \biggl[-\frac{2}{q}(1+n_{E_r}+n_r)(1+2n_{qr})
 \nn & & \;\; 
 +
 \biggl( 
 \frac{1}{q+E_{r}-E_{qr}} +  
 \frac{1}{q-E_{r}-E_{qr}} 
 \biggr)
 (1+n_{E_r}+n_r)(n_{qr}-n_q)
 \nn & & \;\; 
 +
 \biggl(
 \frac{1}{q+E_{r}+E_{qr}} +  
 \frac{1}{q-E_{r}+E_{qr}} 
 \biggr)
 (1+n_{E_r}+n_r)(1 + n_{qr}+n_q)
 \biggr]
 \nn &+& \!\!
 \frac{1}{2E_r} 
 \Bigl[\delta(\omega + E_{r} - r) - \delta(\omega - E_{r}+r) \Bigr]
 \times 
 \nn & & \times 
 \biggl[-\frac{2}{q}(n_{E_r}-n_r)(1+2n_{qr})
 \nn & & \;\; 
 +
 \biggl( 
 \frac{1}{q+E_{r}-E_{qr}} +  
 \frac{1}{q-E_{r}-E_{qr}} 
 \biggr)
 (n_{E_r}-n_r)(n_{qr}-n_q)
 \nn & & \;\; 
 +
 \biggl(
 \frac{1}{q+E_{r}+E_{qr}} +  
 \frac{1}{q-E_{r}+E_{qr}} 
 \biggr)
 (n_{E_r}-n_r)(1 + n_{qr}+n_q)
 \biggr]
 \nn & - & \!\!
 \Bigl[\delta(\omega - q - r -E_{qr}) - \delta(\omega+q+r+E_{qr}) \Bigr]
 \frac{(1+n_{qr})(1+n_q+n_r)+n_q n_r}
      {(q+E_{r}+E_{qr})(q-E_{r}+E_{qr})}
 \nn & - &  \!\!
 \Bigl[ \delta(\omega-q-r+E_{qr}) - \delta(\omega + q + r -E_{qr}) \Bigr]
 \frac{n_{qr}(1+n_q + n_r )-n_qn_r}
      {(q+E_{r}-E_{qr})(q-E_{r}-E_{qr})}
 \nn & - &  \!\!
 \Bigl[\delta(\omega - q + r -E_{qr}) - \delta(\omega+q-r+E_{qr}) \Bigr]
 \frac{n_r(1+n_q+n_{qr})-n_q n_{qr}}
      {(q-E_{r}+E_{qr})(q+E_{r}+E_{qr})}
 \nn & - &  \!\!
 \Bigl[\delta(\omega + q - r -E_{qr}) - \delta(\omega-q+r+E_{qr}) \Bigr]
 \frac{n_q(1+n_r+n_{qr})-n_r n_{qr}}
      {(q-E_{r}-E_{qr})(q+E_{r}-E_{qr})}
 \Biggr\} \;, \hspace*{1cm} \nonumber
\\
 f_{\mathcal{I}^{1}_\rmii{i}} 
 &=& \frac{D-2}{D-1}
 \frac{\omega \pi q }{8 r E_{qr}} \Bigg\{  \la{fIi1} \\
 & & \!\!
 \frac{1}{2E_r} 
 \Bigl[\delta(\omega - E_{r} - r)\Bigr]
 \biggl[\frac{2q}{\omega}(1+n_{E_r}+n_r)(1+2n_{qr})
 \nn & & \;\; 
 -
 \biggl( 
 \frac{\omega+2q}{q+E_{r}-E_{qr}} +  
 \frac{\omega-2q}{q-E_{r}-E_{qr}} 
 \biggr)
 (1+n_{E_r}+n_r)(n_{qr}-n_q)
 \nn & & \;\; 
 -
 \biggl(
 \frac{\omega+2q}{q+E_{r}+E_{qr}} +  
 \frac{\omega-2q}{q-E_{r}+E_{qr}} 
 \biggr)
 (1+n_{E_r}+n_r)(1 + n_{qr}+n_q)
 \biggr]
 \nn &-& \!\!
 \frac{1}{2E_r} 
 \Bigl[\delta(\omega + E_{r}+r) \Bigr] 
 \biggl[\frac{2q}{\omega}(1+n_{E_r}+n_r)(1+2n_{qr})
 \nn & & \;\; 
 -
 \biggl( 
 \frac{\omega-2q}{q+E_{r}-E_{qr}} +  
 \frac{\omega+2q}{q-E_{r}-E_{qr}} 
 \biggr)
 (1+n_{E_r}+n_r)(n_{qr}-n_q)
 \nn & & \;\; 
 -
 \biggl(
 \frac{\omega-2q}{q+E_{r}+E_{qr}} +  
 \frac{\omega+2q}{q-E_{r}+E_{qr}} 
 \biggr)
 (1+n_{E_r}+n_r)(1 + n_{qr}+n_q)
 \biggr]
 \nn &+& \!\! 
 \frac{1}{2E_r} 
 \Bigl[\delta(\omega + E_{r} - r)\Bigr] 
 \biggl[\frac{2q}{\omega}(n_{E_r}-n_r)(1+2n_{qr})
 \nn & & \;\; 
 -
 \biggl( 
 \frac{\omega-2q}{q+E_{r}-E_{qr}} +  
 \frac{\omega+2q}{q-E_{r}-E_{qr}} 
 \biggr)
 (n_{E_r}-n_r)(n_{qr}-n_q)
 \nn & & \;\; 
 -
 \biggl(
 \frac{\omega-2q}{q+E_{r}+E_{qr}} +  
 \frac{\omega+2q}{q-E_{r}+E_{qr}} 
 \biggr)
 (n_{E_r}-n_r)(1 + n_{qr}+n_q)
 \biggr]
  \nn &-& \!\!
 \frac{1}{2E_r} 
 \Bigl[\delta(\omega - E_{r}+r) \Bigr]
 \biggl[\frac{2q}{\omega}(n_{E_r}-n_r)(1+2n_{qr})
 \nn & & \;\; 
 -
 \biggl( 
 \frac{\omega+2q}{q+E_{r}-E_{qr}} +  
 \frac{\omega-2q}{q-E_{r}-E_{qr}} 
 \biggr)
 (n_{E_r}-n_r)(n_{qr}-n_q)
 \nn & & \;\; 
 -
 \biggl(
 \frac{\omega+2q}{q+E_{r}+E_{qr}} +  
 \frac{\omega-2q}{q-E_{r}+E_{qr}} 
 \biggr)
 (n_{E_r}-n_r)(1 + n_{qr}+n_q)
 \biggr]
 \nn & - & \!\!
 \Bigl[\delta(\omega - q - r -E_{qr})(2q-\omega) + \delta(\omega+q+r+E_{qr}) (2q+\omega)\Bigr]
 \frac{(1+n_{qr})(1+n_q+n_r)+n_q n_r}
      {(q+E_{r}+E_{qr})(q-E_{r}+E_{qr})}
 \nn & - &  \!\!
 \Bigl[ \delta(\omega-q-r+E_{qr})(2q-\omega) + \delta(\omega + q + r -E_{qr})(2q+\omega) \Bigr]
 \frac{n_{qr}(1+n_q + n_r )-n_qn_r}
      {(q+E_{r}-E_{qr})(q-E_{r}-E_{qr})}
 \nn & - &  \!\!
 \Bigl[\delta(\omega - q + r -E_{qr})(2q-\omega) + \delta(\omega+q-r+E_{qr}) (2q+\omega)\Bigr]
 \frac{n_r(1+n_q+n_{qr})-n_q n_{qr}}
      {(q-E_{r}+E_{qr})(q+E_{r}+E_{qr})}
 \nn & + &  \!\!
 \Bigl[\delta(\omega + q - r -E_{qr})(2q+\omega) + \delta(\omega-q+r+E_{qr})(2q-\omega) \Bigr]
 \frac{n_q(1+n_r+n_{qr})-n_r n_{qr}}
      {(q-E_{r}-E_{qr})(q+E_{r}-E_{qr})}
 \Biggr\} \;, \nonumber
\\
 f_{\mathcal{I}^{0}_\rmii{j}}& =& 
 \frac{\omega^6 \pi }{4 q r E_{qr}} \Bigg\{ 
 \la{fIj0} \\
 & & \!\!
 \frac{1}{8q^2} 
 \Bigl[\delta(\omega - 2 q) - \delta(\omega+2 q) \Bigr]
 \times 
 \nn & & \times \biggl[
 \biggl( 
 \frac{1}{(q+r-E_{qr})(q+r)} -  
 \frac{1}{(q-r+E_{qr})(q-r)} 
 \biggr)
 (1 + 2 n_q) (n_{qr}-n_r)
 \nn & & \;\; 
 +
 \biggl(
 \frac{1}{(q+r+E_{qr})(q+r)} -  
 \frac{1}{(q-r-E_{qr})(q-r)} 
 \biggr)
 (1 + 2 n_q) (1 + n_{qr}+n_r)
 \biggr]
 \nn & + & \!\!
 \frac{1}{8r^2} 
 \Bigl[\delta(\omega - 2 r) - \delta(\omega+2 r) \Bigr]
 \times 
 \nn & & \times \biggl[
 \biggl( 
 \frac{1}{(q+r-E_{qr})(q+r)} -  
 \frac{1}{(q-r-E_{qr})(q-r)} 
 \biggr)
 (1+2n_r)(n_{qr}-n_q)
 \nn & & \;\; 
 +
 \biggl(
 \frac{1}{(q+r+E_{qr})(q+r)} -  
 \frac{1}{(q-r+E_{qr})(q-r)} 
 \biggr)
 (1+2n_r)(1 + n_{qr}+n_q)
 \biggr]
 \nn & + & \!\!
 \Bigl[\delta(\omega - q - r -E_{qr}) - \delta(\omega+q+r+E_{qr}) \Bigr]
 \frac{(1+n_{qr})(1+n_q+n_r)+n_q n_r}
      {(q+r+E_{qr})^2(q-r+E_{qr})(q-r-E_{qr})}
 \nn & + &  \!\!
 \Bigl[\delta(\omega-q-r+E_{qr}) - \delta(\omega + q + r -E_{qr}) \Bigr]
 \frac{n_{qr}(1+n_q + n_r ) - n_qn_r}
      {(q+r-E_{qr})^2(q-r+E_{qr})(q-r-E_{qr})}
 \nn & + &  \!\!
 \Bigl[\delta(\omega - q + r -E_{qr}) - \delta(\omega+q-r+E_{qr}) \Bigr]
 \frac{n_r(1+n_q+n_{qr})-n_q n_{qr}}
      {(q-r+E_{qr})^2(q+r+E_{qr})(q+r-E_{qr})}
 \nn & + &  \!\!
 \Bigl[\delta(\omega + q - r -E_{qr}) - \delta(\omega-q+r+E_{qr}) \Bigr]
 \frac{n_q(1+n_r+n_{qr})-n_r n_{qr}}
      {(q-r-E_{qr})^2(q+r+E_{qr})(q+r-E_{qr})}
 \Biggr\} \;. \nonumber
\ea

\section{Evaluation of the masters}\label{details}

In this appendix, we provide details on how the master spectral functions of eq.~(\ref{rhoeta1}) were evaluated. We do this case by case, starting from $\rho^{ }_{\mathcal{J}_\rmi{b}}$ and ending with $\rho^{ }_{\It{j}{ }}$. Note that in this process, we will as a rule discard all contributions that would lead to contact terms proportional to $\delta(\omega)$ or derivatives thereof.

\subsection{\texorpdfstring{$\rho^{ }_{\Jt{b}{ }}$}{}}

From section B.2 of ref.~\cite{Laine:2011xm}, we can immediately read off the result
\be
  \rho^{ }_{\mathcal{J}^{0}_\rmii{b}}(\omega)  
  =
 \frac{\omega^4}{16\pi}
 \bigl( 1 + 2 n_{\frac{\omega}{2}} \bigr)
  + \rmO(\epsilon) 
 \;. \la{Jb0_final}
\ee
For the other cases, we note that the relation
\be
 f_{\mathcal{J}^{1}_\rmii{b}} =
 -\frac{D-2}{D-1}\frac{q^2}{\omega^2}f_{\mathcal{J}^{0}_\rmii{b}}
\ee
quickly leads to
\ba
  \rho^{ }_{\mathcal{J}^{1}_\rmii{b}}(\omega)
  &=&
  -\frac{D-2}{D-1}\int_{\vec{q}} \frac{\omega^2\pi}{4 }
 \Bigl[ \delta(\omega - 2 q) - \delta(\omega + 2 q) \Bigr]
 (1+2 n_q)
 \nn & = &
 -\frac{\omega^4}{96\pi}
 \bigl( 1 + 2 n_{\frac{\omega}{2}} \bigr)
 + \rmO(\epsilon)
 \;, \la{Jb1_final}
\ea
while for $\rho^{ }_{\mathcal{J}^{2}_\rmii{b}}$ we similarly obtain
\be
 f_{\mathcal{J}^{2}_\rmii{b}} =
 \frac{D(D-2)}{D^2-1}\frac{q^4}{\omega^4}f_{\mathcal{J}^{0}_\rmii{b}}\;
\ee
and
\ba
   \rho^{ }_{\mathcal{J}^{2}_\rmii{b}}(\omega) 
   &=&
  \frac{D(D-2)}{D^2-1} \int_{\vec{q}} \frac{\pi q^2}{4 }
 \Bigl[ \delta(\omega - 2 q) - \delta(\omega + 2 q) \Bigr]
 (1+2 n_q)
 \nn &=&
 \frac{\omega^4}{480\pi}
 \bigl( 1 + 2 n_{\frac{\omega}{2}} \bigr)
  + \rmO(\epsilon)
 \;. \la{Jb2_final}
\ea
Summing up the different parts contributing to $\rho^{ }_{\mathcal{J}^{}_\rmii{b}}$, we finally reach the compact result
\be
  \rho^{ }_{\mathcal{J}^{}_\rmii{b}}(\omega)
  =
 -\frac{\omega^4}{40\pi}
 \bigl( 1 + 2 n_{\frac{\omega}{2}} \bigr)
  + \rmO(\epsilon)
 \;. \la{Jb_final}
\ee

\subsection{\texorpdfstring{$\rho^{ }_{\It{b}{ }}$}{}}

This time, we start from the relation given in section B.4 of \cite{Laine:2011xm}, 
\be
   \rho^{ }_{\mathcal{I}^{0}_\rmii{b}}(\omega) =
   -\Tint{Q} \frac{1}{Q^2}\frac{\rho^{ }_{\mathcal{J}^{0}_\rmii{b}}(\omega)}{\omega^2}\;.
\ee
Using the known result
\be
 \Tint{Q} \frac{1}{Q^2}=\int_{\vec{q}}\frac{n_q}{q}=\frac{T^2}{12}+\rmO(\epsilon)\;,
 \la{Q1}
\ee
this immediately gives us
\be
 \rho^{ }_{\mathcal{I}^{0}_\rmii{b}}(\omega)
   =
 -\frac{\omega^2T^2}{192\pi}
 \bigl( 1 + 2 n_{\frac{\omega}{2}} \bigr)
 + \rmO(\epsilon)
 \;. \la{Ib0_final}
\ee

For $\rho^{ }_{\mathcal{I}^{1}_\rmii{b}}$, we on the other hand easily obtain
\be
 \rho^{ }_{\mathcal{I}^{1}_\rmii{b}}(\omega) =
   \frac{D-2}{D-1} \Tint{Q} \frac{q^2}{Q^2}
   \frac{\rho^{ }_{\mathcal{J}^{0}_\rmii{b}}(\omega)}{\omega^4}\;,
\ee
which, using the identity
\be
 \Tint{Q} \frac{q^2}{Q^2}=\int_{\vec{q}}q\; n_q=\frac{\pi^2T^4}{30}+\rmO(\epsilon)\;, \la{Qq2}
\ee
leads to the result
\be
 \rho^{ }_{\mathcal{I}^{1}_\rmii{b}}(\omega)
   =
 \frac{\pi T^4 }{720}
 \bigl( 1 + 2 n_{\frac{\omega}{2}} \bigr)
 + \rmO(\epsilon)
 \;. \la{Ib1_final}
\ee

Finally, the corresponding expressions for $\rho^{ }_{\mathcal{I}^{2}_\rmii{b}}$ read
\ba
 \rho^{ }_{\mathcal{I}^{2}_\rmii{b}}(\omega)& =&
   -\Tint{Q} \frac{1}{Q^2}\frac{\rho^{ }_{\mathcal{J}^{1}_\rmii{b}}(\omega)}{\omega^2}
   \nn &=&
 \frac{\omega^2 T^2}{1152\pi}
 \bigl( 1 + 2 n_{\frac{\omega}{2}} \bigr)
 + \rmO(\epsilon)
 \; , \la{Ib2_final}
\ea
providing us with the outcome
\be
   \rho^{ }_{\mathcal{I}^{ }_\rmii{b}}(\omega)
   =
 \bigg[\frac{\pi  T^4}{90}+\frac{ T^2 \omega ^2}{288 \pi } \bigg]
 \bigl( 1 + 2 n_{\frac{\omega}{2}} \bigr)
 + \rmO(\epsilon)
 \;. \la{Ib_final}
\ee

\subsection{\texorpdfstring{$\rho^{ }_{\It{d}{}}$}{}}

Starting with $\rho^{ }_{\mathcal{I}^{0}_\rmii{d}}$, we follow section B.6 of \cite{Laine:2011xm} and obtain
\ba
 \rho^{ }_{\mathcal{I}^{0}_\rmii{d}}(\omega) &=&
 - \frac{1}{2}\, \Tint{Q} \frac{1}{Q^2}  \lim_{m\to 0} \Bigg\{
  \frac{{\rm d}}{{\rm d}m^2}\int_{\vec{r}} \frac{\omega^4\pi}{4 E_r^2}
 \Bigl[ \delta(\omega - 2 E_r) - \delta(\omega + 2 E_r) \Bigr]
 \bigl( 1+2 n_{E_r}\, \bigr)\Biggr\}
 \nn &=&
  \frac{\omega^2T^2}{192\pi}
 \bigl( 1 + 2 n_{\frac{\omega}{2}} \bigr)
  + \rmO(\epsilon)
 \;. \la{Id0_final}
\ea
For $\rho^{ }_{\mathcal{I}^{1}_\rmii{d}}$, we on the other hand first keep $m$ nonzero, obtaining
\ba
 \rho^{ }_{\mathcal{I}^{1}_\rmii{d}}(\omega) &=&
 \frac{D-2}{2(D-1)}\Tint{Q} \frac{q^2}{Q^2}\lim_{m\to 0}
 \Bigg\{
  \frac{{\rm d}}{{\rm d}m^2}\int_{\vec{r}} \frac{\omega^4\pi}{4 E_r^2}
 \Bigl[ \delta(\omega - 2 E_r) - \delta(\omega + 2 E_r) \Bigr]
 \bigl( 1+2 n_{E_r}\, \bigr)\Biggr\}
 \nn &=&
  -\frac{\pi T^4}{720}
 \bigl( 1 + 2 n_{\frac{\omega}{2}} \bigr)
  + \rmO(\epsilon)
 \;, \la{Id1_final}
\ea
while a similar simple exercise produces for $\rho^{ }_{\mathcal{I}^{2}_\rmii{d}}$
\ba
 \rho^{ }_{\mathcal{I}^{2}_\rmii{d}}(\omega) &=&
 \frac{D-2}{2(D-1)} \Tint{Q} \frac{1}{Q^2}\lim_{m\to 0}
 \Bigg\{
  \frac{{\rm d}}{{\rm d}m^2}
 \int_{\vec{r}} \frac{\omega^2\pi ({E_r}^2-m^2)}{4 E_r^2}
 \nn &&\times
 \Bigl[ \delta(\omega - 2 E_r) - \delta(\omega + 2 E_r) \Bigr]
 \bigl( 1+2 n_{E_r}\, \bigr)\Biggr\}\nn
 &=&
 -\frac{\omega^2T^2}{384\pi}
 \bigl( 1 + 2 n_{\frac{\omega}{2}} \bigr) + \rmO(\epsilon)
 \;. \la{Id2_final}
\ea

With $\rho^{ }_{\mathcal{I}^{3}_\rmii{d}}$, we finally get
\ba
 \rho^{ }_{\mathcal{I}^{3}_\rmii{d}}(\omega) &=&
- \frac{D(D-2)}{D^2-1} \Tint{Q} \frac{1}{Q^2}\lim_{m\to 0}
 \Bigg\{
\frac{{\rm d}}{{\rm d}m^2}
\int_{\vec{r}} \frac{\pi ({E_r}^2-m^2)^2}{4 E_r^2}
 \nn &&\times
 \Bigl[ \delta(\omega - 2 E_r) - \delta(\omega + 2 E_r) \Bigr]
 \bigl( 1+2 n_{E_r}\, \bigr)\Biggr\}\nn
 &=&
 \frac{\omega^2T^2}{1152\pi}
 \bigl( 1 + 2 n_{\frac{\omega}{2}} \bigr) + \rmO(\epsilon)
 \;, \la{Id3_final}
\ea
which leads to the desired result
\be
 \rho^{ }_{\mathcal{I}^{ }_\rmii{d}}(\omega) =
-\bigg[\frac{\pi  T^4}{90}+\frac{5 T^2 \omega ^2}{144 \pi } \bigg]
\bigl( 1 + 2 n_{\frac{\omega}{2}} \bigr)
+ \rmO(\epsilon)
 \;. \la{Id_final}
\ee

\subsection{\texorpdfstring{$\rho^{ }_{\It{h}{}}$}{}}

Next, we move on to master integrals of type h, which represent our first truly two-loop topology. We begin from a detailed treatment of $\rho_{\mathcal{I}^{0}_\rmii{h}}$, which has been considered in section B.10 of \cite{Laine:2011xm}, but is now generalized to the presence of the mass parameter $m$. As discussed above, this parameter will not only enable us to derive results for other master integrals containing squared propagators, but its nonzero value in addition serves as an infrared (IR) regulator, implying that we may set the $\lambda$ parameter of \cite{Laine:2011xm} to zero. For brevity, we will write down an explicit final result only for  $\rho_{\mathcal{I}^{0}_\rmii{h}}$, while for the other integrals, we simply indicate, how the calculation proceeds.

\subsubsection*{\texorpdfstring{$\rho^{ }_{\It{h}{0}}$}{}}

The evaluation of $\rho^{ }_{\It{h}{0}}$ begins from eq.~(\ref{fIh0}), obtained by performing the corresponding Matsubara sum with nonzero $m$. From here on, we follow the treatment of \cite{Laine:2011xm} and divide the terms appearing in this expression to three categories, dubbed `factorized powerlike' (fz,p), `factorized exponential' (fz,e), and `phase space' (ps) integrals. For a discussion of the physical interpretation of these contributions, see appendix A.1 of \cite{Laine:2011xm}. 

The (fz,p) contribution corresponds to the UV divergent expression
\be
 \rho^{(\rmi{fz,p})}_{\mathcal{I}^{0}_\rmii{h}}(\omega) \equiv
 \int_{\vec{q,r}}
 \frac{\omega^4 \pi }{16 q r E_r E_{qr}}
 \delta(\omega - E_{r} - r)
 \biggl(
 \frac{1}{q+E_{r}+E_{qr}} +
 \frac{1}{q-E_{r}+E_{qr}}
 \biggr)
 (1+n_{E_r}+n_r)
  \;,  \la{Ih_T}
\ee
where again $E_{qr}\equiv |\mathbf{q}-\mathbf{r}|$ and the $\mathbf{r}$ integral can be easily performed using 
\ba
 \int_{\vec{r}} \delta(\omega - E_{r} - r) & = & \left. \frac{2\;r^{D-2} E_r}{(4\pi)^\frac{D-1}{2}\Gamma(\frac{D-1}{2})\, \omega}\right|_{r=\frac{\omega^2-m^2}{2\omega}} .
\ea
The remaining integral over $\vec{q}$ on the other hand reduces to
\ba
&& \hspace{-3em}\int_{\vec{q}}
 \frac{1 }{4 q E_{qr}}
 \biggl(
 \frac{1}{q+E_{r}+E_{qr}} +
 \frac{1}{q-E_{r}+E_{qr}}
 \biggr) \nonumber 
 =
  \int_Q \frac{1}{Q^2(Q-R)^2}
 \Bigg{|}_{R=(E_r i ,r \vec{e}_r )}
  \la{vacint_h} \\*
 &=&
 \frac{\Lambda^{-2\epsilon}}{(4\pi)^2}
 \biggl(
   \frac{1}{\epsilon} + \ln \frac{\bmu^2}{m^2} + 2 + \rmO(\epsilon)
 \biggr)
 \;,
\ea
leading to the result
\ba
  \rho^{(\rmi{fz,p})}_{\mathcal{I}^{0}_\rmii{h}}(\omega)
 &=&
 \frac{\omega^2(\omega^2-m^2) \Lambda^{-4\epsilon} }{4(4\pi)^3} (1+2n_{\frac{\omega}{2}})
 \bigg(
   \frac{1}{\epsilon} + \ln \frac{\bmu^2}{(\omega-\fr{m^2}{\omega})^2}
   + \ln \frac{\bmu^2}{m^2} + 4
 \bigg)
 \;.
\ea
In the last stage, we have set $n_{E_r}+n_{r}=2n_{\fr{\omega}{2}}$, owing to the identity
\ba
\lim_{m\to 0}
 \bigg\{
\frac{{\rm d}}{{\rm d}m^2}  (n_{E_r}+n_r )
 \Biggr\}  = 0 \; .
 \ea
 
For the UV-finite (fz,e) part, corresponding to the remaining terms on the first six lines of eq.~(\ref{fIh0}), we perform a change of integration variables according to
\be
 \int_{\vec{q}} \frac{1}{q E_{qr}}
 = \frac{1}{4\pi^2r} \int_0^\infty \! {\rm d}q
 \int_{E_{qr}^-}^{E_{qr}^+} \! {\rm d}E_{qr}
 \;, \quad
 E_{qr}^\pm \equiv |q\pm r |
 \;, \la{fz_measure}
\ee
and interchange the order of integrations in the terms including $n_{qr}$,
\be
 \int_0^\infty \! {\rm d}q
 \int_{E_{qr}^-}^{E_{qr}^+} \! {\rm d}E_{qr}
 =
 \int_{0}^{\infty} \! {\rm d}E_{qr}
 \int_{
  |r - E_{qr}|
 }^{
  r + E_{qr}
 } \! {\rm d}q
 \;. \la{fz_order}
\ee
Denoting 
\ba
 \Delta_{ij}&\equiv& q+(-1)^{i}E_r+(-1)^{j}E_{qr}\;,
\ea
we then obtain
\ba
 \rho_{\mathcal{I}^{0}_\rmii{h}}^{(\rmi{fz,e})}(\omega) &\equiv&
 \Bigg[
  \frac{\omega^3}{2(4\pi)^3} (1+n_{E_r}+n_r )
  \Bigg\{
  \int_0^\infty \! {\rm d}q\, n_q
  \int_{E_{qr}^-}^{E_{qr}^+} \! {\rm d}E_{qr}   \,
      \mathbb{P}\left(\frac{1}{\Delta_{00}} +\frac{1}{\Delta_{10}} -\frac{1}{\Delta_{01}} -\frac{1}{\Delta_{11}}\right)
  \nn 
  &+&
  \int_{0}^\infty \! {\rm d}E_{qr}\, n_{qr}
  \int_{
  |r - E_{qr}|
 }^{
  r + E_{qr}} \! {\rm d}q  \, 
    \mathbb{P}\left(\frac{1}{\Delta_{00}} +\frac{1}{\Delta_{01}} +\frac{1}{\Delta_{10}} +\frac{1}{\Delta_{11}}\right)
       \Biggr\}
  \Bigg]_{r=\frac{\omega^2-m^2}{2\omega}} \nn
&=&  
  \frac{\omega^3}{(4\pi)^3} (1+2n_{\fr{\omega}{2}} )
  \int_0^\infty \! {\rm d}q \, n_q \,
  \ln\left|\fr{2q\,\omega-m^2}{2q\,\omega+m^2}\times\fr{2q+\omega}{2q-\omega}\right|
  \;,  \la{Ih0_fz_e_l0}
\ea
where $\mathbb{P}$ stands for a principal value integral and we have used the fact that the integrals on the two first rows of eq.~(\ref{Ih0_fz_e_l0}) are clearly identical upon the redefinition $q\leftrightarrow E_{qr}$.

The (ps) part of the integral, corresponding to the last four rows of eq.~(\ref{fIh0}), is technically the most complicated one. For it, we begin by writing the integration measure in the form
\be
 \int_{\vec{q},\vec{r}} \frac{\pi}{4qrE_{qr}}
 = \frac{2}{(4\pi)^3} \int_0^\infty \! {\rm d}q \int_0^\infty \! {\rm d}r
 \int_{E_{qr}^-}^{E_{qr}^+} \! {\rm d}E_{qr}
 \;, 
\ee
where $E_{qr}^\pm$ are defined as above. Following then the steps laid out in section A.4 of \cite{Laine:2011xm}, we reach (after some labor) the result
\ba
 \rho_{\mathcal{I}^{0}_\rmii{h}}^{(\rmi{ps})}(\omega) & = &
 \frac{\omega^4}{(4\pi)^3} (1+2 n_{\frac{\omega}{2}})  \Bigg\{
 \nn
 & \mbox{(\i)}  &  + \,
 \fr12\int_{0}^{\frac{\omega}{2}}
 \! {\rm d}q
 \int_{0}^{\omega/2-q}
 \! {\rm d}r \;
  \mathbb{P}\Bigg(\frac{F_{\mathcal{I}^{0}_\rmii{h}}(\frac{\omega}{2}-q,\frac{\omega}{2}-r,q+r)}
  {-2\omega r+m^2} \la{Ih0_ps} \\
  &&+\, \frac{F_{\mathcal{I}^{0}_\rmii{h}}(\frac{\omega}{2}-r,\frac{\omega}{2}-q,q+r)}
  {-2\omega q+m^2} \Bigg)
  \left[ 1 + n_{q+r} +  n_{\fr{\omega}2-q}
  +(1+n_{\fr{\omega}2-r})
  \frac{n_{q+r} n_{\fr{\omega}2-q}}{n_r^2}\right]
 \nn
 & \mbox{(\v)}  & + \,
 \fr12\int_{ 0 }^{\infty}
 \! {\rm d}q
 \int_{0}^{ \infty }
 \! {\rm d}r \;
\mathbb{P}\Bigg(\frac{F_{\mathcal{I}^{0}_\rmii{h}}(\frac{\omega}{2}+q,\frac{\omega}{2}+r,q+r)}
  {2\omega r+m^2} \nn
 &&+\, \frac{F_{\mathcal{I}^{0}_\rmii{h}}(\frac{\omega}{2}+r,\frac{\omega}{2}+q,q+r)}
  {2\omega q+m^2} \Bigg)
  \left[ n_{q+r} - n_{q+\frac{\omega}{2}} +
    (1 + n_{q+\frac{\omega}{2}})
    \frac{n_{q+r}n_{r+\frac{\omega}{2}}}{n_r^2} \right]
 \nn
 & \mbox{(\iv)}  & + \,
 \int_{ \frac{\omega}{2} }^{\infty}
 \! {\rm d}q
 \int_{0}^{q-\omega/2}
 \! {\rm d}r \;
 \mathbb{P} \Bigg(\frac{F_{\mathcal{I}^{0}_\rmii{h}}(q-\frac{\omega}{2},\frac{\omega}{2}+r,q-r)}
  {2\omega r+m^2} \nn
  && +\, \frac{F_{\mathcal{I}^{0}_\rmii{h}}(\frac{\omega}{2}+r,q-\frac{\omega}{2},q-r)}
  {-2\omega q+m^2} \Bigg)
  \biggl[
 n_{q-\frac{\omega}{2}} - n_{q}
   - n_{q-\frac{\omega}{2}}
    \frac{(1+n_{q-r})(n_q - n_{r+\frac{\omega}{2}})}
    {n_r n_{-\frac{\omega}{2}}}  \biggr]
 \nn
 &&\hspace{1em}\Bigg\} \;, \nonumber
\ea
where we have made use of the symmetries of the integrand to write
\ba
 &\mbox{(\i)}&: \frac{1}
  {-2\omega r+m^2}
  \rightarrow
  \fr12\left(\frac{1}
  {-2\omega r+m^2} + \frac{1}
  {-2\omega q+m^2} \right),
 \nn &\mbox{(\v)}&:
 \frac{1}
  {2\omega r+m^2}
  \rightarrow
  \fr12\left(\frac{1}
  {2\omega r+m^2} + \frac{1}
  {2\omega q+m^2} \right)  \la{symIh0}
\ea
in the first two parts of the expression. In addition to this, we have for further reference introduced the function $F(x,y,z)$ in all of the integrands, which in the present case obtains the value $F_{\mathcal{I}^{0}_\rmii{h}}(x,y,z)=1$ . It is straightforward to see that eq.~(\ref{Ih0_ps}) reduces to eq.~(B.42) of ref.~\cite{Laine:2011xm}, when the limit $m\rightarrow 0$ is taken.

Next, we must perform the integrals in the above three parts (i)--(iii). In each case, we separate the result (with the overall normalization factor $\frac{\omega^4}{(4\pi)^3} (1+2 n_{\frac{\omega}{2}})$ left out) to three subcontributions based on the distribution functions they contain, denoting by $A$ those parts either containing one $n_{q+r}$ but no other $n$'s or unity; by $B$ those with distribution functions independent of $r$; and by $C$ the remaining piece, proportional to a negative power of $n_r$. In some cases, we will also find it convenient to perform a change of variables from $q$ and $r$ to $q+r\equiv x$ and $q-r \equiv y$.

\paragraph{(\i)}
Taking advantage of the symmetry of the integral, it is easy to see that the $A$ contribution can be written in the form

\ba
A^{(\i)}_{\mathcal{I}^{0}_\rmii{h}} &=&
  \fr12 \int_{0}^{\frac{\omega}{2}}
 \! {\rm d}x
 \int_{0}^{x}
 \! {\rm d}y \;
\mathbb{P} \left(\frac{1}
  {-(x-y)\omega +m^2} + \frac{1}
  {-(x+y)\omega +m^2} \right)
  \left( 1 + n_{x} \right)
  \nonumber \\* &=&
 \fr1{2\omega}\int_{0}^{\frac{\omega}{2}}
 \! {\rm d}q
 \left( 1 + n_{q} \right)
 \ln \frac{m^2}{|2q\omega -m^2|}
 \;,
\ea
in which we have at the last stage renamed $x=q$ and where the `vacuum' contribution is clearly analytically integrable. For the $B$ and $C$ parts, we use the symmetry of the integrand to restrict the integration region to the $q>r$ part, ending up with
\ba
B^{(\i)}_{\mathcal{I}^{0}_\rmii{h}} &=&
\fr1{2\omega}\int_{0}^{\frac{\omega}{4}}
 \! {\rm d}q  \; n_q
  \left( \ln \frac{m^2}{|2q\omega -m^2|}
  + \frac{q} {-\fr{\omega}{2}+q+\frac{m^2}{2\omega}}
 \right)\nonumber \\
&+&  \fr1{2\omega}
    \int_{\frac{\omega}{4}}^{\frac{\omega}{2}}
 \! {\rm d}q \; n_q\;
 \mathbb{P}\left( \ln \frac{m^2}{|\omega^2-2q\omega -m^2|}
  - \frac{q-\fr{\omega}{2}} {-\fr{\omega}{2}+q+\frac{m^2}{2\omega}}
 \right)
\; , \\
C^{(\i)}_{\mathcal{I}^{0}_\rmii{h}} &=&
  -\fr1{2\omega}\int_{0}^{\frac{\omega}{2}}
 \! {\rm d}q
 \int_{0}^{\fr{\omega}{4}-|q-\fr{\omega}{4}|}
 \! {\rm d}r \;
 \mathbb{P} \left(\frac{1}
  {r-\fr{m^2}{2\omega}} + \frac{1}
  {q-\fr{m^2}{2\omega}} \right)\nonumber \\
  &\times&
  (1+n_{\fr{\omega}2-r})
  \frac{n_{q+r} n_{\fr{\omega}2-q}}{n_r^2}
  \;.
\ea
The latter of these expressions must be evaluated as a two-dimensional numerical integral.

\paragraph{(\v)}
For these integrals, the exact same steps as above lead to the results
\ba
 A^{(\v)}_{\mathcal{I}^{0}_\rmii{h}}&=&
- \fr1{2\omega}\int_{0}^{\infty}
 \! {\rm d}q \; n_q
 \ln  \frac{m^2}{2q\omega + m^2}
 \;, \\
 B^{(\v)}_{\mathcal{I}^{0}_\rmii{h}}&=&
  \fr1{2\omega}
  \int_{\frac{\omega}{2}}^{\infty}
 \! {\rm d}q \; n_q
 \left( \ln \frac{m^2}{-\omega^2+2q\omega + m^2}
  - \frac{q-\fr{\omega}{2}} {-\fr{\omega}{2}+q+\frac{m^2}{2\omega}}
 \right)
 \; , \\
 C^{(\v)}_{\mathcal{I}^{0}_\rmii{h}}&=& 
 \fr1{2\omega}\int_{0}^{\infty}
 \! {\rm d}q
 \int_{0}^{q}
 \! {\rm d}r \;
 \left(\frac{1}
  { r+\fr{m^2}{2\omega}} + \frac{1}
  { q+\fr{m^2}{2\omega}} \right)
    (1 + n_{q+\frac{\omega}{2}})
    \frac{n_{q+r}n_{r+\frac{\omega}{2}}}{n_r^2}
  \;.
\ea

\paragraph{(\iv)}
This time, the $A$ contribution clearly vanishes, while the two others produce
\ba
B^{(\iv)}_{\mathcal{I}^{0}_\rmii{h}} &=&
  -\fr1{2\omega} \int_{0 }^{\infty}
 \! {\rm d}q \; n_{q}
 \left( \ln \frac{m^2}{2q\omega + m^2}
  + \frac{q} {\fr{\omega}{2}+q -\frac{m^2}{2\omega}} \right)\nn 
&+& \fr1{2\omega} \int_{ \frac{\omega}{2} }^{\infty}
 \! {\rm d}q \;n_{q}
 \left( \ln \frac{m^2}{-\omega^2+2q\omega + m^2}
  + \frac{q-\fr{\omega}{2}} {q-\frac{m^2}{2\omega}} \right)
  \;, \\
C^{(\iv)}_{\mathcal{I}^{0}_\rmii{h}} &=&
 -\fr1{2\omega}\int_{ \frac{\omega}{2} }^{\infty}
 \! {\rm d}q
 \int_{ 0}^{
    q-\omega/2}
 \! {\rm d}r \;
 \left(\frac{1}
  { r+\fr{m^2}{2\omega}} - \frac{1}
  {q-\fr{m^2}{2\omega}} \right)\nonumber \\
  &\times &
  n_{q-\frac{\omega}{2}}
    \frac{(1+n_{q-r})(n_q - n_{r+\frac{\omega}{2}})}
    {n_r n_{-\frac{\omega}{2}}}
   \;.
\ea

Collecting all of the above results together and organizing the terms somewhat, we finally obtain a lengthy expression for the function $\rho_{\mathcal{I}^{0}_\rmii{h}}^{ }(\omega)$
\ba
&&\hspace{-3em}{\Lambda}^{4\epsilon}\fr{2(4\pi)^3\rho_{\mathcal{I}^{0}_\rmii{h}}^{ }(\omega)} {\omega^3 (1+2n_{\fr{\omega}{2}} )} =
  \frac{\omega^2-m^2}{2\omega}
 \biggl(
   \frac{1}{\epsilon} + \ln \frac{\bmu^2}{(\omega-\fr{m^2}{\omega})^2}
   + \ln \frac{\bmu^2}{m^2} + 4
 \biggr)
 \nn
&+& \fr{\omega}{2}\left[
 1+\left(1-\fr{m^2}{\omega^2}\right)
 \ln\left(\fr{m^2}{\omega^2-m^2}\right)
 \right]
 \nn 
&+&\int_{0}^{\frac{\omega}{4}}
 \! {\rm d}q  \; n_q \Bigg\{
 2\ln\frac{2q+\omega}{-2q+\omega}
+\left( \fr{\omega}{2}-\frac{m^2}{2\omega} \right)
 \left( \frac{1} {q+\fr{\omega}{2}-\frac{m^2}{2\omega}}
   + \frac{1} {q-\fr{\omega}{2}+\frac{m^2}{2\omega}}\right)
 \Biggr\}
  \nn 
&+& \int_{\frac{\omega}{4}}^{\frac{\omega}{2}}
 \! {\rm d}q \; n_q \Bigg\{
 2\ln\frac{2q+\omega}{-2q+\omega}
+ \ln\fr{2q\omega-m^2}{|\omega^2-2q\omega+m^2|}
 \nn 
&-&2+\fr{\omega}{2} \frac{1} {q+\fr{\omega}{2}-\frac{m^2}{2\omega}}
 +\frac{m^2}{2\omega} \mathbb{P} \left( \frac{1} {q-\fr{\omega}{2}+\frac{m^2}{2\omega}}
   - \frac{1} {q+\fr{\omega}{2}-\frac{m^2}{2\omega}}\right)
 \Biggr\}
 \nn
&+& \int_{ \frac{\omega}{2} }^{\infty}
 \! {\rm d}q \; n_q \Bigg\{
 2\ln\frac{2q+\omega}{2q-\omega}-2\ln\frac{2q\omega-\omega^2+m^2}{2q\omega-m^2}
 + \fr{\omega}{2} \left(
  \frac{1} {q+\fr{\omega}{2}-\frac{m^2}{2\omega}}
  -\frac{1} {q-\frac{m^2}{2\omega}}\right)
 \nn 
&-& 1 -\frac{m^2}{2\omega} \left(
  \frac{1} {q+\fr{\omega}{2}-\frac{m^2}{2\omega}}
  -\frac{1} {q-\fr{\omega}{2}+\frac{m^2}{2\omega}}
  -\frac{1} {q-\frac{m^2}{2\omega}}\right)
 \Biggr\}
 \nn 
&+&\int_{0}^{\frac{\omega}{2}}
 \! {\rm d}q
 \int_{0}^{\fr{\omega}{4}-|q-\fr{\omega}{4}|}
 \! {\rm d}r \;\mathbb{P}
 \left(\frac{1}
  {- r+\fr{m^2}{2\omega}} + \frac{1}
  {- q+\fr{m^2}{2\omega}} \right)
  (1+n_{\fr{\omega}2-r})
  \frac{n_{q+r} n_{\fr{\omega}2-q}}{n_r^2}
 \nn 
&+&\int_{0}^{\infty}
 \! {\rm d}q
 \int_{0}^{q}
 \! {\rm d}r \;
 \left(\frac{1}
  { r+\fr{m^2}{2\omega}} + \frac{1}
  { q+\fr{m^2}{2\omega}} \right)
    (1 + n_{q+\frac{\omega}{2}})
    \frac{n_{q+r}n_{r+\frac{\omega}{2}}}{n_r^2}
 \nn 
&-&\int_{ \frac{\omega}{2} }^{\infty}
 \! {\rm d}q
 \int_{ 0}^{
    \frac{2q-\omega}{2}}
 \! {\rm d}r \;
 \left(\frac{1}
  { r+\fr{m^2}{2\omega}} + \frac{1}
  { -q+\fr{m^2}{2\omega}} \right)
  n_{q-\frac{\omega}{2}}
    \frac{(1+n_{q-r})(n_q - n_{r+\frac{\omega}{2}})}
    {n_r n_{-\frac{\omega}{2}}}
  \;. \la{rho_Ih0_m}
\ea
Unlike in the (fz,p), (fz,e) and (ps) parts separately, taking the $m\to 0$ limit of this expression leads to a finite result, which in fact is seen to agree with that derived in ref.~\cite{Laine:2011xm}. We will nevertheless keep $m$ nonzero for the time being, as this will turn out useful in the following.

\subsubsection*{\texorpdfstring{$\rho^{ }_{\It{h}{1}}$}{}}

As discussed in sec.~\ref{calculations}, the integrand of $\rho^{ }_{\It{h}{1}}$ is related to that of $\rho^{ }_{\It{h}{0}}$ through
\ba
 f_{\mathcal{I}^{1}_\rmii{h}} =
 -\frac{D-2}{D-1}\frac{q^2}{\omega^2}f_{\mathcal{I}^{0}_\rmii{h}}\;,
\ea
from which it is easy to derive the (fz,p) result
\ba
  \rho^{(\rmi{fz,p})}_{\mathcal{I}^{1}_\rmii{h}}(\omega)
 &=&
  -\frac{D-2}{D-1}\frac{\omega \pi \Lambda^{-2\epsilon} r^{D-3}(1+n_{E_r}+n_r)}{2(4\pi)^{\frac{D-1}{2}} \Gamma(\frac{D-1}{2})}
 \left. \int_Q \frac{q^2}{Q^2(Q-R)^2}
 \right|_{R=(E_r i ,r \vec{e}_r )}\nn
 &=&
 \frac{\omega^4\Lambda^{-4\epsilon}} {(4\pi)^3}(1+2n_{\frac{\omega}{2}})
 \Bigg\{
 \frac{1}{72} \left(-1+\frac{m^6}{\omega ^6}\right)\left(\frac{1}{\epsilon}+\ln \frac{\bmu^2}{m^2}+\ln \frac{\bmu^2}{(\omega-\fr{m^2}{\omega})^2}\right)
 \nn &&
   -\frac{23}{432}+\frac{23 m^6}{432 \omega ^6}-\frac{m^4}{144 \omega ^4}+\frac{m^2}{144 \omega ^2}
 \Biggr\} \;. \la{Ih1_fzp}
\ea
For the (fz,e) part, we similarly obtain
\ba
 \rho_{\mathcal{I}^{1}_\rmii{h}}^{(\rmi{fz,e})}(\omega) &=&
 -\frac{2}{3}\Biggl[
  \frac{\omega}{2(4\pi)^3} (1+n_{E_r}+n_r ) \nn
  &\times&  \Bigg\{
  \int_0^\infty \! {\rm d}q
  \int_{E_{qr}^-}^{E_{qr}^+} \!\! {\rm d}E_{qr}  \, n_q \, q^2
      \left(\frac{1}{\Delta_{00}} +\frac{1}{\Delta_{10}} -\frac{1}{\Delta_{01}} -\frac{1}{\Delta_{11}}\right)
  \nn &+ &
  \int_{0}^\infty \! {\rm d}E_{qr}
  \int_{
  |r - E_{qr}|
 }^{
  r + E_{qr}} \!\! {\rm d}q   \, n_{qr} q^2
    \left(\frac{1}{\Delta_{00}} +\frac{1}{\Delta_{01}} +\frac{1}{\Delta_{10}} +\frac{1}{\Delta_{11}}\right)
       \Biggr\}
  \Biggl]_{r=\frac{\omega^2-m^2}{2\omega}}
    .\;  \la{Ih1_fz_e}
\ea
Proceeding finally to the (ps) part, we recall that in the derivation of the equivalent of our eq.~(\ref{Ih0_ps}) in \cite{Laine:2011xm}, a change of variables of the type
\ba
 & \mbox{(\i)}  &
 \phi(q,r,E_{qr}) \rightarrow \phi\left(\frac{\omega}{2}-q,\frac{\omega}{2}-r,q+r\right)\;, \nn
 & \mbox{(\v)}  &
 \phi(q,r,E_{qr}) \rightarrow \phi\left(\frac{\omega}{2}+q,\frac{\omega}{2}+r,q+r\right)\;, \la{psshift} \\
 & \mbox{(\iv)}  &
 \phi(q,r,E_{qr}) \rightarrow \phi\left(q-\frac{\omega}{2},\frac{\omega}{2}+r,q-r\right) \nonumber 
\ea
was carried out. It is then straightforward to see that the (ps) contribution to $\It{h}{1}$ takes the form of eq.~(\ref{Ih0_ps}) with 
\ba
F_{\It{h}{1}}(x,y,z)&=& -\frac{2x^2}{3\omega^2}\;.
\ea
All integrals encountered in evaluating this expression can be dealt with using the methods discussed above.

\subsubsection*{\texorpdfstring{$\rho^{}_{\mathcal{I}^{2}_\rmii{h}}$}{}}
For $\rho^{ }_{\mathcal{I}^{2}_\rmii{h}}(\omega)$, we have
\ba
  f_{\mathcal{I}^{2}_\rmii{h}} =
 -\frac{D-2}{D-1}\frac{r^2}{\omega^2}f_{\mathcal{I}^{0}_\rmii{h}}\;,
\ea
from which one straightforwardly obtains the relations
\ba
  \rho^{(\rmi{fz,p})}_{\mathcal{I}^{2}_\rmii{h}}(\omega)
  &=&
  -\frac{D-2}{D-1}\left(\fr{\omega^2-m^2}{2\omega^2}\right)^2
 \rho_{\mathcal{I}^{0}_\rmii{h}}^{(\rmi{fz,p})}(\omega)
 \nn &=& 
 \frac{\omega^4\Lambda^{-4\epsilon}} {(4\pi)^3}
 \frac{\left(m^2-\omega ^2\right)^3}{24 \omega ^6}(1+2n_{\frac{\omega}{2}})
 \Bigg\{
 \frac{1}{\epsilon}+\ln \frac{\bmu^2}{m^2}+\ln \frac{\bmu^2}{(\omega-\fr{m^2}{\omega})^2}+\fr{11}{3}
 \Biggr\} 
 \;, \la{Ih2_fzp}\\
 \rho_{\mathcal{I}^{2}_\rmii{h}}^{(\rmi{fz,e})}(\omega) &=&
 -\frac{2}{3}\left(\fr{\omega^2-m^2}{2\omega^2}\right)^2
 \rho_{\mathcal{I}^{0}_\rmii{h}}^{(\rmi{fz,e})}(\omega)\la{Ih2_fz_e}
  \\ &=&  
  -
  \frac{2\omega}{3(4\pi)^3} 
  \left(\fr{\omega^2-m^2}{2\omega}\right)^2
  (1+2n_{\fr{\omega}{2}} )
  \int_0^\infty \! {\rm d}q \, n_q \, 
  \ln\left|\fr{2q\omega-m^2}{2q\omega+m^2}\times\fr{2q+\omega}{2q-\omega}\right|
    \;,  \nonumber
\ea
as well as
\ba
F_{\It{h}{2}}(x,y,z)&=& -\frac{2y^2}{3\omega^2}\;.
\ea
Again, all integrals encountered are carried out with methods used in the previous cases.

\subsubsection*{\texorpdfstring{$\rho^{ }_{\mathcal{I}^{3}_\rmii{h}}$}{}}

With $\It{h}{3}$, we encounter the first master integral with a squared propagator. To this end, we take a mass derivative of $\rho^{}_{\mathcal{I}^{2}_\rmii{h}}(\omega)$, obtaining
\ba
\rho^{ }_{\mathcal{I}^{3}_\rmii{h}}(\omega) 
 &=&\omega^2\lim_{m\to 0} 
 \Bigg\{
  \frac{{\rm d}}{{\rm d}m^2} \rho^{}_{\mathcal{I}^{2}_\rmii{h}}(\omega)
  \Biggr\}
 \;.
\ea
As we have above computed $\rho^{ }_{\mathcal{I}^{2}_\rmii{h}}(\omega)$ while keeping $m$ nonzero, taking the derivative is in principle a very straightforward task. The only somewhat problematic issue is related to the IR divergences that appear in individual parts of the integral once one proceeds to set $m\to 0$ in the end. Our strategy with them is to identify those parts of the integrals that diverge in this limit, and subseqently add and subtract from the integrands terms that are analytically computable (with finite $m$), yet render the original integral convergent. The divergent terms are expected to cancel against each other once we assemble the result for the full master spectral function, while the remaining finite parts are dealt with using the methods described above.

In the one-dimensional integrals, originating from the $A$ and $B$ parts defined above, the divergences in the $m\to 0$ limit may in principle appear due to one of two reasons: A pole at $q=\fr{\omega}{2}$, coming from a factor $1/(q-\fr{\omega}{2}+\fr{m^2}{2\omega})$, or an explicit $\ln m^2$ term originating from the analytic $r$ integral. For ${\mathcal{I}^{3}_\rmii{h}}$, the latter does not occur, while the former can be dealt with by writing
\ba
\fr{(4\pi )^3\rho^{(\rmi{1d,div}) }_{\mathcal{I}^{3}_\rmii{h}}(\omega)} { 1+2n_{\fr{\omega}{2}}}
&=& \int_{ \frac{\omega}{2} }^{\infty}
 \! {\rm d}q \;n_{q}
 \Bigg[
 \frac{-m^4+\omega ^4}{24 \left(q+\frac{m^2}{2 \omega }-\frac{\omega }{2}\right)}+\frac{m^6-2 m^4 \omega ^2+m^2 \omega ^4}{48 \left(q+\frac{m^2}{2 \omega }-\frac{\omega }{2}\right)^2 \omega }
 \Bigg] 
 \hspace{7em}\\
 &=&
\int_{ \frac{\omega}{2} }^{\infty}
 \! {\rm d}q \, \Bigg\{n_{q}
 \Bigg[
 \frac{-m^4+\omega ^4}{24 \left(q+\frac{m^2}{2 \omega }-\frac{\omega }{2}\right)}+\frac{m^6-2 m^4 \omega ^2+m^2 \omega ^4}{48 \left(q+\frac{m^2}{2 \omega }-\frac{\omega }{2}\right)^2 \omega }
 \Bigg]
 \nn &&
 -n_{\fr{\omega}{2}} \frac{\omega }{2q}
 \Bigg[
 \frac{-m^4+\omega ^4}{24 \left(q+\frac{m^2}{2 \omega }-\frac{\omega }{2}\right)}+\frac{m^6-2 m^4 \omega ^2+m^2 \omega ^4}{48 \left(q+\frac{m^2}{2 \omega }-\frac{\omega }{2}\right)^2 \omega }
 \Bigg]
 \Bigg\}
 \nn &&
 +\int_{ \frac{\omega}{2} }^{\infty}
 \! {\rm d}q \;n_{\fr{\omega}{2}} \frac{\omega }{2q}
 \Bigg[
 \frac{-m^4+\omega ^4}{24 \left(q+\frac{m^2}{2 \omega }-\frac{\omega }{2}\right)}+\frac{m^6-2 m^4 \omega ^2+m^2 \omega ^4}{48 \left(q+\frac{m^2}{2 \omega }-\frac{\omega }{2}\right)^2 \omega }
 \Bigg]
 \;. \nonumber
\ea
Here, the integral containing the curly brackets is seen to be finite, while the divergent term on the last row can be evaluated analytically, producing
\ba
\int_{ \frac{\omega}{2} }^{\infty}
 \! {\rm d}q \;n_{\fr{\omega}{2}} \frac{\omega }{2q}
 \Bigg[
 \frac{-m^4+\omega ^4}{24 \left(q+\frac{m^2}{2 \omega }-\frac{\omega }{2}\right)}+\frac{m^6-2 m^4 \omega ^2+m^2 \omega ^4}{48 \left(q+\frac{m^2}{2 \omega }-\frac{\omega }{2}\right)^2 \omega }
 \Bigg]
 =
 -\frac{\omega ^4}{24}  \left(\frac{m^2}{\omega^2 }-1 +  \ln\fr{m^2}{\omega ^2}\right) n_{\fr{\omega}{2}}
 \,. \nonumber
\ea

Turning our attention next to the two-dimensional part, we observe that the only divergence originates from the $r=0$ limit of $C^{(\iv)}_{\mathcal{I}^{3}_\rmii{h}}$. This can be regulated by subtracting from the integrand a term of the type $\alpha/(r+\frac{m^2}{2\omega})$, where $\alpha$ is a suitably chosen residue. This amounts to writing $C^{(\iv)}_{\mathcal{I}^{3}_\rmii{h}}$ in the somewhat complicated form
\ba
\fr{2}{3}C^{(\iv)}_{\mathcal{I}^{3}_\rmii{h}}&=& \frac{ \omega ^4 }{24} n_{\fr{\omega}{2}} \ln \fr{m^2}{\omega^2}  -\frac{\omega^4}{24 T} \int_{ \frac{\omega}{2} }^{\infty}
 \! {\rm d}q n_q(1+n_q) \ln\left|\fr{2 q \omega -\omega ^2}{\omega^2}\right|
\nn 
&+&
\fr{2}{3}C^{(\iv)}_{\mathcal{I}^{3}_\rmii{h}} + 
\frac{\omega^4}{24 T} \int_{ \frac{\omega}{2} }^{\infty}
 \! {\rm d}q\, n_q(1+n_q)\, 
 \int_{ 0}^{
    \frac{2q-\omega}{2}}
 \!\!\!\! {\rm d}r \;
 \frac{1}{\left(r+\fr{m^2}{2\omega} \right)} 
 \;,
\ea
where the two terms on the latter row combine to a finite integral.

Apart from the above terms, all other parts of $\rho^{}_{\mathcal{I}^{3}_\rmii{h}}$ are IR finite and computable with standard methods.

\subsubsection*{\texorpdfstring{$\rho^{}_{\mathcal{I}^{4}_\rmii{h}}$}{}}
This time, we begin with the integrand relation
\be
 f_{\mathcal{I}^{4}_\rmii{h}} =
 \frac{D(D-2)}{D^2-1}
 \lim_{m\to 0} 
 \Bigg\{
  \frac{{\rm d}}{{\rm d}m^2}\frac{q^4}{\omega^2}
 f_{\mathcal{I}^{0}_\rmii{h}}
 \Biggr\}\;,
\ee
using which we may define a non-differentiated two argument version of the master integral $\rho^{ }_{\mathcal{I}^{4}_\rmii{h}}(\omega,m)$,
\ba
 \rho^{ }_{\mathcal{I}^{4}_\rmii{h}}(\omega) \equiv
 \lim_{m\to 0} 
 \Bigg\{ 
  \frac{{\rm d}}{{\rm d}m^2} 
  \rho^{ }_{\mathcal{I}^{4}_\rmii{h}}(\omega,m)
  \Biggr\}
 \;.
\ea
It is then a straightforward exercise to derive the results
\ba
  \rho^{(\rmi{fz,p})}_{\mathcal{I}^{4}_\rmii{h}}(\omega,m)
 &=& 
 \frac{\omega^4\Lambda^{-4\epsilon}} {(4\pi)^3}(1+2 n_{\frac{\omega}{2}}) 
 \Bigg\{
 \frac{-m^{10}+\omega ^{10}}{600 \omega ^8} \left(\frac{1}{\epsilon}+\ln \frac{\bmu^2}{m^2}+\ln \frac{\bmu^2}{(\omega-\fr{m^2}{\omega})^2}\right)
 \nn &&
 -\frac{m^2}{800}-\frac{27 m^{10}}{4000 \omega ^8}+\frac{m^8}{800 \omega ^6}+\frac{m^6}{3600 \omega ^4}-\frac{m^4}{3600 \omega ^2}+\frac{27 \omega ^2}{4000}
 \Biggr\} 
 \;, \la{Ih4_fzp}\\
 \rho_{\mathcal{I}^{4}_\rmii{h}}^{(\rmi{fz,e})}(\omega,m) &=&   
  \frac{8}{15} 
 \biggl[
  \frac{\omega}{2(4\pi)^3} (1+n_{E_r}+n_r ) \\
  &\times&  \Bigg\{  
  \int_0^\infty \! {\rm d}q 
  \int_{E_{qr}^-}^{E_{qr}^+} \!\! {\rm d}E_{qr}  \, n_q \,
      \left(\frac{1}{\Delta_{00}} +\frac{1}{\Delta_{10}} -\frac{1}{\Delta_{01}} -\frac{1}{\Delta_{11}}\right) q^4
  \nn & + & 
  \int_{0}^\infty \! {\rm d}E_{qr}
  \int_{ |r - E_{qr}|}^{r + E_{qr}}
 \!\! {\rm d}q   \, n_{qr}
    \left(\frac{1}{\Delta_{00}} +\frac{1}{\Delta_{01}} +\frac{1}{\Delta_{10}} +\frac{1}{\Delta_{11}}\right) q^4
       \Biggr\} 
  \biggl]_{r=\frac{\omega^2-m^2}{2\omega}}
  \;,  \la{Ih4_fz_e} \nonumber
\ea
as well as
\ba
F_{\It{h}{4}}(x,y,z)&=& \frac{8x^4}{15\omega^2}\, ,
\ea
where the function $F_{\It{h}{4}}(x,y,z)$ refers to the non-differentiated version $\mathcal{I}^{4}_\rmii{h}$ as well. The differentiation of these expressions with respect to $m^2$ and the subsequent subtraction of divergences proceeds exactly as explained in the previous cases.

\subsubsection*{\texorpdfstring{$\rho^{}_{\mathcal{I}^{5}_\rmii{h}}$}{}}
For $\mathcal{I}^{5}_\rmii{h}$, we again use the definition
\be
 f_{\mathcal{I}^{5}_\rmii{h}} =
 \frac{D(D-2)}{D^2-1}
 \lim_{m\to 0} 
 \Bigg\{
  \frac{{\rm d}}{{\rm d}m^2}\frac{r^4}{\omega^2}
 f_{\mathcal{I}^{0}_\rmii{h}}
 \Biggr\}\
\ee
to define the non-differentiated version of the integral, $\rho^{ }_{\mathcal{I}^{5}_\rmii{h}}(\omega,m)$, through
\ba
 \rho^{ }_{\mathcal{I}^{5}_\rmii{h}}(\omega) \equiv
 \lim_{m\to 0} 
 \Bigg\{ 
  \frac{{\rm d}}{{\rm d}m^2} 
  \rho^{ }_{\mathcal{I}^{5}_\rmii{h}}(\omega,m)
  \Biggr\}\;.
\ea
For this function, we easily obtain the results
\ba
  \rho^{(\rmi{fz,p})}_{\mathcal{I}^{5}_\rmii{h}}(\omega,m)
 &=& 
 \frac{\omega^4 \Lambda^{-4\epsilon}} {(4\pi)^3}
 \frac{\left(-m^2+\omega ^2\right)^5}{120 \omega ^8 }
 (1+2 n_{\frac{\omega}{2}}) \nn
 &&\times
 \Bigg\{
  \frac{1}{\epsilon}
 +\ln \frac{\bmu^2}{m^2}+\ln \frac{\bmu^2}{(\omega-\fr{m^2}{\omega})^2}
 +\frac{107}{30}
 \Biggr\} 
 \;, \la{Ih5_fzp} \\
 \rho_{\mathcal{I}^{5}_\rmii{h}}^{(\rmi{fz,e})}(\omega,m) &=& 
 \frac{8}{15}
 \fr{1}{\omega^2}
 \left(\fr{\omega^2-m^2}{2\omega}\right)^4
 \rho_{\mathcal{I}^{0}_\rmii{h}}^{(\rmi{fz,e})}(\omega)
 \nn &=&  
  \frac{8\omega}{15(4\pi)^3} 
  \left(\fr{\omega^2-m^2}{2\omega}\right)^4
  (1+2n_{\fr{\omega}{2}} )
  \nn && \times
  \int_0^\infty \! {\rm d}q \, n_q \, 
  \ln\left|\fr{2q\omega-m^2}{2q\omega+m^2} \times \fr{2q+\omega}{2q-\omega}\right|
  \;,  \la{Ih5_fz_e}
\ea
and finally
\ba
F_{\It{h}{5}}(x,y,z)&=& \frac{8y^4}{15\omega^2}\, .
\ea

\subsubsection*{\texorpdfstring{$\rho^{}_{\mathcal{I}^{6}_\rmii{h}}$}{}} \la{Ih6}

The master integral $\mathcal{I}^{6}_\rmii{h}$ has a rather complicated structure in momentum space. It is related to $\mathcal{I}^{0}_\rmii{h}$ via the relation
\be
 f_{\mathcal{I}^{6}_\rmii{h}} =
 \lim_{m\to 0} 
 \Bigg\{
  \frac{{\rm d}}{{\rm d}m^2}
 \left(\frac{D^2-2 D-2}{D^2-1}q^2r^2+\frac{2}{D^2-1}(\mathbf{q}\cdot\mathbf{r})^2\right) 
\frac{f_{\mathcal{I}^{0}_\rmii{h}}}{\omega^2} 
 \Biggr\}\; ,
\ee
which we again write in the form
\ba
 \rho^{ }_{\mathcal{I}^{6}_\rmii{h}}(\omega) \equiv
 \lim_{m\to 0} 
 \Bigg\{ 
  \frac{{\rm d}}{{\rm d}m^2} 
  \rho^{ }_{\mathcal{I}^{6}_\rmii{h}}(\omega,m)
  \Biggr\}\;.
\ea
From here, we straightforwardly obtain the (fz,p) result
\ba
  \rho^{(\rmi{fz,p})}_{\mathcal{I}^{6}_\rmii{h}}(\omega,m)
 &=& 
 \frac{\omega^4\Lambda^{-4\epsilon}} {(4\pi)^3}
 \fr{\left(-m^2+\omega ^2\right)^3}{720 \omega ^8}(1+2 n_{\frac{\omega}{2}}) 
 \Bigg\{
 \left(2 m^4+m^2 \omega ^2+2 \omega ^4\right)\la{Ih6_fzp} \\
 &&\times \left(\frac{1}{\epsilon}+\ln \frac{\bmu^2}{m^2}+\ln \frac{\bmu^2}{(\omega-\fr{m^2}{\omega})^2}\right)
 +\frac{112 m^4}{15}+\frac{26 m^2 \omega ^2}{15}+\frac{112 \omega ^4}{15}
 \Biggr\} 
 \;, \nonumber
\ea
but to evaluate the other parts requires a little more care. 

To deal with the (fz,e) and (ps) contributions, we first note that the full master can be written in the form
\ba
 \rho^{ }_{\mathcal{I}^{6}_\rmii{h}}(\omega) &=&
 \lim_{m\to 0} 
 \Bigg\{
  \frac{{\rm d}}{{\rm d}m^2}\int_{\vec{q,r}}\frac{1}{\omega^2}
 \left(\frac{D^2-2 D-2}{D^2-1}q^2r^2+\frac{2}{D^2-1}(\mathbf{q}\cdot\mathbf{r})^2\right) 
 f_{\mathcal{I}^{0}_\rmii{h}}\Bigg\}
 \nn &=&
 \frac{1}{2D(D-2)}\rho^{ }_{\mathcal{I}^{4}_\rmii{h}}(\omega)
 +\frac{1}{2D(D-2)}\rho^{ }_{\mathcal{I}^{5}_\rmii{h}}(\omega)
 \nn &+&
 \lim_{m\to 0} 
 \Bigg\{
  \frac{{\rm d}}{{\rm d}m^2}\int_{\vec{q,r}}\frac{1}{\omega^2}
 \left(\frac{D^2-2 D-1}{D^2-1}q^2r^2\right) 
 f_{\mathcal{I}^{0}_\rmii{h}} \Biggr\}
 \nn &+&
 \lim_{m\to 0} 
 \Bigg\{
 \frac{{\rm d}}{{\rm d}m^2}\int_{\vec{q,r}}\frac{1}{\omega^2}
 \left(\frac{1}{D^2-1}\frac{E_{qr}^4}{2}-\frac{1}{D^2-1}E_{qr}^2\left(q^2+r^2\right) \right) 
 f_{\mathcal{I}^{0}_\rmii{h}} \Biggr\}
 \;,
\ea
which prompts us to define
\ba
\rho^{ }_{\mathcal{I}^{6(1)}_\rmii{h}}(\omega) &=&
\lim_{m\to 0} 
 \Bigg\{
  \frac{{\rm d}}{{\rm d}m^2}\int_{\vec{q,r}}\frac{1}{\omega^2}
 \left(\frac{D^2-2 D-1}{D^2-1}q^2r^2\right) 
 f_{\mathcal{I}^{0}_\rmii{h}} \Biggr\}
 \nn &\equiv&
 \lim_{m\to 0} 
 \Bigg\{
  \frac{{\rm d}}{{\rm d}m^2}
 \rho^{ }_{\mathcal{I}^{6(1)}_\rmii{h}}(\omega,m) \Biggr\}
 \;, \\
\rho^{ }_{\mathcal{I}^{6(2)}_\rmii{h}}(\omega) &=&
\lim_{m\to 0} 
 \Bigg\{
  \frac{{\rm d}}{{\rm d}m^2}\int_{\vec{q,r}}\frac{1}{\omega^2}
 \left(\frac{1}{D^2-1}\frac{E_{qr}^4}{2}\right) 
 f_{\mathcal{I}^{0}_\rmii{h}} \Biggr\}
 \nn &\equiv&
 \lim_{m\to 0} 
 \Bigg\{
  \frac{{\rm d}}{{\rm d}m^2}
 \rho^{ }_{\mathcal{I}^{6(2)}_\rmii{h}}(\omega,m) \Biggr\}
 \;,
 \\
\rho^{ }_{\mathcal{I}^{6(3)}_\rmii{h}}(\omega) &=&-
\lim_{m\to 0} 
 \Bigg\{
  \frac{{\rm d}}{{\rm d}m^2}\int_{\vec{q,r}}\frac{1}{\omega^2}
 \left(\frac{1}{D^2-1} q^2 E_{qr}^2\right) 
 f_{\mathcal{I}^{0}_\rmii{h}} \Biggr\}
 \nn &\equiv&
 \lim_{m\to 0} 
 \Bigg\{
  \frac{{\rm d}}{{\rm d}m^2}
 \rho^{ }_{\mathcal{I}^{6(3)}_\rmii{h}}(\omega,m) \Biggr\}
 \;, \\
\rho^{ }_{\mathcal{I}^{6(4)}_\rmii{h}}(\omega) &=&-
\lim_{m\to 0} 
 \Bigg\{
  \frac{{\rm d}}{{\rm d}m^2}\int_{\vec{q,r}}\frac{1}{\omega^2}
 \left(\frac{1}{D^2-1} r^2 E_{qr}^2\right) 
 f_{\mathcal{I}^{0}_\rmii{h}} \Biggr\}
 \nn &\equiv&
 \lim_{m\to 0} 
 \Bigg\{
  \frac{{\rm d}}{{\rm d}m^2}
 \rho^{ }_{\mathcal{I}^{6(4)}_\rmii{h}}(\omega,m) \Biggr\}
 \;.
\ea
For the (fz,e) parts of these integrals, we then obtain
\ba
 \rho^{(\rmi{fz,e})}_{\mathcal{I}^{6(1)}_\rmii{h}}(\omega,m)
 &=&
 \frac{7}{15}\biggl[
  \frac{\omega}{2(4\pi)^3} (1+n_{E_r}+n_r ) \\
  &\times&  \Bigg\{  
  \int_0^\infty \! {\rm d}q 
  \int_{E_{qr}^-}^{E_{qr}^+} \!\! {\rm d}E_{qr}  \, n_q \,
      \left(\frac{1}{\Delta_{00}} +\frac{1}{\Delta_{10}} -\frac{1}{\Delta_{01}} -\frac{1}{\Delta_{11}}\right) q^2r^2
  \nn & + &
  \int_{0}^\infty \! {\rm d}E_{qr}
  \int_{ 
  |r - E_{qr}|
 }^{
  r + E_{qr}
 } \!\! {\rm d}q   \, n_{qr}
    \left(\frac{1}{\Delta_{00}} +\frac{1}{\Delta_{01}} +\frac{1}{\Delta_{10}} +\frac{1}{\Delta_{11}}\right) q^2r^2
       \Biggr\} 
  \biggl]_{r=\frac{\omega^2-m^2}{2\omega}}
  \nn &=&
  -\frac{7}{10}
  \left(\frac{\omega^2-m^2}{2\omega}\right)^2
  \rho^{(\rmi{fz,e})}_{\mathcal{I}^{1}_\rmii{h}}(\omega)
  \;, \nn
 \rho^{(\rmi{fz,e})}_{\mathcal{I}^{6(2)}_\rmii{h}}(\omega,m)
 &=&
 \frac{1}{15} \biggl[
  \frac{\omega}{2(4\pi)^3} (1+n_{E_r}+n_r ) \\
  &\times&  \Bigg\{  
  \int_0^\infty \! {\rm d}q 
  \int_{E_{qr}^-}^{E_{qr}^+} \!\! {\rm d}E_{qr}  \, n_q \,
      \left(\frac{1}{\Delta_{00}} +\frac{1}{\Delta_{10}} -\frac{1}{\Delta_{01}} -\frac{1}{\Delta_{11}}\right) 
      \fr{E_{qr}^4}{2}
  \nn & + &
  \int_{0}^\infty \! {\rm d}E_{qr}
  \int_{ 
  |r - E_{qr}|
 }^{
  r + E_{qr}
 } \!\! {\rm d}q   \, n_{qr}
    \left(\frac{1}{\Delta_{00}} +\frac{1}{\Delta_{01}} +\frac{1}{\Delta_{10}} +\frac{1}{\Delta_{11}}\right) 
    \fr{E_{qr}^4}{2}
       \Biggr\} 
  \biggl]_{r=\frac{\omega^2-m^2}{2\omega}}
  \;, \nn
\rho^{(\rmi{fz,e})}_{\mathcal{I}^{6(3)}_\rmii{h}}(\omega,m)
 &=&
 - \frac{1}{15}
 \biggl[
  \frac{\omega}{2(4\pi)^3} (1+n_{E_r}+n_r ) \\
  &\times&  \Bigg\{  
  \int_0^\infty \! {\rm d}q 
  \int_{E_{qr}^-}^{E_{qr}^+} \!\! {\rm d}E_{qr}  \, n_q \,
      \left(\frac{1}{\Delta_{00}} +\frac{1}{\Delta_{10}} -\frac{1}{\Delta_{01}} -\frac{1}{\Delta_{11}}\right) q^2 E_{qr}^2
  \nn & + &
  \int_{0}^\infty \! {\rm d}E_{qr}
  \int_{ 
  |r - E_{qr}|
 }^{
  r + E_{qr}
 } \!\! {\rm d}q   \, n_{qr}
    \left(\frac{1}{\Delta_{00}} +\frac{1}{\Delta_{01}} +\frac{1}{\Delta_{10}} +\frac{1}{\Delta_{11}}\right) q^2 E_{qr}^2
       \Biggr\} 
  \biggl]_{r=\frac{\omega^2-m^2}{2\omega}}
  \;, \nn
\rho_{\mathcal{I}^{6(4)}_\rmii{h}}^{(\rmi{fz,e})}(\omega,m) & = &
 - \frac{1}{15}
 \biggl[
  \frac{\omega}{2(4\pi)^3} (1+n_{E_r}+n_r )\la{Ih64_fze} \\
  &\times&  \Bigg\{  
  \int_0^\infty \! {\rm d}q 
  \int_{E_{qr}^-}^{E_{qr}^+} \!\! {\rm d}E_{qr}  \, n_q \,
      \left(\frac{1}{\Delta_{00}} +\frac{1}{\Delta_{10}} -\frac{1}{\Delta_{01}} -\frac{1}{\Delta_{11}}\right) r^2 E_{qr}^2
  \nn & + &
  \int_{0}^\infty \! {\rm d}E_{qr}
  \int_{ 
  |r - E_{qr}|
 }^{
  r + E_{qr}
 } \!\! {\rm d}q   \, n_{qr}
    \left(\frac{1}{\Delta_{00}} +\frac{1}{\Delta_{01}} +\frac{1}{\Delta_{10}} +\frac{1}{\Delta_{11}}\right) r^2 E_{qr}^2
       \Biggr\} 
  \biggl]_{r=\frac{\omega^2-m^2}{2\omega}}
  \;,  \nonumber
\ea
while the (ps) contributions are available via
\ba
F_{\It{h}{6(1)}}(x,y,z)&=& \frac{7x^2y^2}{15\omega^2}\, , \\
F_{\It{h}{6(2)}}(x,y,z)&=& \frac{z^4}{30\omega^2}\, , \\
F_{\It{h}{6(3)}}(x,y,z)&=& -\frac{x^2 z^2}{15\omega^2}\, , \\
F_{\It{h}{6(4)}}(x,y,z)&=& -\frac{y^2 z^2}{15\omega^2}\, . \\
\ea
The evaluation of the individual integrals is again a relatively straightforward exercise, after which the final result for the master is available by simply adding the different contributions together.

\subsubsection*{\texorpdfstring{$\rho^{}_{\mathcal{I}^{7}_\rmii{h}}$}{}}
This time, our starting point is 
\be
f_{\mathcal{I}^{7}_\rmii{h}} =
 \lim_{m\to 0} 
 \Bigg\{
  \frac{{\rm d}}{{\rm d}m^2}
 \left(\frac{D^2-2 D-2}{D^2-1}q^2(\mathbf{q-r})^2+\frac{2}{D^2-1}(\mathbf{q}\cdot(\mathbf{q-r}))^2\right) \frac{f_{\mathcal{I}^{0}_\rmii{h}}}{\omega^2}
 \Biggr\}\, ,
\ee
which leads to the rather convenient result
\ba
 \rho^{ }_{\mathcal{I}^{7}_\rmii{h}}(\omega) 
 &= &\rho^{ }_{\mathcal{I}^{6}_\rmii{h}}(\omega)
 - \frac{D (D-2)}{D^2-2D-1}\rho^{ }_{\mathcal{I}^{6(1)}_\rmii{h}}(\omega)
 - D (D-2)\rho^{ }_{\mathcal{I}^{6(3)}_\rmii{h}}(\omega)
 \;. \la{Ih7_exp}
\ea
For this function, the (fz,p) contribution is easily evaluated and reads
\ba
  \rho^{(\rmi{fz,p})}_{\mathcal{I}^{7}_\rmii{h}}(\omega,m)
 &=& 
 \frac{\omega^4\Lambda^{-4\epsilon}} {(4\pi)^3}(1+2 n_{\frac{\omega}{2}}) 
 \Biggl\{ \la{Ih7_fzp}\\
 &&
 \frac{-m^{10}-25 m^6 \omega ^4+25 m^4 \omega ^6+\omega ^{10}}{3600 \omega ^8} \left(\frac{1}{\epsilon}+\ln \frac{\bmu^2}{m^2}+\ln \frac{\bmu^2}{(\omega-\fr{m^2}{\omega})^2}\right)
 \nn &&
 -\frac{7 m^2}{3600}-\frac{47 m^{10}}{54000 \omega ^8}+\frac{7 m^8}{3600 \omega ^6}-\frac{337 m^6}{10800 \omega ^4}+\frac{337 m^4}{10800 \omega ^2}+\frac{47 \omega ^2}{54000}
 \Biggr\} 
 \;, \nonumber
\ea
while the (fz,e) and (ps) contributions are immediately available using previous results.

\subsubsection*{Final result}
\begin{figure}
\centering
\includegraphics[width=10cm]{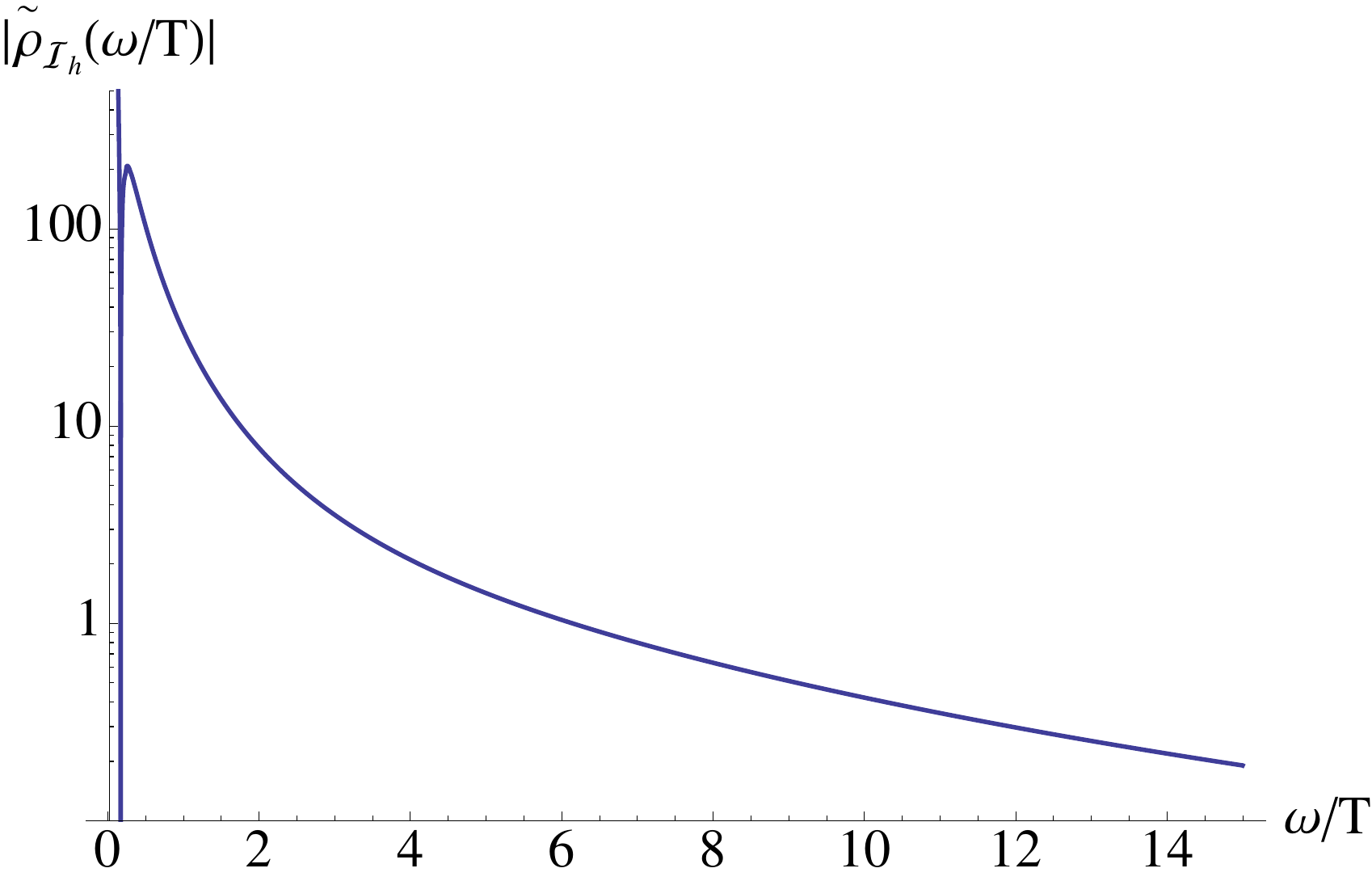}
\caption{The behavior of the absolute value of $\tilde{\rho}_{\It{h}{}} (\omega/T)$, displayed on a logarithmic scale. The spike in the curve corresponds to the sign of the function changing from negative to positive with increasing $\omega$.}
\label{fig2}
\end{figure}
Collecting all of the above pieces, proceeding to the $m\to 0$ limit, and performing the remaining convergent integrals numerically, we arrive at the result
\ba
\Lambda^{4\epsilon}\rho^{ }_{\It{h}{ }} (\omega)&=& \frac{\omega^4} {(4\pi)^3}(1+2 n_{\frac{\omega}{2}})\Bigg\{-\frac{37}{36\epsilon}-\frac{37}{18}\ln\,\frac{\bar{\Lambda}^2}{\omega^2}-\frac{29}{27}+\tilde{\rho}_{\It{h}{}} (\omega/T)\Bigg\}\, .
\ea
Here, $\tilde{\rho}_{\It{h}{}}(\omega/T)$ is a dimensionless quantity, whose behavior as a function of frequency we display in fig.~\ref{fig2}.

\subsection{\texorpdfstring{$\rho^{ }_{\mathcal{I}_\rmii{f}}$}{}}

Moving on to ${\mathcal{I}_\rmii{f}}$, the only master integral needed is $\rho^{}_{\mathcal{I}^{1}_\rmii{f}}(\omega)$, related to the already evaluated $\rho^{}_{\mathcal{I}^{0}_\rmii{f}}(\omega)$ (see section B.8 of ref.~\cite{Laine:2011xm}) through
\ba
  f_{\mathcal{I}^{1}_\rmii{f}} =
 -\frac{D-2}{D-1}\frac{q^2}{\omega^2}f_{\mathcal{I}^{0}_\rmii{f}}\;\;.
\ea
Noting that this integral only contains a (ps) part, we quickly obtain the convergent result 
\begin{figure}
\centering
\includegraphics[width=10cm]{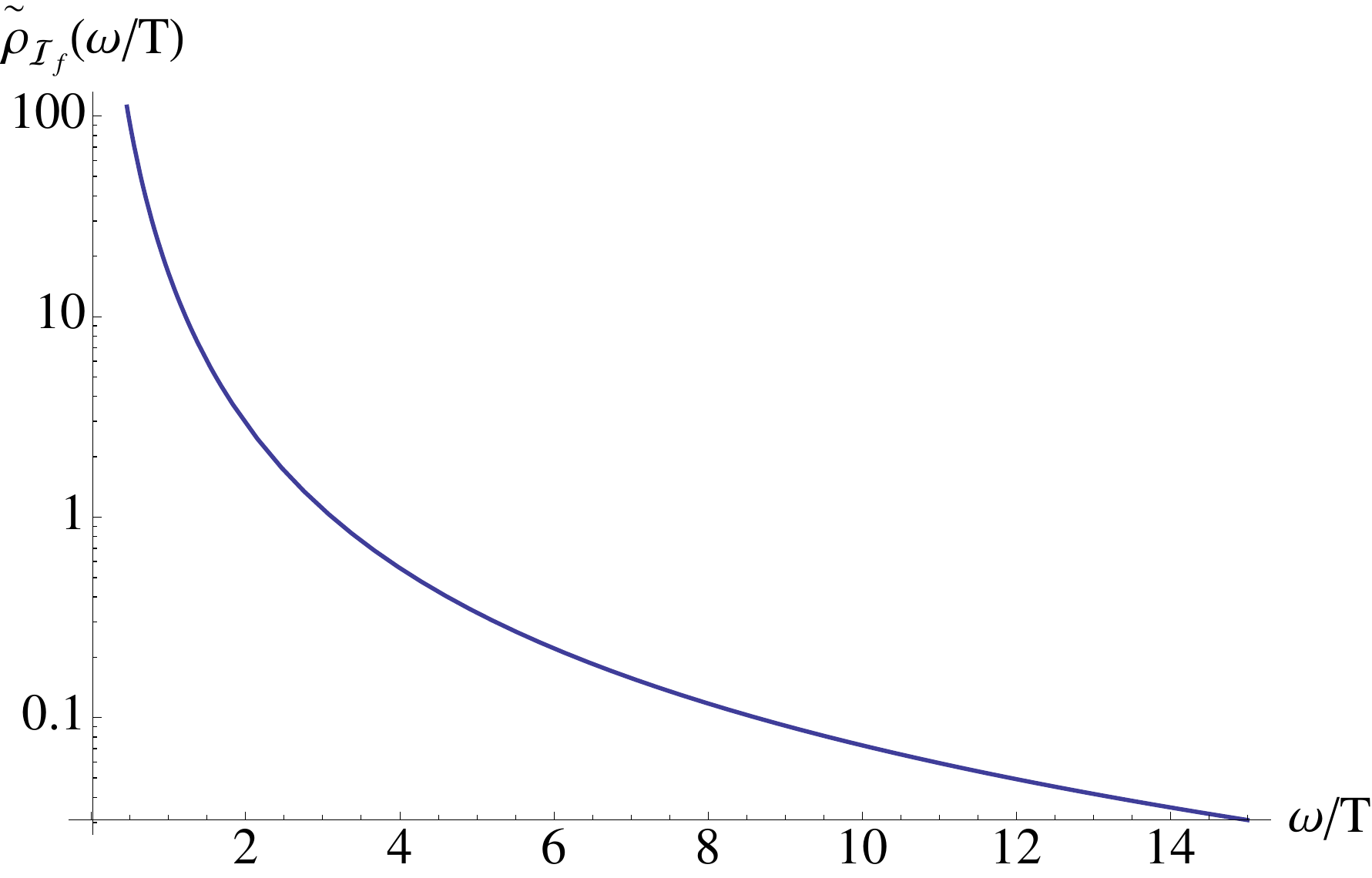}
\caption{The behavior of the function $\tilde{\rho}_{\It{f}{}} (\omega/T)$.}
\label{fig3}
\end{figure}
\ba
 \rho_{\mathcal{I}^{1}_\rmii{f}}^{(\rmi{ps})}(\omega) & = & 
 \frac{2}{3(4\pi)^3} (1+2 n_{\frac{\omega}{2}})  \Biggl\{ 
 \nn 
 & \mbox{(\i)}  &  + \frac{1}{2}\,
 \int_{0}^{\frac{\omega}{2}}
 \! {\rm d}q
 \int_{0}^{\frac{\omega-2q}{2}} 
 \!\!\!\! {\rm d}r \; 
 \left(1 + n_{q+r} +  n_{\fr{\omega}2-q}
  +(1+n_{\fr{\omega}2-r})  
  \frac{n_{q+r} n_{\fr{\omega}2-q}}{n_r^2}\right) 
  \nn & &\times
    \left[\left(q-\fr{\omega}2\right)^2+\left(r-\fr{\omega}2\right)^2\right]
 \nn
 & \mbox{(\v)}  & + \frac{1}{2} \, 
 \int_{ 0 }^{\infty}
 \! {\rm d}q
 \int_{0}^{ \infty }
 \! {\rm d}r \; 
 \left(n_{q+r} - n_{q+\frac{\omega}{2}} + 
    (1 + n_{q+\frac{\omega}{2}}) 
    \frac{n_{q+r}n_{r+\frac{\omega}{2}}}{n_r^2}\right) 
 \nn & &\times
    \left[\left(q+\fr{\omega}2\right)^2+\left(r+\fr{\omega}2\right)^2\right] \nn
 & \mbox{(\iv)}  & + \,  
 \int_{ \frac{\omega}{2} }^{\infty}
 \! {\rm d}q
 \int_{0}^{
    \frac{2q-\omega}{2}}
 \!\!\!\! {\rm d}r \; 
 \left(n_{q-\frac{\omega}{2}} - n_{q} 
   - n_{q-\frac{\omega}{2}} 
    \frac{(1+n_{q-r})(n_q - n_{r+\frac{\omega}{2}})}
    {n_r n_{-\frac{\omega}{2}}}\right) 
    \nn & &\times
    \left[\left(q-\fr{\omega}2\right)^2+\left(r+\fr{\omega}2\right)^2\right]
 \Biggr\} \,, \la{If1_ps}
\ea
where we have made use of the change of variables introduced in eq.~(\ref{psshift}). Utilizing finally the $\epsilon=0$ relation
\ba
\rho^{ }_{\It{f}{ }} (\omega)&=&-6 \rho^{ }_{\It{f}{1}} (\omega),
\ea
this provides us with the final result for $\rho^{ }_{\It{f}{ }} (\omega)$, 
\ba
\rho^{ }_{\It{f}{ }} (\omega)&=& -\frac{\omega^4} {(4\pi)^3}(1+2 n_{\frac{\omega}{2}})\bigg\{\frac{1}{16}+\tilde{\rho}_{\It{f}{}} (\omega/T)\bigg\}\, .
\ea
Here, the constant term corresponds to the $T=0$ limit of the result, and the function $\tilde{\rho}_{\It{f}{}} (\omega/T)$ is displayed in fig.~\ref{fig3}.


\subsection{\texorpdfstring{$\rho^{ }_{\It{i}{ }}$}{}}

With the i type integrals, we start from the i', which does not appear in the expression for $\rho^{ }_{\It{i}{ }}$, but is required in the evaluation of $\rho^{}_{\mathcal{I}^{3}_\rmii{i}}$.

\subsubsection*{\texorpdfstring{$\rho_{\mathcal{I}^{}_\rmii{i'}}^{ }$}{}}

Proceeding as before, we begin by reading off the (fz,p) contribution from eq.~(\ref{fIip}),
\ba
 && \hspace{-1em}
 \rho_{\mathcal{I}^{}_\rmii{i'}}^{(\rmi{fz,p})}(\omega) \equiv
 \int_{\vec{q,r}} 
 \frac{\omega^2 \pi q }{4 r E_r E_{qr}}
 \delta(\omega - E_{r} - r) 
 \nn && \times
 \biggl(
 -\frac{2}{q} +
 \frac{1}{q+E_{r}+E_{qr}} +  
 \frac{1}{q-E_{r}+E_{qr}} 
 \biggr)
 (1+n_{E_r}+n_r)
  \;,  \la{Iip_T} 
\ea
where the first term vanishes in dimensional regularization and the rest produces
\ba
 \rho_{\mathcal{I}^{}_\rmii{i'}}^{(\rmi{fz,p})}(\omega)
 &=& -\fr{4(D-1)}{D-2}\rho_{\mathcal{I}^{1 }_\rmii{h}}^{(\rmi{fz,p})}(\omega)
 \nonumber \\* & = & 
 \frac{\omega^4\Lambda^{-4\epsilon}} {12(4\pi)^3}
 (1+2n_{\frac{\omega}{2}})\bigg(1-\frac{m^6}{\omega^6}\bigg)
 \nn &&\times
  \Bigg(\frac{1}{\epsilon}
 + \ln \frac{\omega^2\bmu^2}{(\omega^2-m^2)^2}
   + \ln \frac{\bmu^2}{m^2}
   +\frac{25\omega^4+22\omega^2 m^2+25m^4}{6(\omega^4+\omega^2 m^2+m^4)}\Bigg)
  \;.  \la{Iip_fzp} 
\ea
Similarly, the (fz,e) part gives 
\ba
 && \hspace{-1.3cm}
 \rho_{\mathcal{I}^{}_\rmii{i'}}^{(\rmi{fz,e})}(\omega) =\biggl[
  \frac{2\omega}{(4\pi)^3} (1+n_{E_r}+n_r ) \nn
  &\times&  \Biggl\{  
  \int_0^\infty \! {\rm d}q 
  \int_{E_{qr}^-}^{E_{qr}^+} \!\! {\rm d}E_{qr}  \, n_q \, q^2
      \left(\frac{1}{\Delta_{00}} +\frac{1}{\Delta_{10}} -\frac{1}{\Delta_{01}} -\frac{1}{\Delta_{11}}\right)
  \nn &+ & 
  \int_{\lambda}^\infty \! {\rm d}E_{qr}
  \int_{ 
  |r - E_{qr}|}^{r + E_{qr}} \!\! {\rm d}q   \, n_{qr} q^2
    \left(-\frac{4}{q} +\frac{1}{\Delta_{00}} +\frac{1}{\Delta_{01}} +\frac{1}{\Delta_{10}} +\frac{1}{\Delta_{11}}\right)
       \Biggr\} 
  \biggl]_{r=\frac{\omega^2-m^2}{2\omega}} 
  \;,  \la{Iip_fz_e}
\ea
while the (ps) contribution can be related to the corresponding part of the $\rho^{}_{\mathcal{I}^{1}_\rmii{h}}$ integral
\ba
 \rho_{\mathcal{I}^{ }_\rmii{i'}}^{(\rmi{ps})}(\omega) & = & 
 -6\rho_{\mathcal{I}^{1 }_\rmii{h}}^{(\rmi{ps})}(\omega)
  \;, \la{Iip_ps}
\ea
leading us to the final result for the spectral function.

\subsubsection*{\texorpdfstring{$\rho^{}_{\mathcal{I}^{3}_\rmii{i}}$}{}}
With $\mathcal{I}^{3}_\rmii{i}$, it is a straightforward exercise to show that the integrand of the spectral function is related to that of $\rho^{}_{\mathcal{I}^{ }_\rmii{i'}}(\omega)$ through
\be
f_{\mathcal{I}^{3}_\rmii{i}}
 = -\frac{D-2}{D-1}
 \lim_{m\to 0} 
 \Biggl\{
  \frac{{\rm d}}{{\rm d}m^2}
 r^2
 f_{\mathcal{I}^{}_\rmii{i'}}
 \Biggr\}
 - f_{\mathcal{I}^{2}_\rmii{h}} 
 - f_{\mathcal{I}^{2}_\rmii{b}}\; .
\ee
Using this, we immediately reach the expressions
\ba
 \rho_{\mathcal{I}^{3}_\rmii{i}}^{(\rmi{fz,p})}(\omega)+\rho_{\mathcal{I}^{2}_\rmii{h}}^{(\rmi{fz,p})}(\omega)+\rho_{\mathcal{I}^{2}_\rmii{b}}^{(\rmi{fz,p})}(\omega) 
  &=& 
 \frac{\omega^4\Lambda^{-4\epsilon}} {36(4\pi)^3}
 (1+2n_{\frac{\omega}{2}})
 \nn 
 &\times&
  \left(\frac{1}{\epsilon}
 + \ln \frac{\bmu^2}{\omega^2}
   + \ln \frac{\bmu^2}{m^2}
   +\frac{25}{12} +\frac{\omega^2}{2m^2}\right)
 \;, \la{Ii3_fzp} \\
 \rho_{\mathcal{I}^{3}_\rmii{i}}^{(\rmi{fz,e})}(\omega)+\rho_{\mathcal{I}^{2}_\rmii{h}}^{(\rmi{fz,e})}(\omega)+\rho_{\mathcal{I}^{2}_\rmii{b}}^{(\rmi{fz,e})}(\omega) &=&
  -\fr{2}{3}
 \lim_{m\to 0}
 \frac{{\rm d}}{{\rm d}m^2}\Biggl\{ 
 \left(\fr{\omega^2-m^2}{2\omega}\right)^2
 \rho_{\mathcal{I}^{ }_\rmii{i'}}^{(\rmi{fz,e})}(\omega)
 \Biggr\} 
  \;,  \la{Ii3_fz_e} \\
 \rho_{\mathcal{I}^{3}_\rmii{i}}^{(\rmi{ps})}(\omega)+\rho_{\mathcal{I}^{2}_\rmii{h}}^{(\rmi{ps})}(\omega)+\rho_{\mathcal{I}^{2}_\rmii{b}}^{(\rmi{ps})}(\omega)  &=&  
 -\fr{40}{7} \rho_{\mathcal{I}^{6(1)}_\rmii{h}}^{(\rmi{ps})}(\omega) \;, \la{Ii3_ps}
\ea
which in combination with our previous results complete the evaluation of the integral.

\subsubsection*{\texorpdfstring{$\rho^{}_{\mathcal{I}^{1}_\rmii{i}}$}{}}

This time, the (fz,p) contribution takes the form
\ba
 \rho_{\mathcal{I}^{1}_\rmii{i}}^{(\rmi{fz,p})}(\omega) &\equiv&
 \fr{D-2}{D-1}\int_{\vec{q,r}} 
 \frac{\omega \pi q }{16 r E_r E_{qr}}
 \delta(\omega - E_{r} - r) 
 \nn & \times &
 \biggl(
 \frac{2q}{\omega} -
 \frac{\omega+2q}{q+E_{r}+E_{qr}} -  
 \frac{\omega-2q}{q-E_{r}+E_{qr}} 
 \biggr)
 (1+n_{E_r}+n_r)
  \;,  \la{Ii1_T} 
\ea
where the first term vanishes, while the two others produce the result
\ba
 && \hspace{-5.5em}
 \rho_{\mathcal{I}^{1}_\rmii{i}}^{(\rmi{fz,p})}(\omega) 
 =
 -\fr{D-2}{D-1}\biggl[ 
 \frac{\pi r^{D-3}}{2(4\pi)^\frac{D-1}{2}\Gamma(\frac{D-1}{2}) }
  \biggl(
 \omega\left.\int_Q \frac{q^2}{Q^2(Q-R)^2}
 \right|_{R=(E_r i ,r \vec{e}_r )}
  \nn &&
 +2\left. \int_Q \frac{q^2iq_n}{Q^2(Q-R)^2} 
 \right|_{R=(E_r i ,r \vec{e}_r )}
 \biggr)
 (1+n_{E_r}+n_r)\biggl]_{r=\frac{\omega^2-m^2}{2\omega}}
   \la{Ii1_fzpp}  \\
   &=&-\frac{\omega^4\Lambda^{-4\epsilon}} {288(4\pi)^3}
 (1+2n_{\frac{\omega}{2}})\bigg(1-\frac{4m^6}{\omega^6}+\frac{3m^8}{\omega^8}\bigg)
 \nn &&\times
  \Bigg(\frac{1}{\epsilon}
+\ln \frac{\omega^2\bmu^2}{(\omega^2-m^2)^2}
   + \ln \frac{\bmu^2}{m^2}
   +\frac{2(5\omega^4+10\omega^2 m^2+18m^4)}{3(\omega^4+2\omega^2 m^2+3m^4)}  \Bigg)
 \;.  \la{Ii1_fzp} 
\ea
The two remaining parts of $\mathcal{I}^{1}_\rmii{i}$ on the other hand read
\ba
 \rho_{\mathcal{I}^{1}_\rmii{i}}^{(\rmi{fz,e})}(\omega) &=&
 -
  \biggl[
  \frac{1}{3(4\pi)^3} (1+n_{E_r}+n_r ) \Biggl\{  
  \int_0^\infty \! {\rm d}q 
  \int_{E_{qr}^-}^{E_{qr}^+} \!\! {\rm d}E_{qr}  \, n_q \, q^2
  \nn
  &\times & 
      \left(\frac{\omega+2q}{\Delta_{00}} +\frac{\omega-2q}{\Delta_{10}} -\frac{\omega+2q}{\Delta_{01}} -\frac{\omega-2q}{\Delta_{11}}\right)
      + \, 
  \int_{\lambda}^\infty \! {\rm d}E_{qr}
  \int_{ 
  |r - \sqrt{E_{qr}^2-\lambda^2}|
 }^{
  r + \sqrt{E_{qr}^2-\lambda^2}
 } \!\! {\rm d}q   \, n_{qr}\, q^2 
  \nn & \times &  
    \left(-\fr{4q}{\omega}+\frac{\omega+2q}{\Delta_{00}} +\frac{\omega-2q}{\Delta_{10}} +\frac{\omega+2q}{\Delta_{01}} +\frac{\omega-2q}{\Delta_{11}}\right)
       \Biggr\} 
  \biggl]_{r=\frac{\omega^2-m^2}{2\omega}}
  \;,  \la{Ii1_fz_e} \\
 \rho_{\mathcal{I}^{1}_\rmii{i}}^{(\rmi{ps})}(\omega) & = &
 \fr{2}{3} 
 \frac{\omega}{(4\pi)^3} (1+2 n_{\frac{\omega}{2}})  \Biggl\{ 
 \nn 
  & \mbox{(\i)}  &  + \,
 \fr12\int_{\frac{\lambda^2}{2\omega}}^{\frac{\omega}{2}}
 \! {\rm d}q
 \int_{\frac{\lambda^2}{4q}}^{\frac{\omega(\omega-2q)+\lambda^2}{2\omega}} 
 \!\!\!\! {\rm d}r \; 
 \left[\frac{-2q\left(\frac{\omega}{2}-q\right)^2}
  {-2\omega r+m^2} + \frac{-2r\left(\frac{\omega}{2}-r\right)^2}
  {-2\omega q+m^2} \right]
  \nn & & \times
  \left[ 1 + n_{q+r} +  n_{\fr{\omega}2-q}
  +(1+n_{\fr{\omega}2-r})  
  \frac{n_{q+r} n_{\fr{\omega}2-q}}{n_r^2}\right]
 \nn
 & \mbox{(\v)}  & + \, 
 \fr12\int_{ 0 }^{\infty}
 \! {\rm d}q
 \int_{\frac{\lambda^2}{4q}}^{ \infty }
 \! {\rm d}r \; 
 \left[\frac{2q\left(\frac{\omega}{2}+q\right)^2}
  {2\omega r+m^2} + \frac{2r\left(\frac{\omega}{2}+r\right)^2}
  {2\omega q+m^2} \right]
  \nn & & \times
  \left[ n_{q+r} - n_{q+\frac{\omega}{2}} + 
    (1 + n_{q+\frac{\omega}{2}}) 
    \frac{n_{q+r}n_{r+\frac{\omega}{2}}}{n_r^2} \right]
 \nn
 & \mbox{(\iv)}  & + \, 
 \int_{ \frac{\omega}{2} }^{\infty}
 \! {\rm d}q
 \int_{ - \frac{\lambda^2}{4q}}^{
    \frac{\omega(2q-\omega)-\lambda^2}{2\omega}}
 \!\!\!\! {\rm d}r \; 
 \left[\frac{-2q\left(\frac{\omega}{2}-q\right)^2}
  {2\omega r+m^2} + \frac{2r\left(\frac{\omega}{2}+r\right)^2}
  {-2\omega q+m^2} \right]
  \nn & & \times
  \biggl[
 n_{q-\frac{\omega}{2}} - n_{q} 
   - n_{q-\frac{\omega}{2}} 
    \frac{(1+n_{q-r})(n_q - n_{r+\frac{\omega}{2}})}
    {n_r n_{-\frac{\omega}{2}}}  \biggr]
 \Biggr\}
 \;, \la{Ii1_ps}
\ea
where we have made use of the shifts of variables described in eq.~(\ref{psshift}).

\subsubsection*{\texorpdfstring{$\rho^{}_{\mathcal{I}^{2}_\rmii{i}}$}{}}
With the results for $\mathcal{I}^{1}_\rmii{i}$ at hand, it is very straightforward to to obtain the corresponding expressions for $\mathcal{I}^{2}_\rmii{i}$, utilizing the relation
\be
 f_{\mathcal{I}^{2}_\rmii{i}} =
 \lim_{m\to 0} 
 \Biggl\{
  \frac{{\rm d}}{{\rm d}m^2}\omega^2
 f_{\mathcal{I}^{1}_\rmii{i}}
 \Biggr\} \; .
\ee
A simple differentiation now results in the expressions
\ba
 \rho^{(\rmi{fz,p})}_{\mathcal{I}^{2}_\rmii{i}}(\omega) &=&
 -\frac{\omega^4\Lambda^{-4\epsilon}} {144(4\pi)^3}
 (1+2n_{\frac{\omega}{2}})
  \left(1
  -\frac{\omega^2}{2m^2}\right)
 \;,  \la{Ii2_fzp}  \\
 \rho_{\mathcal{I}^{2}_\rmii{i}}^{(\rmi{fz,e})}(\omega) &=&
  \lim_{m\to 0} 
 \Biggl\{
  \frac{{\rm d}}{{\rm d}m^2}\int_{\vec{q,r}}\omega^2
 \rho_{\mathcal{I}^{1}_\rmii{i}}^{(\rmi{fz,e})}(\omega)
 \Biggr\} 
  \;,  \la{Ii2_fz_e} \\
 \rho_{\mathcal{I}^{2}_\rmii{i}}^{(\rmi{ps})}(\omega) & = &
 \lim_{m\to 0} 
 \Biggl\{
  \frac{{\rm d}}{{\rm d}m^2}\int_{\vec{q,r}}\omega^2
 \rho_{\mathcal{I}^{1}_\rmii{i}}^{(\rmi{ps})}(\omega)
 \Biggr\} \;. \la{Ii2_ps}
\ea

\subsubsection*{Final result}
\begin{figure}
\centering
\includegraphics[width=10cm]{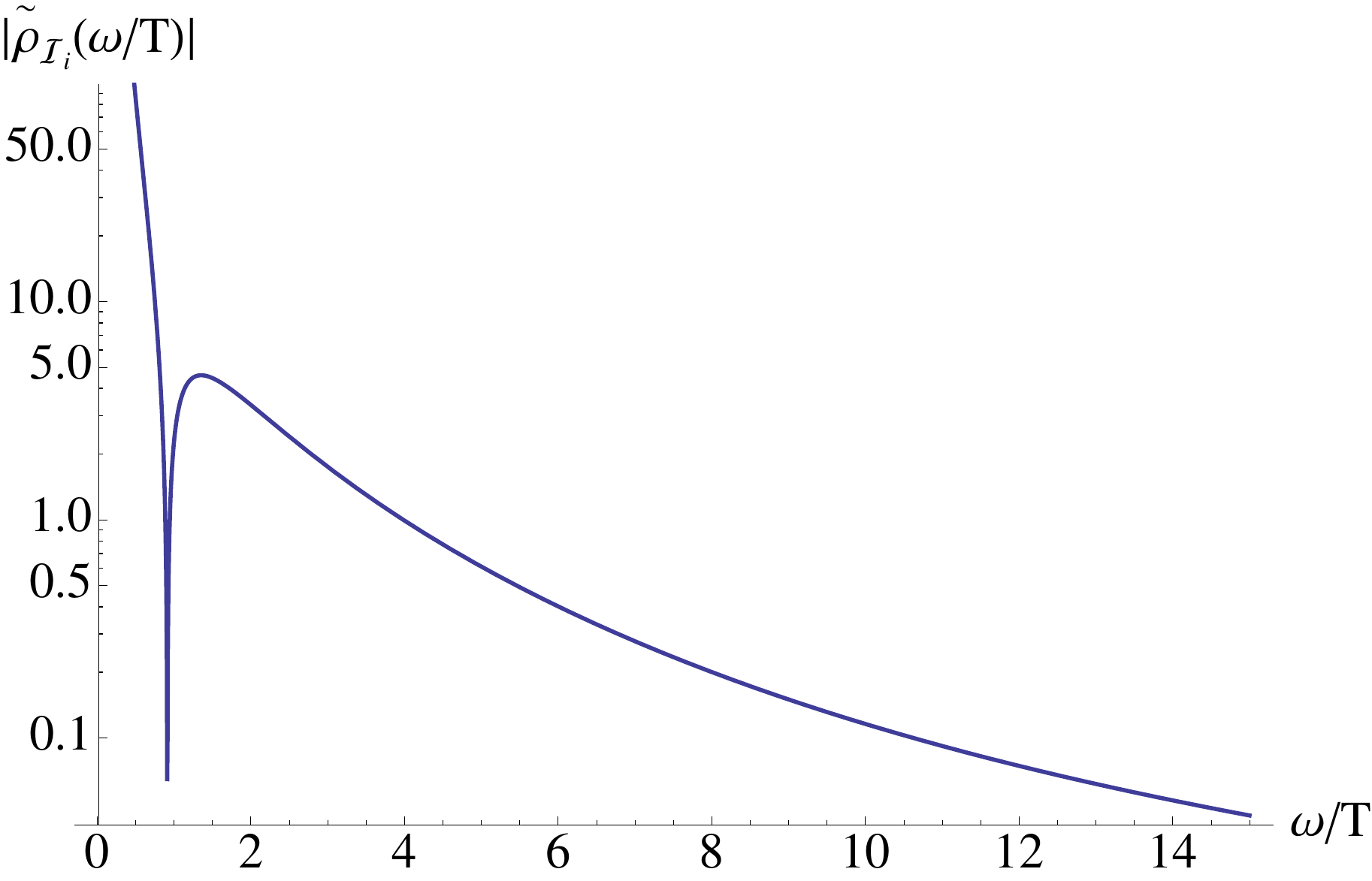}
\caption{The behavior of the function $\tilde{\rho}_{\It{i}{}} (\omega/T)$. The spike corresponds to the sign of the function changing from negative to positive.}
\label{fig4}
\end{figure}

Having evaluated all of the different pieces contributing to $\rho^{}_{\It{i}{ }}$, we are ready to write down our final result for the quantity. Proceeding to the $m\to 0$ limit, we see that all IR divergent terms cancel, leaving the result
\ba
\Lambda^{4\epsilon}\rho^{ }_{\It{i}{ }} (\omega)&=& -\frac{\omega^4} {(4\pi)^3}(1+2 n_{\frac{\omega}{2}})\Bigg\{\frac{1}{4\epsilon}+\fr12\ln\,\frac{\bar{\Lambda}^2}{\omega^2}+\frac{23}{144}-\frac{2\pi^2T^2}{9\omega^2}+\tilde{\rho}_{\It{i}{}} (\omega/T)\Bigg\} , \;\;\;\;
\ea
where the dimensionless function $\tilde{\rho}_{\It{i}{}}$ is displayed in fig.~\ref{fig4}.

\subsection{\texorpdfstring{$\rho^{ }_{\It{j}{ }}$}{}}

The j type integrals represent the most complicated topology encountered in our calculation. A minor simplification occurs, though, due to the fact that no squared propagators appear in these integrals, implying that instead of introducing the mass parameter $m$ as above, we may use the IR regulator $\lambda$ introduced in \cite{Laine:2011xm}. As this parameter appears as the mass of the $Q-R$ propagator, it doesn't break the $Q\leftrightarrow R$ symmetry of the integral, which turns out to simplify its evaluation somewhat. 

We begin the calculation by briefly recalling how the simplest case $\rho_{\mathcal{I}^{0}_\rmii{j}}^{ }$ was dealt with in \cite{Laine:2011xm}; the formulae introduced here will turn out useful also in the other cases considered.

\subsubsection*{\texorpdfstring{$\rho_{\mathcal{I}^{0}_\rmii{j}}^{ }$}{}}

Recalling that the integral $\rho_{\mathcal{I}^{0}_\rmii{j}}^{}$ is UV convergent, we may set $D=4$ from the beginning. The (fz,p), (fz,e) and (ps) parts of the integral can then be directly read off from \cite{Laine:2011xm}, with the results taking the form
\ba
\rho_{\mathcal{I}^{0}_\rmii{j}}^{(\rmi{fz,p})}(\omega)
  &\equiv &
  \int_{\vec{q,r}} 
 \frac{\omega^6 \pi }{16 q r^3 E_{qr}} \Biggl\{ 
 \Bigl[\delta(\omega - 2 r) - \delta(\omega+2 r) \Bigr]
 \nn &\times &  \biggl[
 \biggl(
 \frac{1}{(q+r+E_{qr})(q+r)} -  
 \frac{1}{(q-r+E_{qr})(q-r)} 
 \biggr)
 (1+2n_r)
 \biggr]
  \;, \la{Ij0_fzp}  \\
 \rho_{\mathcal{I}^{0}_\rmii{j}}^{(\rmi{fz,e})}(\omega) & \equiv & 
  \frac{\omega^4}{(4\pi)^3} (1 + 2 n_{\frac{\omega}{2}})
  \int_0^\infty \! {\rm d}q \, n_q
  \int_{ 
    \sqrt{(q - \frac{\omega}{2})^2 + \lambda^2}
   }^{ 
    \sqrt{(q + \frac{\omega}{2})^2 + \lambda^2}
   }
   \!\! {\rm d}E_{qr} \, \Biggl\{ 
  \nn & & \; \; \;
  \mathbbm{P} \, \biggl[ 
  \frac{1}{(q+E_{qr}+\frac{\omega}{2})(q+\frac{\omega}{2})} -  
  \frac{1}{(q+E_{qr}-\frac{\omega}{2})(q-\frac{\omega}{2})}
  \nn & & \hspace*{1cm} + \, 
  \frac{1}{(q-E_{qr}-\frac{\omega}{2})(q-\frac{\omega}{2})} -
  \frac{1}{(q-E_{qr}+\frac{\omega}{2})(q+\frac{\omega}{2})} 
  \biggr] 
  \Biggr\}
  \nn & + &
  \frac{\omega^4}{(4\pi)^3} (1 + 2 n_{\frac{\omega}{2}})
 \int_{\lambda}^{\infty} \! {\rm d}E_{qr} \, n_{E_{qr}} 
 \int_{ 
  |\frac{\omega}{2} - \sqrt{ E_{qr}^2-\lambda^2}|
 }^{
  \frac{\omega}{2} + \sqrt{ E_{qr}^2-\lambda^2}
 }
  \!\! {\rm d}q \, \Biggl\{ 
  \nn & & \; \; \;
  \mathbbm{P} \, \biggl[ 
  \frac{1}{(E_{qr}+q+\frac{\omega}{2})(q+\frac{\omega}{2})} -  
  \frac{1}{(E_{qr}+q-\frac{\omega}{2})(q-\frac{\omega}{2})}
  \nn & & \hspace*{1cm} + \, 
  \frac{1}{(E_{qr}-q+\frac{\omega}{2})(q-\frac{\omega}{2})} -
  \frac{1}{(E_{qr}-q-\frac{\omega}{2})(q+\frac{\omega}{2})} 
  \biggr] 
 \Biggr\}
  \;, \la{Ij0_fze} \\
 \rho_{\mathcal{I}^{0}_\rmii{j}}^{(\rmi{ps})}(\omega) & = & 
 \frac{\omega^4}{2(4\pi)^3} (1 + 2 n_{\frac{\omega}{2}})  \Biggl\{ 
 \nn 
 & \mbox{(\i)}  &  - \,
 \int_{\frac{\lambda^2}{2\omega}}^{\frac{\omega}{2}}
 \! {\rm d}q
 \int_{\frac{\lambda^2}{4q}}^{\frac{\omega(\omega-2q)+\lambda^2}{2\omega}} 
 \!\!\!\! {\rm d}r \; 
 \biggl( \frac{ 
 1 + n_{q+r} +  n_{\fr{\omega}2-q}
  +(1+n_{\fr{\omega}2-r})  
  \frac{n_{q+r} n_{\fr{\omega}2-q}}{n_r^2} } {qr} \biggr)\;
 \nn
 & \mbox{(\v)}  & - \, 
 \int_{ 0 }^{\infty}
 \! {\rm d}q
 \int_{\frac{\lambda^2}{4q}}^{ \infty }
 \! {\rm d}r \; 
 \biggl( \frac{ n_{q+r} - n_{q+\frac{\omega}{2}} + 
    (1 + n_{q+\frac{\omega}{2}}) 
    \frac{n_{q+r}n_{r+\frac{\omega}{2}}}{n_r^2}
 } {qr} \biggr)\;
 \la{Ij0_ps} \\
 & \mbox{(\iv)}  & + \, 
 2\int_{ \frac{\omega}{2} }^{\infty}
 \! {\rm d}q
 \int_{ - \frac{\lambda^2}{4q}}^{
    \frac{\omega(2q-\omega)-\lambda^2}{2\omega}}
 \!\!\!\! {\rm d}r \;
 \biggl( \frac{ n_{q-\frac{\omega}{2}} - n_{q} 
   - n_{q-\frac{\omega}{2}} 
    \frac{(1+n_{q-r})(n_q - n_{r+\frac{\omega}{2}})}
    {n_r n_{-\frac{\omega}{2}}}
  } {qr} \biggr)\;
 \Biggr\} \;, \nonumber
\ea
where we have now denoted $E_{qr}\equiv \sqrt{(\mathbf{q-r})^2+\lambda^2}$. Note also that our notation for the integration variables in the (fz,e) part is somewhat different than that of \cite{Laine:2011xm}.

The evaluation of the above integrals is explained in great detail in appendix A of  \cite{Laine:2011xm}. For the (fz,p) and (fz,e) parts, the expressions can (after some labor) be reduced to one-dimensional integrals over elementary functions, whereas the (ps) part leads to both one- and two-dimensional integrals. The parameter $\lambda$ can be set to zero at the very end, upon an explicit cancelation of the IR divergent terms.

\subsubsection*{\texorpdfstring{$\rho^{}_{\mathcal{I}^{1}_\rmii{j}}$}{}}
In the evaluation of $\rho^{ }_{\mathcal{I}^{1}_\rmii{j}}(\omega)$, our starting point is the simple relation
\ba
f_{\mathcal{I}^{1}_\rmii{j}} =
 -\frac{D-2}{D-1}\frac{q^2}{\omega^2}f_{\mathcal{I}^{0}_\rmii{j}}\;.
\ea
Unfortunately, the altered UV behavior of the integrand has the effect that unlike with $\rho^{ }_{\mathcal{I}^{0}_\rmii{j}}(\omega)$, we cannot set $D=4$ yet, but rather have to first subtract an analytically evaluatable UV divergent part from the rest. To this end, we first consider the (fz,p) contribution, where this subtraction must be implemented, obtaining
\ba
 && \hspace*{-2.0cm} \rho_{\mathcal{I}^{1}_\rmii{j}}^{(\rmi{fz,p})}(\omega)
  \equiv 
  -\frac12\frac{D-2}{D-1} \int_{\vec{q,r}} 
 \frac{\omega^4 \pi (q^2+r^2)}{16 q r^3 E_{qr}} \Biggl\{ 
 \Bigl[\delta(\omega - 2 r) - \delta(\omega+2 r) \Bigr]
 \times 
 \nn & & \times \biggl[
 \biggl(
 \frac{1}{(q+r+E_{qr})(q+r)} -  
 \frac{1}{(q-r+E_{qr})(q-r)} 
 \biggr)
 (1+2n_r)
 \biggr]
 \nn &=& 
  -\fr12\frac{D-2}{D-1} \fr{\pi}{(4\pi)^{\fr{D-1}{2}} \Gamma\left(\fr{D-1}{2}\right)} \left(\frac{\omega}{2}\right)^{D-1}  (1 + 2 n_{\frac{\omega}{2}})
  \int_{\vec{q}}   \Biggl\{ \fr{q}{E_{qr}} \times \la{Ij1_fzp} 
  \\ & & \times
  \, \biggl[
 \frac{1}{(q+\frac{\omega}{2}+E_{qr})(q+\frac{\omega}{2})} -  
 \frac{1}{(q-\frac{\omega}{2}+E_{qr})(q-\frac{\omega}{2})} 
 \biggr]  \Biggr\}  
 -\fr18\frac{D-2}{D-1}\rho_{\mathcal{I}^{0}_\rmii{j}}^{(\rmi{fz,p})}(\omega)
  \;.  \nonumber 
\ea
The most convenient way of evaluating the new integral involves the use of the identity
\ba
&&\hspace{-3em}
\bigg\{\int_{Q}\fr{-i q_n q^2}{Q^2[(Q-R)^2+\lambda^2](Q-P)^2} \bigg\}_{R=(-i\fr{\omega}{2},\fr{\omega}{2}\vec{e_r})}
\nn&=&
\int_{\vec{q}}   \Biggl\{ \fr{q}{8 E_{qr}}   \, \biggl[
 \frac{1}{(q+\frac{\omega}{2}+E_{qr})(q+\frac{\omega}{2})} -  
 \frac{1}{(q-\frac{\omega}{2}+E_{qr})(q-\frac{\omega}{2})} 
 \biggr]
  \Biggr\} \; ,
\ea
where we have taken advantage of the fact that $\mathbf{p}=0$. Using this relation, we obtain
\ba
 && \hspace*{-1cm} \rho_{\mathcal{I}^{1}_\rmii{j}}^{(\rmi{fz,p})}(\omega)
  = 
  -\fr18\frac{D-2}{D-1}\rho_{\mathcal{I}^{0}_\rmii{j}}^{(\rmi{fz,p})}(\omega)
  -\fr12\frac{D-2}{D-1} \fr{8\pi}{(4\pi)^{\fr{D-1}{2}} \Gamma\left(\fr{D-1}{2}\right)} \left(\frac{\omega}{2}\right)^{D-1}  (1 + 2 n_{\frac{\omega}{2}})
  \nn & & \times
  \bigg\{\int_{Q}\fr{-i q_n q^2}{Q^2[(Q-R)^2+\lambda^2](Q-P)^2} \bigg\}
  {\begin{array}{l}
 _{P=(-i\omega,\vec{0})} \\
 _{R=(-i\fr{\omega}{2},\fr{\omega}{2}\vec{e_r})}
\end{array}}
  \;, \la{Ij1_fzpr}  
\ea
where the UV divergence is now contained in a simple one-loop $T=0$ integral. 

To evaluate the above integral, we add and subtract from its integrand the leading terms of a large-$Q^2$ expansion, writing
\ba
&&\int_{Q}\fr{-i q_n q^2}{Q^2[(Q-R)^2+\lambda^2](Q-P)^2}
  = 
 \left\{\int_{Q}\fr{-i q_n q^2}{Q^2[(Q-R)^2+\lambda^2](Q-P)^2}
 -\mathcal{I}^{1,UV}_\rmii{j}
 \right\} +\mathcal{I}^{1,UV}_\rmii{j}, \nonumber \\
 &&
 \; \label{q2Ij}  \\
&&\mathcal{I}^{1,UV}_\rmii{j} = \int_{Q}\fr{-i q_n q^2}{(Q^2+m_r^2)(Q-P)^2}\left[\fr{1}{Q^2+m_r^2}+\fr{2Q\cdot R}{(Q^2+m_r^2)^2}\right]
\;, \la{Ij1uv}
\ea
where $m_r$ is an in principle arbitrary regulatory mass. After this subtraction, the integral inside the curly brackets becomes finite and can be carried out with the same methods we used when dealing with $\rho_{\mathcal{I}^{0}_\rmii{j}}^{(\rmi{fz,p})}(\omega)$, cf.~\cite{Laine:2011xm}. The subtraction term, on the other hand, is evaluated analytically. Here, a potential problem originates from a pinch singularity at $q=\fr{\omega^2-m_r^2}{2\omega}$, which can however be moved to the limit $q=0$ by the choice $m_r=\omega$, where it is protected by the integration measure. After a straightforward exercise, these steps lead us to the result
\be
\left[ \mathcal{I}^{1,UV}_\rmii{j} \right] {\begin{array}{l}
 _{P=(-i\omega,\vec{0})} \\
 _{R=(-i\fr{\omega}{2},\fr{\omega}{2}\vec{e_r})}
\end{array}}
=-\frac{3 \omega \Lambda^{-2\epsilon} }{128  \pi ^2}\left(\fr{1}{\epsilon}+\ln\fr{\bar{\Lambda}^2}{\omega ^2}
\right)
\;. \la{q2IjUV}
\ee
 
With the (fz,e) and (ps) parts, no divergences appear, and we may proceed as with $\mathcal{I}^{0}_\rmii{j}$. This quickly leads us to the results
\ba
 \rho_{\mathcal{I}^{1}_\rmii{j}}^{(\rmi{fz,e})}(\omega) & \equiv & 
  -\fr13\frac{\omega^2}{(4\pi)^3} (1 + 2 n_{\frac{\omega}{2}})
  \int_0^\infty \! {\rm d}q \, n_q
  \int_{ 
    \sqrt{(q - \frac{\omega}{2})^2 + \lambda^2}
   }^{ 
    \sqrt{(q + \frac{\omega}{2})^2 + \lambda^2}
   }
   \!\! {\rm d}E_{qr} \, q^2 \Biggl\{ 
  \nn & & \; \; \;
  \mathbbm{P} \, \biggl[ 
  \frac{1}{(q+E_{qr}+\frac{\omega}{2})(q+\frac{\omega}{2})} -  
  \frac{1}{(q+E_{qr}-\frac{\omega}{2})(q-\frac{\omega}{2})}
  \nn & & \hspace*{1cm} + \, 
  \frac{1}{(q-E_{qr}-\frac{\omega}{2})(q-\frac{\omega}{2})} -
  \frac{1}{(q-E_{qr}+\frac{\omega}{2})(q+\frac{\omega}{2})} 
  \biggr] 
  \Biggr\}
  \nn &  &
  -\fr13\frac{\omega^2}{(4\pi)^3} (1 + 2 n_{\frac{\omega}{2}})
 \int_{\lambda}^{\infty} \! {\rm d}E_{qr} \, n_{E_{qr}} 
 \int_{ 
  |\frac{\omega}{2} - \sqrt{ E_{qr}^2-\lambda^2}|
 }^{
  \frac{\omega}{2} + \sqrt{ E_{qr}^2-\lambda^2}
 }
  \!\! {\rm d}q \, q^2 \Biggl\{ 
  \nn & & \; \; \;
  \mathbbm{P} \, \biggl[ 
  \frac{1}{(E_{qr}+q+\frac{\omega}{2})(q+\frac{\omega}{2})} -  
  \frac{1}{(E_{qr}+q-\frac{\omega}{2})(q-\frac{\omega}{2})}
  \nn & & \hspace*{1cm} + \, 
  \frac{1}{(E_{qr}-q+\frac{\omega}{2})(q-\frac{\omega}{2})} -
  \frac{1}{(E_{qr}-q-\frac{\omega}{2})(q+\frac{\omega}{2})} 
  \biggr] 
 \Biggr\}
   \nn &  &
  -\frac{1}{12}\rho_{\mathcal{I}^{0}_\rmii{j}}^{(\rmi{fz,e})}(\omega)
  \;, \la{Ij1_fze} 
\\
 \rho_{\mathcal{I}^{1}_\rmii{j}}^{(\rmi{ps})}(\omega) & = & 
 -\frac{\omega^2}{3(4\pi)^3} (1 + 2 n_{\frac{\omega}{2}})  \Biggl\{ 
 \nn 
 & \mbox{(\i)}  &  - \,
 \int_{\frac{\lambda^2}{2\omega}}^{\frac{\omega}{2}}
 \! {\rm d}q
 \int_{\frac{\lambda^2}{4q}}^{\frac{\omega(\omega-2q)+\lambda^2}{2\omega}} 
 \!\!\!\! {\rm d}r \; 
 \biggl( \frac{ 
 1 + n_{q+r} +  n_{\fr{\omega}2-q}
  +(1+n_{\fr{\omega}2-r})  
  \frac{n_{q+r} n_{\fr{\omega}2-q}}{n_r^2} } {qr} \biggr)\;
  \left(q-\fr{\omega}{2}\right)^2
 \nn
 & \mbox{(\v)}  & - \, 
 \int_{ 0 }^{\infty}
 \! {\rm d}q
 \int_{\frac{\lambda^2}{4q}}^{ \infty }
 \! {\rm d}r \; 
 \biggl( \frac{ (n_{q+r} - n_{q+\frac{\omega}{2}} + 
    (1 + n_{q+\frac{\omega}{2}}) 
    \frac{n_{q+r}n_{r+\frac{\omega}{2}}}{n_r^2}
 } {qr} \biggr)\;
 \left(q+\fr{\omega}{2}\right)^2
 \nn
 & \mbox{(\iv)}  & + \, 
 \int_{ \frac{\omega}{2} }^{\infty}
 \! {\rm d}q
 \int_{ - \frac{\lambda^2}{4q}}^{
    \frac{\omega(2q-\omega)-\lambda^2}{2\omega}}
 \!\!\!\! {\rm d}r \;
 \biggl( \frac{ n_{q-\frac{\omega}{2}} - n_{q} 
   - n_{q-\frac{\omega}{2}} 
    \frac{(1+n_{q-r})(n_q - n_{r+\frac{\omega}{2}})}
    {n_r n_{-\frac{\omega}{2}}}
  } {qr} \biggr)\;
  \nn && \times
  \left[\left(q-\fr{\omega}{2}\right)^2+\left(r+\fr{\omega}{2}\right)^2\right]
 \Biggr\} \;. \la{Ij1_ps}
\ea

\subsubsection*{\texorpdfstring{$\rho^{ }_{\mathcal{I}^{2}_\rmii{j}}$}{}}
With $\rho^{ }_{\mathcal{I}^{2}_\rmii{j}}$, we start from the relation
\ba
f_{\mathcal{I}^{2}_\rmii{j}} =
 -\frac{D-2}{D-1}\frac{E_{qr}^2-\lambda^2}{\omega^2}f_{\mathcal{I}^{0}_\rmii{j}}\;,
\ea
which immediately produces the (fz,p) contribution
\ba
 && \hspace*{-1.6cm} \rho_{\mathcal{I}^{2}_\rmii{j}}^{(\rmi{fz,p})}(\omega)
  \equiv  
  -\frac{D-2}{D-1}\frac{\omega^2}{(4\pi)^3}  (1 + 2 n_{\frac{\omega}{2}})
  \int_0^\infty \! {\rm d}q 
  \int_{E_{qr}^-}^{E_{qr}^+} \!\! {\rm d}E_{qr} \left(E_{qr}^2-\lambda^2\right) \Biggl\{ 
  \nn & & \; \; \;
 \mathbbm{P} \, \biggl[
 \frac{1}{(q+\frac{\omega}{2}+E_{qr})(q+\frac{\omega}{2})} -  
 \frac{1}{(q-\frac{\omega}{2}+E_{qr})(q-\frac{\omega}{2})} 
 \biggr]
  \Biggr\}  
  \nn &&
  =-\frac{D-2}{D-1} \fr{8\pi}{(4\pi)^{\fr{D-1}{2}} \Gamma\left(\fr{D-1}{2}\right)} \left(\frac{\omega}{2}\right)^{D-1}  (1 + 2 n_{\frac{\omega}{2}})
    \times
  \nn & & \times
  \bigg\{\int_{Q}\fr{-i q_n (q^2-2\vec{q}\cdot\vec{r}+r^2)}{Q^2[(Q-R)^2+\lambda^2](Q-P)^2} \bigg\}
  {\begin{array}{l}
 _{P=(-i\omega,\vec{0})} \\
 _{R=(-i\fr{\omega}{2},\fr{\omega}{2}\vec{e_r})}
\end{array}}
  \;. \la{Ij2_fzp}  
\ea
Of the three terms on the last line here, only the middle one is both UV divergent and new. For it, we again write

\ba
&&\hspace{-4.5em}\int_{Q}\fr{-i q_n \vec{q}\cdot\vec{r}}{Q^2[(Q-R)^2+\lambda^2](Q-P)^2} 
\nonumber \\* &=&
\bigg\{\int_{Q}\fr{-i q_n \vec{q}\cdot\vec{r}}{Q^2[(Q-R)^2+\lambda^2](Q-P)^2}
-
\mathcal{I}^{2,UV}_\rmii{j} \bigg\}
+\mathcal{I}^{2,UV}_\rmii{j}
  \;, \la{qrIj} \\
\mathcal{I}^{2,UV}_\rmii{j}&=&\int_{Q}\fr{-i q_n \vec{q}\cdot\vec{r}}{(Q^2+\omega^2)(Q-P)^2}\fr{1}{Q^2+\omega^2}
\;, \la{Ij2uv}
\ea
where the UV subtraction term clearly vanishes in the limit of our interest,
\ba
  \left[\mathcal{I}^{2,UV}_\rmii{j}\right]
  {\begin{array}{l}
 _{P=(-i\omega,\vec{0})} \\
 _{R=(-i\fr{\omega}{2},\fr{\omega}{2}\vec{e_r})}
\end{array}} =0 \;. \la{qrIjUV}
\ea

The evaluation of the (fz,e) and (ps) parts of the integral proceeds as before. After some work, we obtain the results 
\ba
 \rho_{\mathcal{I}^{2}_\rmii{j}}^{(\rmi{fz,e})}(\omega) & \equiv & 
  -\frac{2\omega^2}{3(4\pi)^3} (1 + 2 n_{\frac{\omega}{2}})
  \int_0^\infty \! {\rm d}q \, n_q
  \int_{ 
    \sqrt{(q - \frac{\omega}{2})^2 + \lambda^2}
   }^{ 
    \sqrt{(q + \frac{\omega}{2})^2 + \lambda^2}
   }
   \!\! {\rm d}E_{qr} \,\left(E_{qr}^2-\lambda^2\right) \Biggl\{ 
  \nn & & \; \; \;
  \mathbbm{P} \, \biggl[ 
  \frac{1}{(q+E_{qr}+\frac{\omega}{2})(q+\frac{\omega}{2})} -  
  \frac{1}{(q+E_{qr}-\frac{\omega}{2})(q-\frac{\omega}{2})}
  \nn & & \hspace*{1cm} + \, 
  \frac{1}{(q-E_{qr}-\frac{\omega}{2})(q-\frac{\omega}{2})} -
  \frac{1}{(q-E_{qr}+\frac{\omega}{2})(q+\frac{\omega}{2})} 
  \biggr] 
  \Biggr\}
  \nn &  &
  -\frac{2\omega^2}{3(4\pi)^3} (1 + 2 n_{\frac{\omega}{2}})
 \int_{\lambda}^{\infty} \! {\rm d}E_{qr} \, n_{E_{qr}} 
 \int_{ 
  |\frac{\omega}{2} - \sqrt{ E_{qr}^2-\lambda^2}|
 }^{
  \frac{\omega}{2} + \sqrt{ E_{qr}^2-\lambda^2}
 }
  \!\! {\rm d}q \, \left(E_{qr}^2-\lambda^2\right)\Biggl\{ 
  \nn & & \; \; \;
  \mathbbm{P} \, \biggl[ 
  \frac{1}{(E_{qr}+q+\frac{\omega}{2})(q+\frac{\omega}{2})} -  
  \frac{1}{(E_{qr}+q-\frac{\omega}{2})(q-\frac{\omega}{2})}
  \nn & & \hspace*{1cm} + \, 
  \frac{1}{(E_{qr}-q+\frac{\omega}{2})(q-\frac{\omega}{2})} -
  \frac{1}{(E_{qr}-q-\frac{\omega}{2})(q+\frac{\omega}{2})} 
  \biggr] 
  \Biggr\}
  \;, \la{Ij2_fze} \\
 \rho_{\mathcal{I}^{2}_\rmii{j}}^{(\rmi{ps})}(\omega) & = & 
 -\frac{\omega^2}{3(4\pi)^3} (1 + 2 n_{\frac{\omega}{2}})  \Biggl\{ 
 \nn 
 & \mbox{(\i)}  &  - \,
 \int_{\frac{\lambda^2}{2\omega}}^{\frac{\omega}{2}}
 \! {\rm d}q
 \int_{\frac{\lambda^2}{4q}}^{\frac{\omega(\omega-2q)+\lambda^2}{2\omega}} 
 \!\!\!\! {\rm d}r \; 
 \biggl( \frac{ 
 1 + n_{q+r} +  n_{\fr{\omega}2-q}
  +(1+n_{\fr{\omega}2-r})  
  \frac{n_{q+r} n_{\fr{\omega}2-q}}{n_r^2} } {qr} \biggr)\;\nn
  &&\times
  \left[\left(q+r\right)^2-\lambda^2\right]
 \nn
 & \mbox{(\v)}  & - \, 
 \int_{ 0 }^{\infty}
 \! {\rm d}q
 \int_{\frac{\lambda^2}{4q}}^{ \infty }
 \! {\rm d}r \; 
 \biggl( \frac{ (n_{q+r} - n_{q+\frac{\omega}{2}} + 
    (1 + n_{q+\frac{\omega}{2}}) 
    \frac{n_{q+r}n_{r+\frac{\omega}{2}}}{n_r^2}
 } {qr} \biggr)\;
 \left[\left(q+r\right)^2-\lambda^2\right]
 \nn
 & \mbox{(\iv)}  & + \, 
 2\int_{ \frac{\omega}{2} }^{\infty}
 \! {\rm d}q
 \int_{ - \frac{\lambda^2}{4q}}^{
    \frac{\omega(2q-\omega)-\lambda^2}{2\omega}}
 \!\!\!\! {\rm d}r \;
 \biggl( \frac{ n_{q-\frac{\omega}{2}} - n_{q} 
   - n_{q-\frac{\omega}{2}} 
    \frac{(1+n_{q-r})(n_q - n_{r+\frac{\omega}{2}})}
    {n_r n_{-\frac{\omega}{2}}}
  } {qr} \biggr)\;
  \nn && \times
  \left[\left(q-r\right)^2-\lambda^2\right]
 \Biggr\} \;, \la{Ij2_ps}
\ea
which can again be straightforwardly evaluated with the methods introduced in \cite{Laine:2011xm}.

\subsubsection*{\texorpdfstring{$\rho^{}_{\mathcal{I}^{3}_\rmii{j}}$}{}}
Starting with 
\ba
f_{\mathcal{I}^{3}_\rmii{j}} =
 \frac{D(D-2)}{D^2-1}\frac{q^4}{\omega^4}f_{\mathcal{I}^{0}_\rmii{j}}\;,
\ea
we obtain the (fz,p) contribution to $\rho^{}_{\mathcal{I}^{3}_\rmii{j}}$
\ba
\rho_{\mathcal{I}^{3}_\rmii{j}}^{(\rmi{fz,p})}(\omega)
  &=& 
  \fr12\frac{D(D-2)}{D^2-1} \fr{8\pi}{(4\pi)^{\fr{D-1}{2}} \Gamma\left(\fr{D-1}{2}\right) \omega^2} \left(\frac{\omega}{2}\right)^{D-1}  (1 + 2 n_{\frac{\omega}{2}})
    \times
  \nn &\times & 
  \bigg\{\int_{Q}\fr{-i q_n q^4}{Q^2[(Q-R)^2+\lambda^2](Q-P)^2} \bigg\}
  {\begin{array}{l}
 _{P=(-i\omega,\vec{0})} \\
 _{R=(-i\fr{\omega}{2},\fr{\omega}{2}\vec{e_r})}
\end{array}}
  \Biggr\} 
  \nn &+&
  \frac{1}{32}\frac{D(D-2)}{D^2-1} \rho_{\mathcal{I}^{0}_\rmii{j}}^{(\rmi{fz,p})}(\omega)
  \;. \la{Ij3_fzp}  
\ea
Here, the new UV divergent integral reads
\ba
&&\int_{Q}\fr{-i q_n q^4}{Q^2[(Q-R)^2+\lambda^2](Q-P)^2}
 =
 \left\{\int_{Q}\fr{-i q_n q^4}{Q^2[(Q-R)^2+\lambda^2](Q-P)^2}
 -\mathcal{I}^{3,UV}_\rmii{j} \right\}
 +\mathcal{I}^{3,UV}_\rmii{j}
 \;, \la{q4Ij} \nn
 \\
&&\mathcal{I}^{3,UV}_\rmii{j}=\int_{Q}\fr{-i q_n q^4}{(Q^2+\omega^2)(Q-P)^2}\bigg[\fr{1}{Q^2+\omega^2}+\fr{2Q\cdot R+2\omega^2}{(Q^2+\omega^2)^2}
 + \fr{4(Q\cdot R)^2+6\omega^2Q\cdot R}{(Q^2+\omega^2)^3}
\nn &&\hspace{14.5em}+\fr{8(Q\cdot R)^3}{(Q^2+\omega^2)^4}\bigg]
\;, \la{Ij3uv}
\ea
where the UV part gives
\be
\left[\mathcal{I}^{3,UV}_\rmii{j}\right]
  {\begin{array}{l}
 _{P=(-i\omega,\vec{0})} \\
 _{R=(-i\fr{\omega}{2},\fr{\omega}{2}\vec{e_r})}
\end{array}} =
-\frac{25 \omega ^3 \Lambda^{-2\epsilon}}{3072  \pi ^2}
\left(\fr{1}{\epsilon}+\ln\frac{\bar{\Lambda}^2}{\omega ^2}
-\fr{19}{5}\right)
 \;, \la{q4IjUV}
\ee
and the integrals inside the curly brackets are again finite and can be treated with numerical methods \cite{Laine:2011xm}.

Finally, the remaining contributions to this master obtain the forms
\ba
 \rho_{\mathcal{I}^{3}_\rmii{j}}^{(\rmi{fz,e})}(\omega) & \equiv & 
  \frac{4}{15(4\pi)^3} (1 + 2 n_{\frac{\omega}{2}})
  \int_0^\infty \! {\rm d}q \, n_q
  \int_{ 
    \sqrt{(q - \frac{\omega}{2})^2 + \lambda^2}
   }^{ 
    \sqrt{(q + \frac{\omega}{2})^2 + \lambda^2}
   }
   \!\! {\rm d}E_{qr} \, q^4 \Biggl\{ 
  \nn & & \; \; \;
  \mathbbm{P} \, \biggl[ 
  \frac{1}{(q+E_{qr}+\frac{\omega}{2})(q+\frac{\omega}{2})} -  
  \frac{1}{(q+E_{qr}-\frac{\omega}{2})(q-\frac{\omega}{2})}
  \nn & & \hspace*{1cm} + \, 
  \frac{1}{(q-E_{qr}-\frac{\omega}{2})(q-\frac{\omega}{2})} -
  \frac{1}{(q-E_{qr}+\frac{\omega}{2})(q+\frac{\omega}{2})} 
  \biggr] 
  \Biggr\}
  \nn & + &
 \frac{4}{15(4\pi)^3} (1 + 2 n_{\frac{\omega}{2}})
 \int_{\lambda}^{\infty} \! {\rm d}E_{qr} \, n_{E_{qr}} 
 \int_{ 
  |\frac{\omega}{2} - \sqrt{ E_{qr}^2-\lambda^2}|
 }^{
  \frac{\omega}{2} + \sqrt{ E_{qr}^2-\lambda^2}
 }
  \!\! {\rm d}q \, q^4   \Biggl\{ 
  \nn & & \; \; \;
  \mathbbm{P} \, \biggl[ 
  \frac{1}{(E_{qr}+q+\frac{\omega}{2})(q+\frac{\omega}{2})} -  
  \frac{1}{(E_{qr}+q-\frac{\omega}{2})(q-\frac{\omega}{2})}
  \nn & & \hspace*{1cm} + \, 
  \frac{1}{(E_{qr}-q+\frac{\omega}{2})(q-\frac{\omega}{2})} -
  \frac{1}{(E_{qr}-q-\frac{\omega}{2})(q+\frac{\omega}{2})} 
  \biggr] 
 \Biggr\}
  \nn & + &
  \frac{1}{60} \rho_{\mathcal{I}^{0}_\rmii{j}}^{(\rmi{fz,e})}(\omega)
  \;, \la{Ij3_fze}
\\
 \rho_{\mathcal{I}^{3}_\rmii{j}}^{(\rmi{ps})}(\omega) & = & 
 \frac{4}{15(4\pi)^3} (1 + 2 n_{\frac{\omega}{2}})  \Biggl\{ 
 \nn 
 & \mbox{(\i)}  &  - \,
 \int_{\frac{\lambda^2}{2\omega}}^{\frac{\omega}{2}}
 \! {\rm d}q
 \int_{\frac{\lambda^2}{4q}}^{\frac{\omega(\omega-2q)+\lambda^2}{2\omega}} 
 \!\!\!\! {\rm d}r \; 
 \biggl( \frac{ 
 1 + n_{q+r} +  n_{\fr{\omega}2-q}
  +(1+n_{\fr{\omega}2-r})  
  \frac{n_{q+r} n_{\fr{\omega}2-q}}{n_r^2} } {qr} \biggr)\;
  \left(q-\fr{\omega}{2}\right)^4
 \nn
 & \mbox{(\v)}  & - \, 
 \int_{ 0 }^{\infty}
 \! {\rm d}q
 \int_{\frac{\lambda^2}{4q}}^{ \infty }
 \! {\rm d}r \; 
 \biggl( \frac{ (n_{q+r} - n_{q+\frac{\omega}{2}} + 
    (1 + n_{q+\frac{\omega}{2}}) 
    \frac{n_{q+r}n_{r+\frac{\omega}{2}}}{n_r^2}
 } {qr} \biggr)\;
 \left(q+\fr{\omega}{2}\right)^4
 \nn
 & \mbox{(\iv)}  & + \, 
 \int_{ \frac{\omega}{2} }^{\infty}
 \! {\rm d}q
 \int_{ - \frac{\lambda^2}{4q}}^{
    \frac{\omega(2q-\omega)-\lambda^2}{2\omega}}
 \!\!\!\! {\rm d}r \;
 \biggl( \frac{ n_{q-\frac{\omega}{2}} - n_{q} 
   - n_{q-\frac{\omega}{2}} 
    \frac{(1+n_{q-r})(n_q - n_{r+\frac{\omega}{2}})}
    {n_r n_{-\frac{\omega}{2}}}
  } {qr} \biggr)\;
  \nn && \times
  \left[\left(q-\fr{\omega}{2}\right)^4+\left(r+\fr{\omega}{2}\right)^4\right]
 \Biggr\} \;, \la{Ij3_ps}
\ea
which we deal with as before.

\subsubsection*{\texorpdfstring{$\rho^{ }_{\mathcal{I}^{5}_\rmii{j}}$}{}}
Next, consider $\rho^{ }_{\mathcal{I}^{5}_\rmii{j}}$. For this integral, the defining relation reads
\ba
f_{\mathcal{I}^{5}_\rmii{j}} =
 \frac{1}{\omega^4}\left(\frac{D^2-2 D-2}{D^2-1}q^2r^2+\frac{2}{D^2-1}(\mathbf{q}\cdot\mathbf{r})^2\right)f_{\mathcal{I}^{0}_\rmii{j}}  
 \;,
\ea
leading to the expression
\ba
 && \hspace*{-1.6cm} \rho_{\mathcal{I}^{5}_\rmii{j}}^{(\rmi{fz,p})}(\omega)
  \equiv 
 \frac{D^2-2 D-2}{D^2-1} \fr{8\pi}{(4\pi)^{\fr{D-1}{2}} \Gamma\left(\fr{D-1}{2}\right) \omega^2} \left(\frac{\omega}{2}\right)^{D+1}  (1 + 2 n_{\frac{\omega}{2}})
    \times
  \nn & & \times
  \bigg\{\int_{Q}\fr{-i q_n q^2}{Q^2[(Q-R)^2+\lambda^2](Q-P)^2} \bigg\}_{R=(-i\fr{\omega}{2},\fr{\omega}{2}\vec{e_r})}
  \nn &&
  +\frac{2}{D^2-1} \fr{8\pi}{(4\pi)^{\fr{D-1}{2}} \Gamma\left(\fr{D-1}{2}\right) \omega^2} \left(\frac{\omega}{2}\right)^{D-1}  (1 + 2 n_{\frac{\omega}{2}})
    \times
  \nn & & \times
  \bigg\{\int_{Q}\fr{-i q_n (\vec{q.r})^2}{Q^2[(Q-R)^2+\lambda^2](Q-P)^2} \bigg\}_{R=(-i\fr{\omega}{2},\fr{\omega}{2}\vec{e_r})}
  \;. \la{Ij5_fzp}  
\ea
Here, the first UV divergent integral was evaluated in connection with $\mathcal{I}^{1}_\rmii{j}$, while the second leads to the expression
\ba
&&\int_{Q}\fr{-i q_n (\vec{q.r})^2}{Q^2[(Q-R)^2+\lambda^2](Q-P)^2}
 = 
 \left\{\int_{Q}\fr{-i q_n (\vec{q.r})^2}{Q^2[(Q-R)^2+\lambda^2](Q-P)^2}
 -\mathcal{I}^{5,UV}_\rmii{j}
 \right\}
 +\mathcal{I}^{5,UV}_\rmii{j} \;, \nonumber \\
 && \la{qr2Ij} \\
&&\mathcal{I}^{5,UV}_\rmii{j}=\int_{Q}\fr{-i q_n (\vec{q.r})^2}{(Q^2+\omega^2)(Q-P)^2}\left[\fr{1}{Q^2+\omega^2}+\fr{2Q\cdot R}{(Q^2+\omega^2)^2}\right]
\;, \la{Ij5uv}
\ea
and ultimately to the result
\be
\left[\mathcal{I}^{5,UV}_\rmii{j}\right]
  {\begin{array}{l}
 _{P=(-i\omega,\vec{0})} \\
 _{R=(-i\fr{\omega}{2},\fr{\omega}{2}\vec{e_r})}
\end{array}} =
-\frac{ \omega ^3\Lambda^{-2\epsilon}}{512  \pi ^2}
\left(\fr{1}{\epsilon}+\ln\frac{\bar{\Lambda}^2}{\omega ^2}
+\fr{2}{3}\right)
 \;. \la{qr2IjUV}
\ee

Finally, the (fz,e) and (ps) contributions to the integral read
\ba
 \rho_{\mathcal{I}^{5}_\rmii{j}}^{(\rmi{fz,e})}(\omega) & \equiv & 
   \frac{7\omega^2}{15(4\pi)^3} (1 + 2 n_{\frac{\omega}{2}})
  \int_0^\infty \! {\rm d}q \, n_q
  \int_{ 
    \sqrt{(q - \frac{\omega}{2})^2 + \lambda^2}
   }^{ 
    \sqrt{(q + \frac{\omega}{2})^2 + \lambda^2}
   }
   \!\! {\rm d}E_{qr} \,
   \nn &\times& \left[\frac{2}{5}q^2r^2+\frac{2}{15}\left(\fr{q^2+r^2+\lambda^2-E_{qr}^2}{2}\right)^2\right] \Biggl\{ 
  \nn & & \; \; \;
  \mathbbm{P} \, \biggl[ 
  \frac{1}{(q+E_{qr}+\frac{\omega}{2})(q+\frac{\omega}{2})} -  
  \frac{1}{(q+E_{qr}-\frac{\omega}{2})(q-\frac{\omega}{2})}
  \nn & & \hspace*{1cm} + \, 
  \frac{1}{(q-E_{qr}-\frac{\omega}{2})(q-\frac{\omega}{2})} -
  \frac{1}{(q-E_{qr}+\frac{\omega}{2})(q+\frac{\omega}{2})} 
  \biggr] 
  \Biggr\}
  \nn & + &
  \frac{7\omega^2}{15(4\pi)^3} (1 + 2 n_{\frac{\omega}{2}})
 \int_{\lambda}^{\infty} \! {\rm d}E_{qr} \, n_{E_{qr}} 
 \int_{ 
  |\frac{\omega}{2} - \sqrt{ E_{qr}^2-\lambda^2}|
 }^{
  \frac{\omega}{2} + \sqrt{ E_{qr}^2-\lambda^2}
 }
  \!\! {\rm d}q \, 
  \nn &\times& \left[\frac{2}{5}q^2r^2+\frac{2}{15}\left(\fr{q^2+r^2+\lambda^2-E_{qr}^2}{2}\right)^2\right] \Biggl\{ 
  \nn & & \; \; \;
  \mathbbm{P} \, \biggl[ 
  \frac{1}{(E_{qr}+q+\frac{\omega}{2})(q+\frac{\omega}{2})} -  
  \frac{1}{(E_{qr}+q-\frac{\omega}{2})(q-\frac{\omega}{2})}
  \nn & & \hspace*{1cm} + \, 
  \frac{1}{(E_{qr}-q+\frac{\omega}{2})(q-\frac{\omega}{2})} -
  \frac{1}{(E_{qr}-q-\frac{\omega}{2})(q+\frac{\omega}{2})} 
  \biggr] 
 \Biggr\}
   \;, \la{Ij5_fze}\\
 &&\hspace{-5.2em}\rho_{\mathcal{I}^{5}_\rmii{j}}^{(\rmi{ps})}(\omega)  =  
 \frac{1}{30(4\pi)^3} (1 + 2 n_{\frac{\omega}{2}})  \Biggl\{ 
 \la{Ij5_ps} \\ 
 & \mbox{(\i)}  &  - \,
 \int_{\frac{\lambda^2}{2\omega}}^{\frac{\omega}{2}}
 \! {\rm d}q
 \int_{\frac{\lambda^2}{4q}}^{\frac{\omega(\omega-2q)+\lambda^2}{2\omega}} 
 \!\!\!\! {\rm d}r \; 
 \biggl( \frac{ 
 1 + n_{q+r} +  n_{\fr{\omega}2-q}
  +(1+n_{\fr{\omega}2-r})  
  \frac{n_{q+r} n_{\fr{\omega}2-q}}{n_r^2} } {qr} \biggr)\;
 \nn && \times
 \bigg[6 \left(q-\fr{\omega}{2}\right)^2 \left(r-\fr{\omega}{2}\right)^2 
 +\fr12\left[ \lambda^2- \fr{\omega}{2}\left(r-\fr{\omega}{2}\right)-q\left(r+\fr{\omega}{2}\right) \right]^2 \bigg]
 \nn
 & \mbox{(\v)}  & - \, 
 \int_{ 0 }^{\infty}
 \! {\rm d}q
 \int_{\frac{\lambda^2}{4q}}^{ \infty }
 \! {\rm d}r \; 
 \biggl( \frac{ (n_{q+r} - n_{q+\frac{\omega}{2}} + 
    (1 + n_{q+\frac{\omega}{2}}) 
    \frac{n_{q+r}n_{r+\frac{\omega}{2}}}{n_r^2}
 } {qr} \biggr)\;
 \nn && \times
 \bigg[6 \left(q+\fr{\omega}{2}\right)^2 \left(r+\fr{\omega}{2}\right)^2 
 +\fr12\left[ \lambda^2+ \fr{\omega}{2}\left(r+\fr{\omega}{2}\right)-q\left(r-\fr{\omega}{2}\right) \right]^2 \bigg]
 \nn
 & \mbox{(\iv)}  & + \, 
 2\int_{ \frac{\omega}{2} }^{\infty}
 \! {\rm d}q
 \int_{ - \frac{\lambda^2}{4q}}^{
    \frac{\omega(2q-\omega)-\lambda^2}{2\omega}}
 \!\!\!\! {\rm d}r \;
 \biggl( \frac{ n_{q-\frac{\omega}{2}} - n_{q} 
   - n_{q-\frac{\omega}{2}} 
    \frac{(1+n_{q-r})(n_q - n_{r+\frac{\omega}{2}})}
    {n_r n_{-\frac{\omega}{2}}}
  } {qr} \biggr)\;
   \nn && \times
 \bigg[6 \left(q-\fr{\omega}{2}\right)^2 \left(r+\fr{\omega}{2}\right)^2 
 +\fr12\left[ \lambda^2+ \fr{\omega}{2}\left(r+\fr{\omega}{2}\right)+q\left(r-\fr{\omega}{2}\right) \right]^2 \bigg]
  \Biggr\} \;. \nonumber
\ea

\subsubsection*{\texorpdfstring{$\rho^{ }_{\mathcal{I}^{6}_\rmii{j}}$}{}}
The evaluation of $\rho^{ }_{\mathcal{I}^{6}_\rmii{j}}(\omega)$, defined by
\be
f_{\mathcal{I}^{6}_\rmii{j}} =
 \frac{1}{\omega^4}\left(\frac{D^2-2 D-2}{D^2-1}q^2(E_{qr}^2-\lambda^2)+\frac{2}{D^2-1}(\mathbf{q}\cdot(\mathbf{q-r}))^2\right)f_{\mathcal{I}^{0}_\rmii{j}} \;,
\ee
is somewhat simplified by separating from it two known contributions according to
\ba
 \rho^{ }_{\mathcal{I}^{6}_\rmii{j}}(\omega) = \rho^{ }_{\mathcal{I}^{3}_\rmii{j}}(\omega)
 +\rho^{ }_{\mathcal{I}^{5}_\rmii{j}}(\omega)
 - \frac{2D (D-2)}{D^2-1}
 \int_{\vec{q,r}}  \frac{q^2\vec{q.r}}{\omega^4} f_{\mathcal{I}^{0}_\rmii{j}}
 \la{Ij6_exp}\;.
\ea
This quickly leads to the (fz,p) result
\ba
\rho_{\mathcal{I}^{6}_\rmii{j}}^{(\rmi{fz,p})}(\omega)
  &=& 
  \rho_{\mathcal{I}^{3}_\rmii{j}}^{(\rmi{fz,p})}(\omega)
 +\rho_{\mathcal{I}^{5}_\rmii{j}}^{(\rmi{fz,p})}(\omega)
 \nn & -&  
   \frac{D (D-2)}{D^2-1}
 \fr{8\pi}{(4\pi)^{\fr{D-1}{2}} \Gamma\left(\fr{D-1}{2}\right) \omega^2} \left(\frac{\omega}{2}\right)^{D-1}  (1 + 2 n_{\frac{\omega}{2}})
    \times
  \nn &\times & 
  \bigg\{\int_{Q}\fr{-i q_n (q^2\vec{q.r}+r^2\vec{q.r})}{Q^2[(Q-R)^2+\lambda^2](Q-P)^2} \bigg\}_{R=(-i\fr{\omega}{2},\fr{\omega}{2}\vec{e_r})} 
 \;, \la{Ij6_fzp}  
\ea
where the second term on the last row is furthermore UV convergent. The first term, on the other hand, produces
\ba
&&\int_{Q}\fr{-i q_n q^2 \vec{q.r}}{Q^2[(Q-R)^2+\lambda^2](Q-P)^2}
 =
 \left\{\int_{Q}\fr{-i q_n q^2 \vec{q.r}}{Q^2[(Q-R)^2+\lambda^2](Q-P)^2}
  -\mathcal{I}^{6,UV}_\rmii{j}
 \right\}
 +\mathcal{I}^{6,UV}_\rmii{j}\; , \nonumber \\
 && \la{q2qrIj} \\
&&\mathcal{I}^{6,UV}_\rmii{j}=\int_{Q}\fr{-i q_n q^2 \vec{q.r}}{(Q^2+\omega^2)(Q-P)^2}\left[\fr{1}{Q^2+\omega^2}+\fr{2Q\cdot R+2\omega^2}{(Q^2+\omega^2)^2}+\fr{4(Q\cdot R)^2}{(Q^2+\omega^2)^3}\right]
\;, \la{Ij6uv}
\ea
with the UV divergent part giving 
\be
\left[\mathcal{I}^{6,UV}_\rmii{j}\right]
  {\begin{array}{l}
 _{P=(-i\omega,\vec{0})} \\
 _{R=(-i\fr{\omega}{2},\fr{\omega}{2}\vec{e_r})}
\end{array}} =
-\frac{5 \omega ^3\Lambda^{-2\epsilon}}{1536  \pi ^2}
\left(\fr{1}{\epsilon}+\ln\frac{\bar{\Lambda}^2}{\omega^2}
+\frac{1}{10}\right)
 \;. \la{q2qrIjUV}
\ee

Once again, it is a simple exercise to write down the two last parts of our integral,  
\ba
 \rho_{\mathcal{I}^{6}_\rmii{j}}^{(\rmi{fz,e})}(\omega) & \equiv &
 \rho_{\mathcal{I}^{3}_\rmii{j}}^{(\rmi{fz,e})}(\omega)
 +\rho_{\mathcal{I}^{5}_\rmii{j}}^{(\rmi{fz,e})}(\omega)
 \nn &+& 
  \frac{4}{15(4\pi)^3} (1 + 2 n_{\frac{\omega}{2}})
  \int_0^\infty \! {\rm d}q \, n_q
  \int_{ 
    \sqrt{(q - \frac{\omega}{2})^2 + \lambda^2}
   }^{ 
    \sqrt{(q + \frac{\omega}{2})^2 + \lambda^2}
   }
   \!\! {\rm d}E_{qr} \, 
  \nn & &
  (q^2+\fr{\omega^2}{4})(E_{qr}^2-q^2-\fr{\omega^2}{4}-\lambda^2)\nonumber
  \Biggl\{ 
  \nn &&
  \mathbbm{P} \, \biggl[ 
  \frac{1}{(q+E_{qr}+\frac{\omega}{2})(q+\frac{\omega}{2})} -  
  \frac{1}{(q+E_{qr}-\frac{\omega}{2})(q-\frac{\omega}{2})}
  \nn & & \hspace*{1cm} + \, 
  \frac{1}{(q-E_{qr}-\frac{\omega}{2})(q-\frac{\omega}{2})} -
  \frac{1}{(q-E_{qr}+\frac{\omega}{2})(q+\frac{\omega}{2})} 
  \biggr] 
  \Biggr\}
\nn
  & + &
  \frac{4}{15(4\pi)^3} (1 + 2 n_{\frac{\omega}{2}})
 \int_{\lambda}^{\infty} \! {\rm d}E_{qr} \, n_{E_{qr}} 
 \int_{ 
  |\frac{\omega}{2} - \sqrt{ E_{qr}^2-\lambda^2}|
 }^{
  \frac{\omega}{2} + \sqrt{ E_{qr}^2-\lambda^2}
 }
  \!\! {\rm d}q \, 
   \nn && 
  (q^2+\fr{\omega^2}{4})(E_{qr}^2-q^2-\fr{\omega^2}{4}-\lambda^2)\Biggl\{ 
  \nn & & \; \; \;
  \mathbbm{P} \, \biggl[ 
  \frac{1}{(E_{qr}+q+\frac{\omega}{2})(q+\frac{\omega}{2})} -  
  \frac{1}{(E_{qr}+q-\frac{\omega}{2})(q-\frac{\omega}{2})}
  \nn & & \hspace*{1cm} + \, 
  \frac{1}{(E_{qr}-q+\frac{\omega}{2})(q-\frac{\omega}{2})} -
  \frac{1}{(E_{qr}-q-\frac{\omega}{2})(q+\frac{\omega}{2})} 
  \biggr] 
 \Biggr\}
  \;, \la{Ij6_fze} \\
 \rho_{\mathcal{I}^{6}_\rmii{j}}^{(\rmi{ps})}(\omega)
  &=& 
  \rho_{\mathcal{I}^{3}_\rmii{j}}^{(\rmi{ps})}(\omega)
 +\rho_{\mathcal{I}^{5}_\rmii{j}}^{(\rmi{ps})}(\omega)
 - \frac{8}{15}
 \int_{\vec{q,r}}\frac{q^2}{\omega^4} \left(q^2+r^2+\lambda ^2-E_{qr}^2\right)
  f_{\mathcal{I}^{0}_\rmii{j}}
  \nn & = &
  \rho_{\mathcal{I}^{3}_\rmii{j}}^{(\rmi{ps})}(\omega)
 +\rho_{\mathcal{I}^{5}_\rmii{j}}^{(\rmi{ps})}(\omega)
 -\frac{4}{15(4\pi)^3}(1 + 2 n_{\frac{\omega}{2}})   \Biggl\{ 
 \nn 
 & \mbox{(\i)}  &  - \,
 \int_{\frac{\lambda^2}{2\omega}}^{\frac{\omega}{2}}
 \! {\rm d}q
 \int_{\frac{\lambda^2}{4q}}^{\frac{\omega(\omega-2q)+\lambda^2}{2\omega}} 
 \!\!\!\! {\rm d}r \; 
 \biggl( \frac{ 
 1 + n_{q+r} +  n_{\fr{\omega}2-q}
  +(1+n_{\fr{\omega}2-r})  
  \frac{n_{q+r} n_{\fr{\omega}2-q}}{n_r^2} } {qr} \biggr)\;
  \nn && \times
  \left(q-\fr{\omega}{2}\right)^2 \left[ \lambda^2- \fr{\omega}{2}\left(r-\fr{\omega}{2}\right)-q\left(r+\fr{\omega}{2}\right) \right]
 \nn
 & \mbox{(\v)}  & - \, 
 \int_{ 0 }^{\infty}
 \! {\rm d}q
 \int_{\frac{\lambda^2}{4q}}^{ \infty }
 \! {\rm d}r \; 
 \biggl( \frac{ (n_{q+r} - n_{q+\frac{\omega}{2}} + 
    (1 + n_{q+\frac{\omega}{2}}) 
    \frac{n_{q+r}n_{r+\frac{\omega}{2}}}{n_r^2}
 } {qr} \biggr)\;
  \nn && \times
  \left(q+\fr{\omega}{2}\right)^2 \left[ \lambda^2+ \fr{\omega}{2}\left(r+\fr{\omega}{2}\right)-q\left(r-\fr{\omega}{2}\right) \right]
 \nn
 & \mbox{(\iv)}  & + \, 
 2\int_{ \frac{\omega}{2} }^{\infty}
 \! {\rm d}q
 \int_{ - \frac{\lambda^2}{4q}}^{
    \frac{\omega(2q-\omega)-\lambda^2}{2\omega}}
 \!\!\!\! {\rm d}r \;
 \biggl( \frac{ n_{q-\frac{\omega}{2}} - n_{q} 
   - n_{q-\frac{\omega}{2}} 
    \frac{(1+n_{q-r})(n_q - n_{r+\frac{\omega}{2}})}
    {n_r n_{-\frac{\omega}{2}}}
  } {qr} \biggr)\;
  \nn && \times
  \biggl[\left(q-\fr{\omega}{2}\right)^2 +  \left(r+\fr{\omega}{2}\right)^2 \biggr]
   \left[ \lambda^2+ \fr{\omega}{2}\left(r+\fr{\omega}{2}\right)+q\left(r-\fr{\omega}{2}\right) \right]
 \Biggr\} \;. \la{Ij6_ps}
\ea

\subsubsection*{\texorpdfstring{$\rho^{ }_{\mathcal{I}^{4}_\rmii{j}}$}{}}
The reason we have left the evaluation of $\rho^{}_{\mathcal{I}^{4}_\rmii{j}}(\omega)$ as the last one becomes obvious when one writes down its definition
\ba
f_{\mathcal{I}^{4}_\rmii{j}} =
 \frac{D(D-2)}{D^2-1}\frac{(E_{qr}^2-\lambda^2)^2}{\omega^4}f_{\mathcal{I}^{0}_\rmii{j}}\;.
\ea
in terms of the variables $\mathbf{q}$ and $\mathbf{r}$. For the (fz,p) part, we namely obtain
\ba
 \rho_{\mathcal{I}^{4}_\rmii{j}}^{(\rmi{fz,p})}(\omega)
  &\equiv&  
  \frac{D(D-2)}{D^2-1}\frac{1}{(4\pi)^3}  (1 + 2 n_{\frac{\omega}{2}})
  \int_0^\infty \! {\rm d}q 
  \int_{E_{qr}^-}^{E_{qr}^+} \!\! {\rm d}E_{qr} \Biggl\{ \left(E_{qr}^2-\lambda^2\right)^2
  \nn & & \; \; \;
 \mathbbm{P} \, \biggl[
 \frac{1}{(q+\frac{\omega}{2}+E_{qr})(q+\frac{\omega}{2})} -  
 \frac{1}{(q-\frac{\omega}{2}+E_{qr})(q-\frac{\omega}{2})} 
 \biggr]
  \Biggr\} 
  \nonumber
  \nn
  &=&
  \frac{D(D-2)}{D^2-1} \fr{8\pi}{(4\pi)^{\fr{D-1}{2}} \Gamma\left(\fr{D-1}{2}\right) \omega^2} \left(\frac{\omega}{2}\right)^{D-1}  (1 + 2 n_{\frac{\omega}{2}})
    \la{Ij4_fzp} \\
    & & \times
  \bigg\{\int_{Q}\fr{-i q_n (q^4-4 q^2 \vec{q}\cdot \vec{r}+4 (\vec{q}\cdot \vec{r})^2+2 q^2 r^2-4 \vec{q}\cdot \vec{r}\; r^2+r^4)}{Q^2[(Q-R)^2+\lambda^2](Q-P)^2} \bigg\}
  {\begin{array}{l}
 _{P=(-i\omega,\vec{0})} \\
 _{R=(-i\fr{\omega}{2},\fr{\omega}{2}\vec{e_r})}
\end{array}}
  \;,   \nonumber 
\ea
where each of the UV divergent integrals in the one-loop expression inside the curly backets has been evaluated above, cf.~eqs.~(\ref{q2Ij}),  (\ref{qrIj}), (\ref{q4Ij}), (\ref{qr2Ij}), and (\ref{q2qrIj}).

With the (fz,e) and (ps) contributions, the calculation proceeds as in the previous cases, producing
\ba
 \rho_{\mathcal{I}^{4}_\rmii{j}}^{(\rmi{fz,e})}(\omega) & \equiv & 
  \frac{8}{15(4\pi)^3} (1 + 2 n_{\frac{\omega}{2}})
  \int_0^\infty \! {\rm d}q \, n_q
  \int_{ 
    \sqrt{(q - \frac{\omega}{2})^2 + \lambda^2}
   }^{ 
    \sqrt{(q + \frac{\omega}{2})^2 + \lambda^2}
   }
   \!\! {\rm d}E_{qr} \, \left(E_{qr}^2-\lambda^2\right)^2 \Biggl\{ 
  \nn & & \; \; \;
  \mathbbm{P} \, \biggl[ 
  \frac{1}{(q+E_{qr}+\frac{\omega}{2})(q+\frac{\omega}{2})} -  
  \frac{1}{(q+E_{qr}-\frac{\omega}{2})(q-\frac{\omega}{2})}
  \nn & & \hspace*{1cm} + \, 
  \frac{1}{(q-E_{qr}-\frac{\omega}{2})(q-\frac{\omega}{2})} -
  \frac{1}{(q-E_{qr}+\frac{\omega}{2})(q+\frac{\omega}{2})} 
  \biggr] 
  \Biggr\}
  \nn & + &
 \frac{8}{15(4\pi)^3} (1 + 2 n_{\frac{\omega}{2}})
 \int_{\lambda}^{\infty} \! {\rm d}E_{qr} \, n_{E_{qr}} 
 \int_{ 
  |\frac{\omega}{2} - \sqrt{ E_{qr}^2-\lambda^2}|
 }^{
  \frac{\omega}{2} + \sqrt{ E_{qr}^2-\lambda^2}
 }
  \!\! {\rm d}q \, \left(E_{qr}^2-\lambda^2\right)^2 \Biggl\{ 
  \nn & & \; \; \;
  \mathbbm{P} \, \biggl[ 
  \frac{1}{(E_{qr}+q+\frac{\omega}{2})(q+\frac{\omega}{2})} -  
  \frac{1}{(E_{qr}+q-\frac{\omega}{2})(q-\frac{\omega}{2})}
  \nn & & \hspace*{1cm} + \, 
  \frac{1}{(E_{qr}-q+\frac{\omega}{2})(q-\frac{\omega}{2})} -
  \frac{1}{(E_{qr}-q-\frac{\omega}{2})(q+\frac{\omega}{2})} 
  \biggr] 
 \Biggr\}
  \;, \la{Ij4_fze} \\
 \rho_{\mathcal{I}^{4}_\rmii{j}}^{(\rmi{ps})}(\omega) & = & 
 \frac{4}{15(4\pi)^3} (1 + 2 n_{\frac{\omega}{2}})  \Biggl\{ 
 \nn 
 & \mbox{(\i)}  &  - \,
 \int_{\frac{\lambda^2}{2\omega}}^{\frac{\omega}{2}}
 \! {\rm d}q
 \int_{\frac{\lambda^2}{4q}}^{\frac{\omega(\omega-2q)+\lambda^2}{2\omega}} 
 \!\!\!\! {\rm d}r \; 
 \biggl( \frac{ 
 1 + n_{q+r} +  n_{\fr{\omega}2-q}
  +(1+n_{\fr{\omega}2-r})  
  \frac{n_{q+r} n_{\fr{\omega}2-q}}{n_r^2} } {qr} \biggr)\;
  \nn && \times
  \left[\left(q+r\right)^2-\lambda^2\right]^2
 \nn
 & \mbox{(\v)}  & - \, 
 \int_{ 0 }^{\infty}
 \! {\rm d}q
 \int_{\frac{\lambda^2}{4q}}^{ \infty }
 \! {\rm d}r \; 
 \biggl( \frac{ (n_{q+r} - n_{q+\frac{\omega}{2}} + 
    (1 + n_{q+\frac{\omega}{2}}) 
    \frac{n_{q+r}n_{r+\frac{\omega}{2}}}{n_r^2}
 } {qr} \biggr)\;
 \nn &&\times
 \left[\left(q+r\right)^2-\lambda^2\right]^2
 \nn
 & \mbox{(\iv)}  & + \, 
 2\int_{ \frac{\omega}{2} }^{\infty}
 \! {\rm d}q
 \int_{ - \frac{\lambda^2}{4q}}^{
    \frac{\omega(2q-\omega)-\lambda^2}{2\omega}}
 \!\!\!\! {\rm d}r \;
 \biggl( \frac{ n_{q-\frac{\omega}{2}} - n_{q} 
   - n_{q-\frac{\omega}{2}} 
    \frac{(1+n_{q-r})(n_q - n_{r+\frac{\omega}{2}})}
    {n_r n_{-\frac{\omega}{2}}}
  } {qr} \biggr)\;
  \nn && \times
  \left[\left(q-r\right)^2-\lambda^2\right]^2
 \Biggr\} \;. \la{Ij4_ps}
\ea

\subsubsection*{Final result}
\begin{figure}
\centering
\includegraphics[width=10cm]{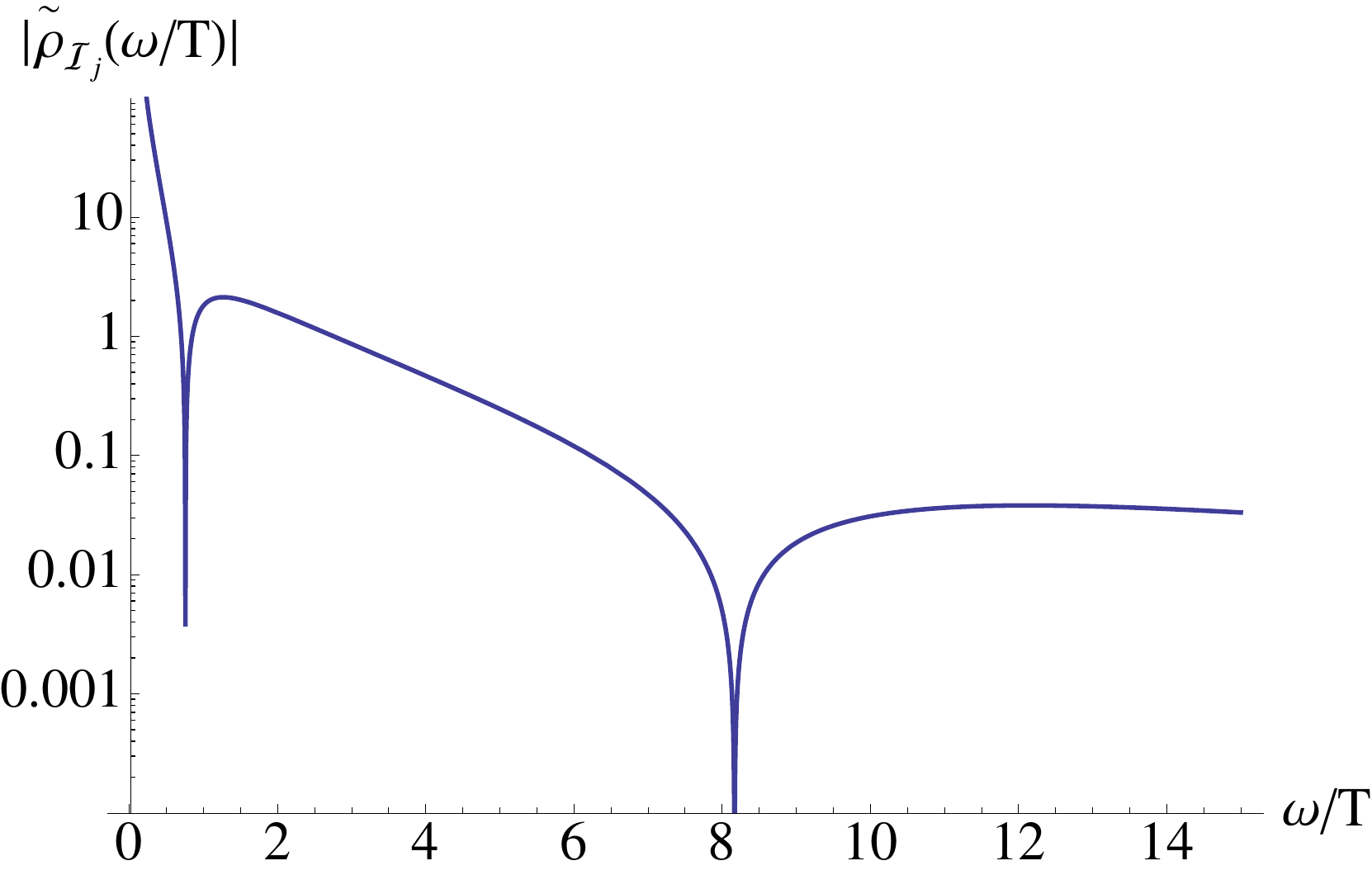}
\caption{The behavior of the absolute value of the function $\tilde{\rho}_{\It{j}{}} (\omega/T)$. The two spikes correspond to the sign of $\tilde{\rho}_{\It{j}{}}$ changing first from positive to negative, and then positive again.}
\label{fig5}
\end{figure}

In analogy with the other master integrals, the only thing left is to collect our final result for the function $\rho^{}_{\It{j}{ }}$. Adding up all the analytic parts and performing the numerics for the remaining integrals, we get
\ba
\Lambda^{4\epsilon}\rho^{ }_{\It{j}{ }} (\omega)&=& \frac{\omega^4} {(4\pi)^3}(1+2 n_{\frac{\omega}{2}})\Bigg\{\frac{23}{18\epsilon}+\frac{46}{18}\ln\,\frac{\bar{\Lambda}^2}{\omega^2}+\frac{41}{27}-\tilde{\rho}_{\It{j}{}} (\omega/T)\Bigg\}\, ,
\ea
while the behavior of $\tilde{\rho}_{\It{j}{}}$ is displayed in fig.~\ref{fig5}.

\end{appendix}

\end{document}